\documentclass[journal]{IEEEtran}

\usepackage{graphicx}%
\usepackage{multirow}%
\usepackage{amsmath,amssymb,amsfonts}%
\usepackage{amsthm}%
\usepackage{mathrsfs}%
\usepackage{xcolor}%
\usepackage{textcomp}%
\usepackage{manyfoot}%
\usepackage{booktabs}%
\usepackage{algorithm}%
\usepackage{algorithmicx}%
\usepackage{algpseudocode}%
\usepackage{listings}%

\usepackage{amsthm,url}
\newtheoremstyle{mytheoremstyle} 
{\topsep}                    
{\topsep}                    
{}                           
{}                           
{\it}                   
{}                           
{.5em}                       
{\textcolor{black}{\thmname{#1}\thmnumber{ #2}\thmnote{ (#3):}}}  
\theoremstyle{mytheoremstyle}

\newtheorem{proposition}{Proposition}

\newtheorem{definition}{Definition}%

\usepackage{xfrac}
\usepackage{graphicx}
\usepackage{subfig}
\usepackage{bm}
\usepackage{wrapfig}
\usepackage{cancel}
\usepackage{float}
\usepackage{tikz}
\usepackage{ifthen}
\usepackage{xparse}
\usepackage{xfp}
\usetikzlibrary{calc,shadows,positioning}

\ExplSyntaxOn
\NewDocumentCommand{\NewWeirdCommand}{mm}
{
	\cs_new:Npn #1 ##1
	{
		\tl_set:Nn \l__simon_args_tl { ##1 }
		#2
	}
}
\NewDocumentCommand{\Arg}{m}
{
	\tl_item:Nn \l__simon_args_tl { #1 }
}

\ExplSyntaxOff

\newcommand{\pc}[1]{\ifthenelse{\equal{#1}{-}}{#1}{#1}}
\newcommand\nc[1]{\ifthenelse{\equal{#1}{0}}{#1}{#1}}

\definecolor{colNull}{RGB}{235 235 235}
\definecolor{colFirst}{RGB}{0 255 0}
\definecolor{colSecond}{RGB}{120 255 120}
\definecolor{colThird}{RGB}{240 255 170}
\definecolor{colFourth}{RGB}{255 150 145}
\definecolor{colFifth}{RGB}{255 30 30}

\newcommand{\setcolor}[1]{%
    \ifnum\fpeval{#1 =0} = 1
	colNull
	\else
	\ifnum\fpeval{#1 =1} = 1
	colFirst
	\else
	\ifnum\fpeval{#1 = 2} = 1
	colSecond
	\else
	\ifnum\fpeval{#1 = 3} = 1
	colThird
	\else
	\ifnum\fpeval{#1 = 4} = 1
	colFourth
	\else
	\ifnum\fpeval{#1 = 5} = 1
	colFifth
	\else
	white
	\fi
	\fi
	\fi
	\fi
	\fi
    \fi
}
\NewWeirdCommand{\createcell}{%
	\begin{tikzpicture}[baseline={([yshift=-.8ex]current bounding box.center)},cellstyle/.style={minimum width=10mm,minimum height=1cm,inner sep=0pt,rounded corners=2},pstyle/.style={font =\tiny,inner sep=0pt,minimum height=0.25cm,minimum width=0.9cm,align=right},nstyle/.style={font=\tiny,inner sep=0pt,minimum height=0.25cm,minimum width=0.6cm}]%
		\node[cellstyle, fill opacity=0.35,fill=\setcolor{\Arg{11}},text opacity=1] (main) at (0,0) {\Arg{1}};%
		\node[pstyle,anchor=north west] (p1) at ($(main.north east)!0!(main.south east)$) {\pc{\Arg{2}}};%
		\node[pstyle,anchor=north west] (p2) at ($(main.north east)!0.5!(main.south east)$) {\pc{\Arg{3}}};%
		\node[pstyle,anchor=north west] (p3) at ($(main.north east)!0.75!(main.south east)$) {\pc{\Arg{4}}};%
		\node[pstyle,anchor=north west] (p4) at ($(main.north east)!0.25!(main.south east)$) {\pc{\Arg{5}}};%
		\node[nstyle,anchor=north west] (n1) at (p1.north east) {\nc{\Arg{6}}};%
		\node[nstyle,anchor=north west] (n2) at (p2.north east) {\nc{\Arg{7}}};%
		\node[nstyle,anchor=north west] (n3) at (p3.north east) {\nc{\Arg{8}}};%
		\node[nstyle,anchor=north west] (n4) at (p4.north east) {\bf {\color{purple} \nc{\Arg{9}}}};%
		\node[draw=none,anchor=north east,font=\tiny,inner sep=1,outer sep=1] (r) at (main.north east) {\Arg{10}};
		\draw[] (p1.north east) -- (p3.south east);%
	\end{tikzpicture}%
}
\NewWeirdCommand{\createoutlinecell}{%
	\begin{tikzpicture}[baseline={([yshift=-.8ex]current bounding box.center)},cellstyle/.style={minimum width=10mm,minimum height=1cm,inner sep=0pt,rounded corners=2},pstyle/.style={font =\tiny,inner sep=0pt,minimum height=0.25cm,minimum width=0.9cm,align=right},nstyle/.style={font=\tiny,inner sep=0pt,minimum height=0.25cm,minimum width=0.6cm}]%
		\node[cellstyle, line width = 1mm, opacity=0.35, draw=\setcolor{\Arg{2}},text opacity=1] (main) at (0,0) {\Arg{1}};%
	\end{tikzpicture}%
}
\NewWeirdCommand{\createcellannotated}{%
	\begin{tikzpicture}[baseline={([yshift=-.8ex]current bounding box.center)},cellstyle/.style={minimum width=10mm,minimum height=1cm,inner sep=0pt,rounded corners=2},pstyle/.style={font =\tiny,inner sep=0pt,minimum height=0.25cm,minimum width=0.9cm,align=right},nstyle/.style={font=\tiny,inner sep=0pt,minimum height=0.25cm,minimum width=0.6cm},descstyle/.style={font=\tiny,draw=none,minimum height=0.25cm, minimum width=1cm, inner sep=1pt}]%
		\node[cellstyle, fill opacity=0.35,fill=\setcolor{\Arg{11}},text opacity=1] (main) at (0,0) {\Arg{1}};%
		\node[pstyle,anchor=north west,shift={(2mm,0)}] (p1) at ($(main.north east)!0!(main.south east)$) {\pc{\Arg{2}}};%
		\node[pstyle,anchor=north west,shift={(2mm,0)}] (p2) at ($(main.north east)!0.5!(main.south east)$) {\pc{\Arg{3}}};%
		\node[pstyle,anchor=north west,shift={(2mm,0)}] (p3) at ($(main.north east)!0.75!(main.south east)$) { \pc{\Arg{4}} };%
		\node[pstyle,anchor=north west,shift={(2mm,0)}] (p4) at ($(main.north east)!0.25!(main.south east)$) {\pc{\Arg{5}}};%
		\node[nstyle,anchor=north west] (n1) at (p1.north east) {\nc{\Arg{6}}};%
		\node[nstyle,anchor=north west] (n2) at (p2.north east) {\nc{\Arg{7}}};%
		\node[nstyle,anchor=north west] (n3) at (p3.north east) {\nc{\Arg{8}}};%
		\node[nstyle,anchor=north west] (n4) at (p4.north east) {\bf {\color{purple} \nc{\Arg{9}} } };%
		\node[draw=none,anchor=north east,font=\tiny,inner sep=1,outer sep=1] (r) at (main.north east) {\Arg{10}};
		\draw[] (p1.north east) -- (p3.south east);%
		\node[descstyle,right=4mm of n1,font=\small] (dn1) {\scalebox{0.56}{number of proper estimates $ N^{(c)}\big(\pi^{0:K}_{\star} \big)$~\eqref{eq:number-of-properly-estimated-objects}}};
		\node[descstyle,right=4mm of n2,font=\small] (dn2) {\scalebox{0.56}{number of missed objects $\mathcal{M}\big(\mathbf{X},\!\mathbf{Y},\pi^{0:K}_{\star} \big)$~\eqref{eq:TGOSPA:MissObjTerm}}};
		\node[descstyle,right=4mm of n3,font=\small] (dn3) {\scalebox{0.56}{number of false alarms $\mathcal{F}\big(\mathbf{X},\!\mathbf{Y},\pi^{0:K}_{\star} \big)$~\eqref{eq:TGOSPA:FalseAlarmTerm}}}; 
		\node[descstyle,right=4mm of n4,font=\small] (dn4) {\scalebox{0.56}{number of switches $S\big(\pi^{0:K}_{\star} \big)$~\eqref{eq:TGOSPA:SwitchTerm}}};
		\foreach \i in {1,...,4}{
			\draw[gray,->] (dn\i) -- (n\i);
		}
		\node[descstyle,left=14mm of p1,font=\small] (dp1) {\scalebox{0.56}{localization term $L_p^{(c)}\big(\mathbf{X},\!\mathbf{Y},\pi^{0:K}_{\star}\big)$~\eqref{eq:TGOSPA:LocTerm}}};
		\node[descstyle,left=14mm of p2,font=\small] (dp2) {\scalebox{0.56}{missed targets term \scalebox{1}{$\tfrac{c^p}{2}$}$ \cdot \big|\mathcal{M}\big(\mathbf{X},\!\mathbf{Y},\pi^{0:K}_{\star}\big) \big|$~\eqref{eq:TGOSPA:withDecomposition}}};
		\node[descstyle,left=14mm of p3,font=\small] (dp3) {\scalebox{0.56}{false alarms term \scalebox{1}{$\tfrac{c^p}{2}$}$ \cdot \big|\mathcal{F}\big(\mathbf{X},\!\mathbf{Y},\pi^{0:K}_{\star}\big) \big|$~\eqref{eq:TGOSPA:withDecomposition}}}; 
		\node[descstyle,left=14mm of p4,font=\small] (dp4) {\scalebox{0.56}{switch term $\gamma^p S\big(\pi^{0:K}_{\star} \big)$~\eqref{eq:TGOSPA:definition}, \eqref{eq:TGOSPA:withDecomposition}}};
		\foreach \i in {1,...,4}{
			\draw[gray] (dp\i) -- (main.west|-dp\i);
			\draw[gray,dotted] (main.west|-dp\i) -- (main.east|-dp\i);
			\draw[gray,->] (main.east|-dp\i) -- ($(p\i.west)+(0.0mm,0mm)$);
		}
		\coordinate (mainl) at ($(main.north)+(-3.2mm,0)$);
		\node[descstyle,above=1.6mm of mainl,inner sep=0pt, outer sep =-0mm,font=\small] (dmain) {\hspace{-28mm}\scalebox{0.56}{value of TGOSPA metric $d_p^{(c,\gamma)}(\mathbf{X},\!\mathbf{Y})$~\eqref{eq:TGOSPA:definition}}};
		\draw[gray,->](dmain) -- ($(mainl.center)-(0,3.5mm)$);
		\node[descstyle,above=2.8mm of r.east,font=\small] (rl) {\hspace{27mm}\scalebox{0.56}{$p$-average localization error $\avgLocError\big(\mathbf{X},\!\mathbf{Y},\pi^{0:K}_{\star}\big)$~\eqref{eq:p-average-loc-error}}};
		\draw[gray,->](rl.south-|r) -- (r);  
		\node[below=1mm of dp3.south west,shift={(19.0mm,0)},descstyle] (cclabel) {column order of algorithms:};
		\node[right=2mm of cclabel,fill opacity=0.35,rounded corners=1,fill=\setcolor{1}] (cc1) {} rectangle ++(2mm,2mm);
		\node[descstyle,right=-3.5mm of cc1] (cc1desc) {1st}; 
		\node[right=-1.5mm of cc1desc,fill opacity=0.35,rounded corners=1,fill=\setcolor{2}] (cc2) {} rectangle ++(2mm,2mm);
		\node[descstyle,right=-3.5mm of cc2] (cc2desc) {2nd}; 
		\node[right=-1.5mm of cc2desc,fill opacity=0.35,rounded corners=1,fill=\setcolor{3}] (cc3) {} rectangle ++(2mm,2mm);
		\node[descstyle,right=-3.5mm of cc3] (cc3desc) {3rd}; 
		\node[right=-1.5mm of cc3desc,fill opacity=0.35,rounded corners=1,fill=\setcolor{4}] (cc4) {} rectangle ++(2mm,2mm);
		\node[descstyle,right=-3.5mm of cc4] (cc4desc) {4th}; 
		\draw[draw=gray,->] (cc2) -- (cc2|-main.south);
	\end{tikzpicture}%
}

\newcommand{\X}{\mathcal{X}}

\newcommand{\baseMetric}{d}
\newcommand{\GOSPAatMostOne}{\baseMetric_p^{(c)}}
\newcommand{\avgLocError}{\overline{\baseMetric_p}}

\newcommand{\T}{^{\top}}

\newcommand{\FT}{\mathcal{F}( \mathcal{T}(\X) )}
\newcommand{\kk}{\mathrm{t}}
\newcommand{\ksp}{\hspace{-0.02cm}}

\newcommand{\dimIndex}{ \mathrm{i} }

\newcommand{\centerPoint}[2]{ {c}_{#2}^{#1} }
\newcommand{\widthCoord}[1]{ {w}^{#1} }
\newcommand{\heightCoord}[1]{ {h}^{#1} }
\newcommand{\leftEndPoint}[2]{ {l}_{#2}^{#1} }
\newcommand{\rightEndPoint}[2]{ {r}_{#2}^{#1} }

\newcommand{\pistarHOTA}{ \pi^{0:K}_{*,\alpha} }

\newcommand{\elementOfTrajectory}{object instance}
\newcommand{\seeminglySwitch}[1]{track change#1}
\newcommand{\seeminglySwitchCAPITAL}[1]{Track Change#1}

\def\GammaZero{\textbf{Gamma zero}}
\def\GammaSmall{\textbf{Gamma small}}
\def\GammaLarge{\textbf{Gamma large}}
\def\GammaExtreme{\textbf{Gamma extreme}}

\theoremstyle{mytheoremstyle}

\newtheorem{ObservationX}[definition]{Observation}

\begin{document}

\IEEEoverridecommandlockouts
\title{
	TGOSPA Metric Parameters Selection and Evaluation for Visual Multi-object Tracking
	\thanks{
		This research was partially supported by the European Union under the project ROBOPROX (reg. no. CZ.02.01.01/00/22\_008/0004590).
	}
}

\author{
	\IEEEauthorblockN{Jan Krejčí\IEEEauthorrefmark{1}, Oliver Kost\IEEEauthorrefmark{1}, Ondřej Straka\IEEEauthorrefmark{1}, Yuxuan Xia\IEEEauthorrefmark{2}, Lennart Svensson\IEEEauthorrefmark{3}, Ángel F. García-Fernández\IEEEauthorrefmark{4}}\\
	\IEEEauthorblockA{
		\IEEEauthorrefmark{1} \textit{Department of Cybernetics, University of West Bohemia in Pilsen}, Pilsen, Czech Republic
	}\\
	\IEEEauthorblockA{
		\IEEEauthorrefmark{2}\textit{Department of Automation, Shanghai Jiaotong University}, Shanghai, China
	}\\
	\IEEEauthorblockA{
		\IEEEauthorrefmark{3}\textit{Signal Processing Group, Chalmers University of Technology}, Göteborg, Sweden
	}\\
	\IEEEauthorblockA{
		\IEEEauthorrefmark{4}\textit{IPTC, ETSI de Telecomunicaci\'on, Universidad Polit\'ecnica de Madrid}, Madrid, Spain
	}\\
	\IEEEauthorblockA{
		Email: \{ jkrejci, kost, straka30 \}@kky.zcu.cz, yuxuan.xia@sjtu.edu.cn, lennart.svensson@chalmers.se, angel.garcia.fernandez@upm.es
	}
}

\maketitle


%
%
%
%
%
%
%
%
%


\begin{abstract}
	Multi-object tracking algorithms are deployed in various applications, each with different performance requirements. For example, track switches pose significant challenges for offline scene understanding, as they hinder the accuracy of data interpretation. Conversely, in online surveillance applications, their impact is often minimal. This disparity underscores the need for application-specific performance evaluations that are both simple and mathematically sound. The trajectory generalized optimal sub-pattern assignment (TGOSPA) metric offers a principled approach to evaluate multi-object tracking performance. It accounts for localization errors, the number of missed and false objects, and the number of track switches, providing a comprehensive assessment framework. This paper illustrates the effective use of the TGOSPA metric in computer vision tasks, addressing challenges posed by the need for application-specific scoring methodologies. By exploring the TGOSPA parameter selection, we enable users to compare, comprehend, and optimize the performance of algorithms tailored for specific tasks, such as target tracking and training of detector or re-ID modules. 
\end{abstract}



\maketitle

\section{Introduction}\label{sec:introduction}
Estimating the number and locations of objects appearing in a given surveillance area is addressed by algorithms for object detection and tracking, see~\cite{BlackmanPopoli:1999,Bar-Shalom-et.al:2011,MTT-Survey:2015,SetsOfTrajectories:2016}.
Their development and implementation have significant potential in various fields, including aerial and naval security as discussed by~\cite{BlackmanPopoli:1999}, medical applications by~\cite{Krejci:Feature-based:2022}, and space situational awareness by~\cite{Faber:MTT-Space:2016}, among others.
This paper mainly focuses on applications that utilize computer vision (CV), though most results presented are generally applicable.
In particular, this paper considers the case when a single monocular camera is used to perform detection/tracking with objects represented by two-dimensional bounding boxes.
Such applications are essential for public safety monitoring, autonomous driving, and many others.

Evaluating the detection and tracking algorithms is key for their convenient selection for particular applications and, thus, their development.
The selection should, however, consider application-specific needs that usually differ among applications.
While there are many aspects one could consider when evaluating algorithm performance as~\cite{Coraluppi:Evaluation:2023}, such as estimation consistency, computational demands, or numerical stability, this paper focuses solely on the evaluation based on empirical data.
Algorithm results, specifically point estimates (mere bounding boxes), must be obtained for applications where ground truth data, also known as annotations, are available.
The results of the algorithm and the ground truth data are then \emph{compared} to each other using a performance evaluation function that yields a single \emph{value}.
To be able to reason among multiple algorithms based on the corresponding evaluation results, the evaluation should capture the efficiency of the algorithms.
In particular, the evaluation should be able to differentiate between algorithms producing different results and clearly justify the difference.

In the CV field, the performance evaluation is usually based on computing \emph{scores}, further called CV scores.
The scores measure similarity, i.e., the higher the score an algorithm achieves the better, for convenience denoted by ``($\uparrow$)''.
The \emph{Multiple object tracking accuracy} (MOTA), \emph{Higher order tracking accuracy} (HOTA), and \emph{Identity F1} (IDF1) scores are often considered authoritative in CV literature, see~\cite{MOTA-Bernardin:2008aa,MOT-16:2016,HOTA:2021,MPNTrack:2020}.
They are also listed as the first three scores in the MOT17 benchmark website in~\cite{MOT17-webpage:2023}.
In particular, HOTA was shown to solve several known problems encountered in the other CV scores~\cite{HOTA:2021}.

In the radar tracking field, performance evaluation often relies on \emph{(mathematical) metrics}.
Metrics measure dissimilarity, i.e., the lower the metric value an algorithm achieves, the better ``($\downarrow$)''.
Metrics satisfy the identity, symmetry, and triangle inequality axioms~\cite{Apostol_book74}.
All the axioms can be useful in practice.
The identity guarantees that reaching the ultimate goal (designing an algorithm whose outputs mach the ground truth exactly) yields a particular metric value: zero.
If multiple annotators are employed to yield independent "ground truth" data, their mutual consistency can be measured by a metric thanks to its symmetry property.
Alternatively, two tracking algorithms can be compared to each other without the need to interpret either of them as ground truth thanks to symmetry.
The triangle inequality is perhaps the most significant axiom as it offers the notion of transitiveness~\cite{OSPA-T:2011}: if some algorithm \emph{A} performs \emph{well} (i.e., its results are close in the metric to the ground truth) and the output of some other algorithm \emph{B} is \emph{close} to that of \emph{A}, we can conclude that \emph{B} performs also \emph{well}.
Although the CV scores such as MOTA, IDF1 and HOTA are commonly referred to as \emph{metrics} in the CV community, they fail to fulfill the identity and triangle inequality even if re-defined to measure dissimilarity ($\downarrow$), see~\cite[Supplementary material]{Nguyen:Visual-metrics-trustworthy:2023}.

To address the inconvenience, this paper proposes to use the \emph{Trajectory generalized optimal sub-pattern assignment} (TGOSPA) metric introduced by~\cite{TrajectoryGOSPA:2020,TimeWeightedTGOSPA:2021}.
Similarly to the CV scores, TGOSPA assigns ground truth objects to estimates at each time step and penalizes \emph{(i)} the distance between pairs of assigned objects and estimates, \emph{(ii)} the number of missed objects, \emph{(iii)} number of false estimates, and \emph{(iv)} the number of track switches\footnote{
	In the CV community, the term \emph{identity} switch is used more often.
}.
Although most CV scores capture some of the TGOSPA metric properties, their definitions are rather heuristic. 
TGOSPA, on the other hand, penalizes all these different quantities in a principled mathematical manner by being a metric as proven by~\cite{TrajectoryGOSPA:2020,TimeWeightedTGOSPA:2021}.

Different applications may allow, e.g., different distance errors or different tolerances for track switches.
TGOSPA introduces \mbox{(hyper-)parameters} that can reflect various user preferences to \emph{tailor} the evaluation to an application at hand.
The parameters include \emph{(I)} a \emph{cut-off} parameter setting the maximum possible distance between ground truth and an estimate, \emph{(II)} an \emph{exponent} parameter that penalizes outliers, and \emph{(III)} a \emph{switching penalty} that penalizes track switches.
The parameters must be selected before the evaluation.
In the literature, however, the effects of the parameter selection are rarely discussed, except for their general interpretation.
To alleviate this, this paper explores several rules for the convenient general selection of the parameters.

Note that there are several alternatives to TGOSPA in the literature, see~\cite{BentoMetrics2020,OSPA2:2020,OSPA-T:2011,BetterOSPA-T:2014}.
In particular, the favorable properties of TGOSPA compared to the~\cite{BentoMetrics2020} metrics were analyzed by~\cite{TrajectoryGOSPA:2020}.
The version of \emph{Optimal sub-pattern assignment metric} (OSPA) called ``OSPA$^{\!(2)}$'' by~\cite{OSPA2:2020} does not penalize all quantities \emph{(i)}-\emph{(iv)} mentioned above. 
The same is true for the OSPA for tracks (OSPAT) by~\cite{OSPA-T:2011}, which is, moreover, not a metric and was analyzed by~\cite{BetterOSPA-T:2014}. 
In addition, the OSPA for multiple tracks (OSPAMT) was introduced by~\cite{BetterOSPA-T:2014}, which is, however, computationally intractable for most practical problems as indicated by~\cite{OSPA2:2020} and does not have a clear interpretation in terms of quantities \emph{(i)}-\emph{(iv)}. 

In this paper, we introduce TGOSPA as a principled metric for CV multi-object tracking evaluation and provide guidelines for selecting its parameters.
The key contributions of the paper are as follows:
\begin{itemize}
    \item It is shown that HOTA and ``1--HOTA'' are not mathematical metrics.
	\item The TGOSPA metric is introduced in the context of CV.
	\item The effects of TGOSPA parameters are revealed in general, including their graphical interpretation. 
	\item A \emph{method} for the TGOSPA parameters selection is proposed and exemplified for CV.
	\item Evaluation examples illustrate the impact of the different TGOSPA parameters, facilitating easier parameter selection in practice.
    \item Three setups of TGOSPA parameters are recommended for practice for \emph{1)} detector training, \emph{2)} online surveillance and \emph{3)} offline scene understanding. Example evaluation of state-of-the-art tracking algorithms is included.
	\item Illustrative examples highlight the differences between TGOSPA and HOTA.
\end{itemize}

The outline of the paper is summarized as follows.
Section~\ref{sec:motivation} introduces a visual tracking example and motivates the need for a convenient performance evaluation metric.
The TGOSPA metric is then introduced in Section~\ref{sec:TGOSPA_metric}, together with the general explanation of its parameter effects.
Application-dependent selection of the parameters is then discussed in Section~\ref{sec:TGOSPA:CV} and performance evaluation examples are given in Section~\ref{sec:numerical:examples}.
Recommendations for practice are given in Section~\ref{sec:practical-evaluation}, including example evaluations.
TGOSPA is then compared to HOTA in Section~\ref{sec:HOTA:vs:TGOSPA} and the paper concludes in Section~\ref{sec:Conclusion}.

\section{Motivation}\label{sec:motivation}
This section first presents the scenario and algorithms that will be used to analyze the performance measures. 
Drawbacks of current CV scores follow.

\subsection{Scenario}\label{sec:scenario}
Consider the MOT17-09 video from the MOT17 dataset by~\cite{MOT-16:2016} available online at~\cite{MOT17-webpage:2023}, see also~\cite{Dendorfer:MOTChallenge:2021}.
Between time steps $k_0{=}382$ and $k_F{=}442$ in that video (61 frames, 2 seconds\footnote{
	The frame rate for the MOT17-09 video is $30$ fps.
}), two pedestrians being annotated with the ground truth IDs $2$ and $6$, denoted as \emph{gt2} and \emph{gt6}, respectively, cross each other.
This leads to a challenging occlusion scenario depicted in Fig.~\ref{fig:MO17-09-scenario}.
In this paper, five selected algorithms are to be evaluated in this scenario using several different performance scores.
Note that only the two-dimensional bounding boxes in the image frame are considered in the evaluation.
The algorithms and a short description of their corresponding tracking results follow.

\newlength{\cropLeft}\setlength{\cropLeft}{8.35cm} 
\newlength{\cropBottom}\setlength{\cropBottom}{2.25cm} 
\newlength{\cropRight}\setlength{\cropRight}{8.38cm} 
\newlength{\cropTop}\setlength{\cropTop}{0.9cm} 
\newlength{\snapWidth}\setlength{\snapWidth}{0.235\textwidth} 
\begin{figure*}
	\vspace{-0.4cm}
	\centering
	\subfloat[$k=382$ (start of the scenario)]{
		\includegraphics[width=\snapWidth,trim={\cropLeft, \cropBottom, \cropRight, \cropTop},clip]{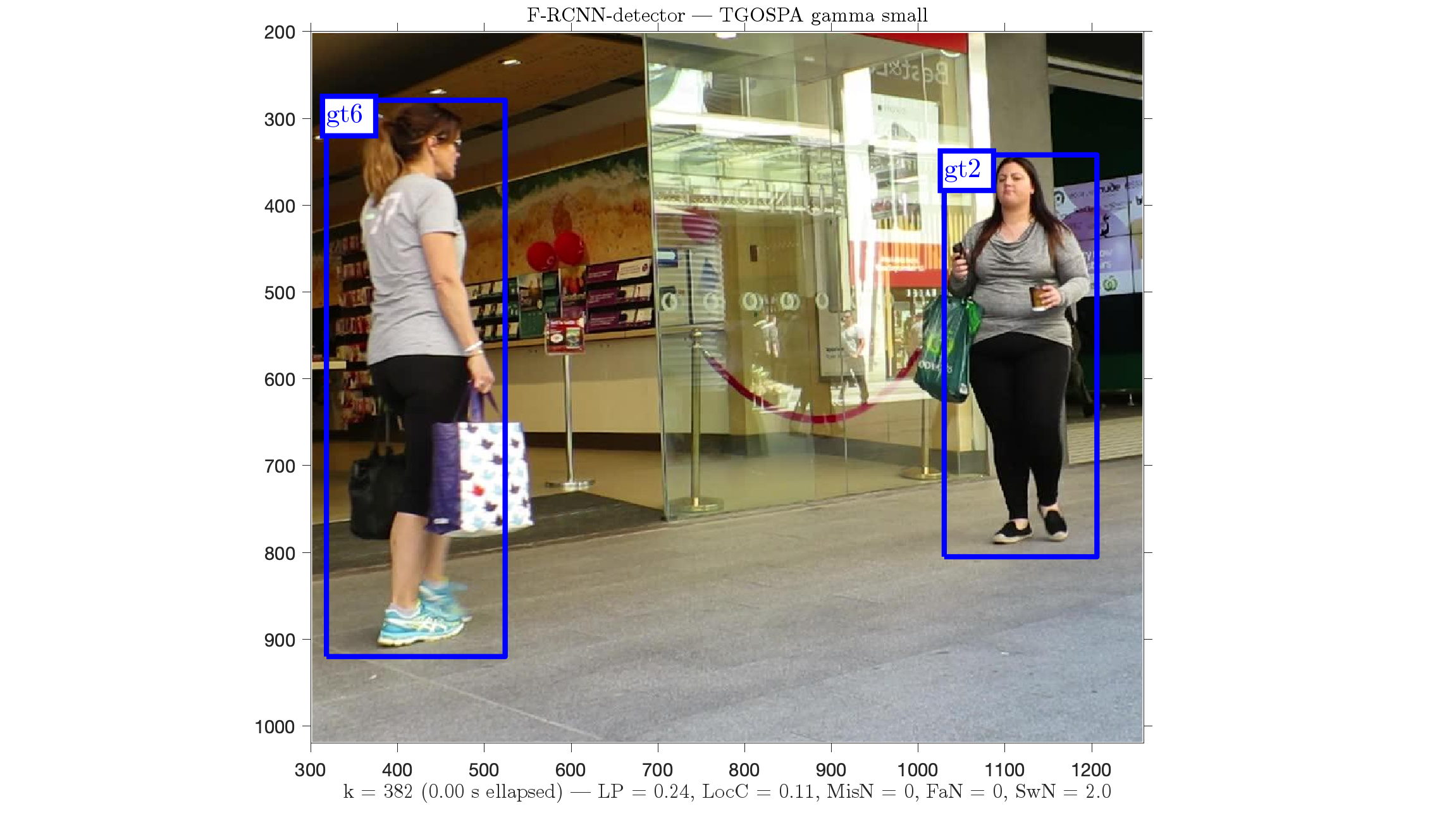}    }
	\subfloat[$k=403$]{ 
		\includegraphics[width=\snapWidth,trim={\cropLeft, \cropBottom, \cropRight, \cropTop},clip]{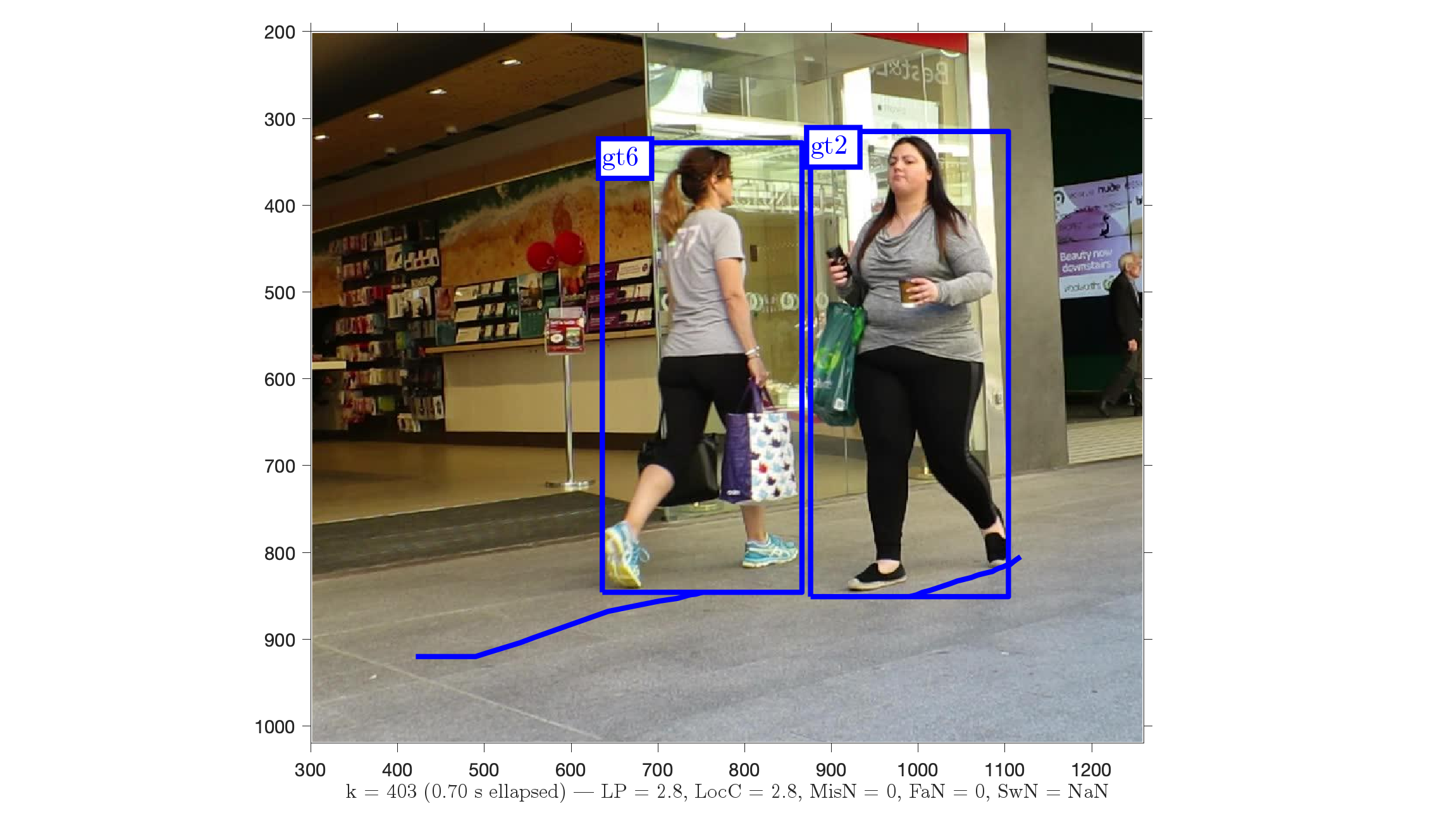}    }
	\subfloat[$k=427$]{ 
		\includegraphics[width=\snapWidth,trim={\cropLeft, \cropBottom, \cropRight, \cropTop},clip]{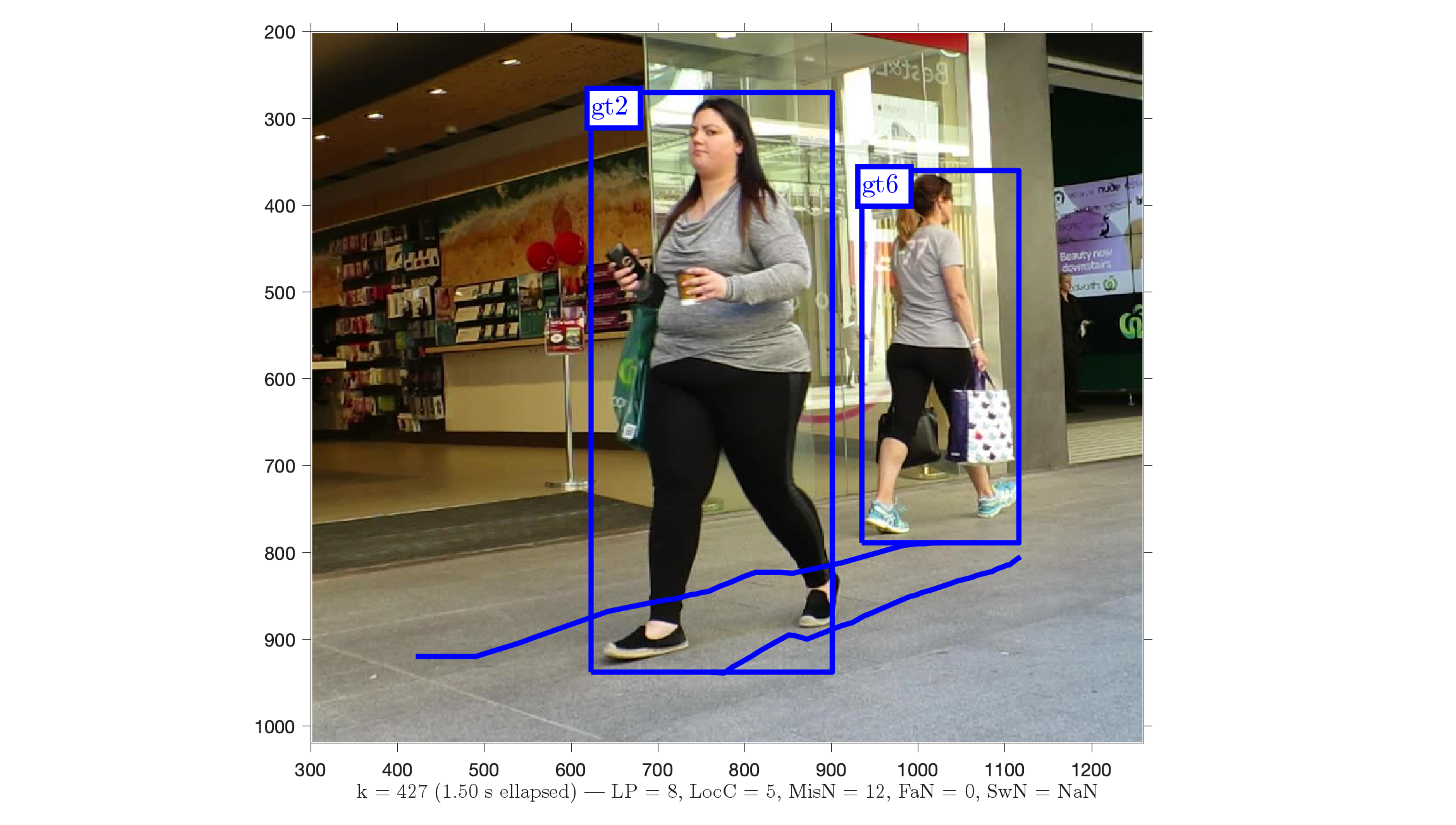}    }
	\subfloat[$k=442$ (end of the scenario)]{ 
		\includegraphics[width=\snapWidth,trim={\cropLeft, \cropBottom, \cropRight, \cropTop},clip]{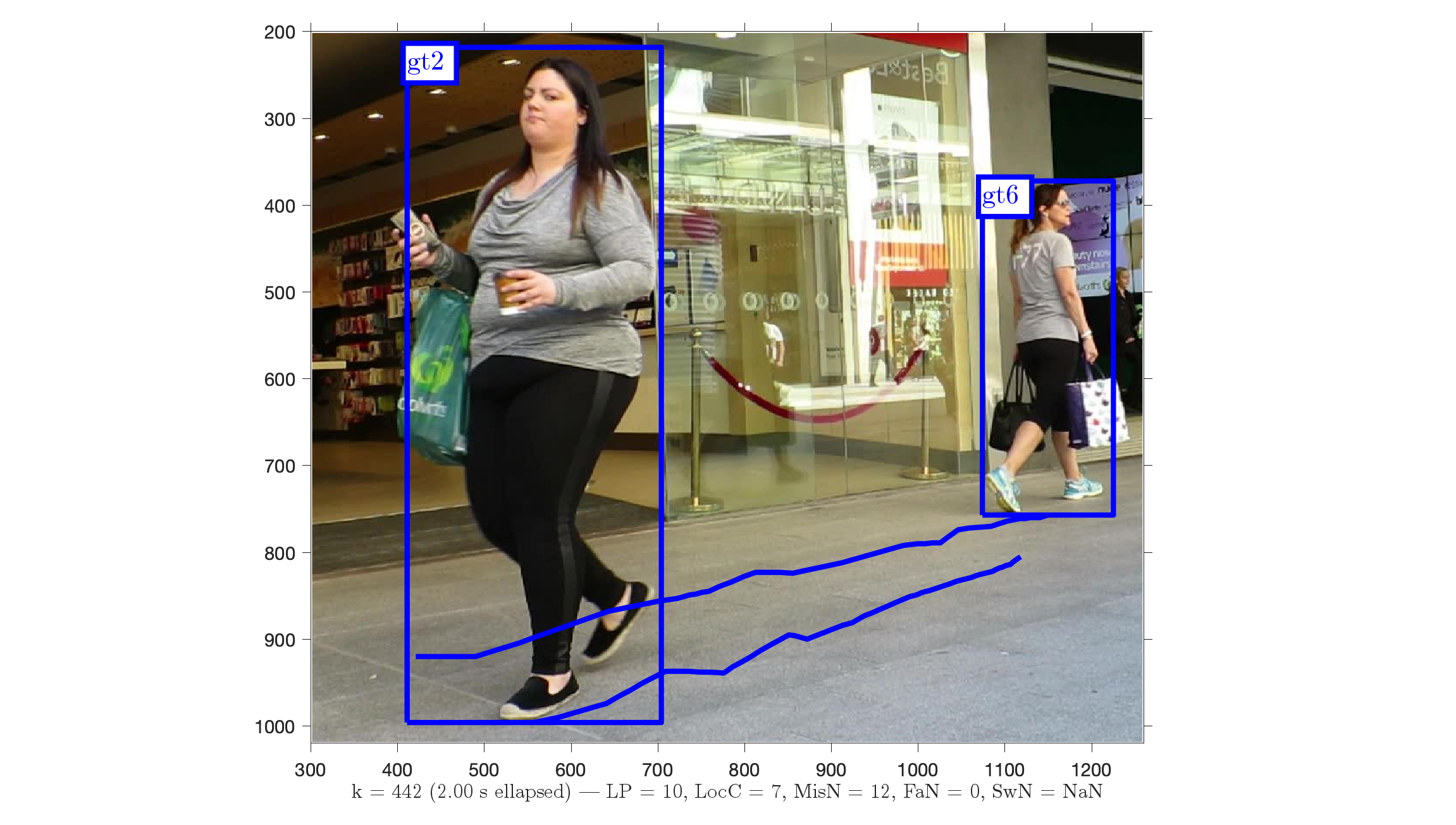}    }
	\caption{Part of the publicly available MOT17-09 scenario studied in this paper. The ground truth objects are depicted in blue. The blue traces depict past locations of the bottom-center point.}
	\label{fig:MO17-09-scenario}
	\vspace{-0.5cm}
\end{figure*}

\subsection{Algorithms}\label{sec:algorithms}
\begin{figure}
	\vspace{-0.4cm}
	\centering
	\subfloat[$k=403$]{
		\includegraphics[width=\snapWidth,trim={\cropLeft, \cropBottom, \cropRight, \cropTop},clip]{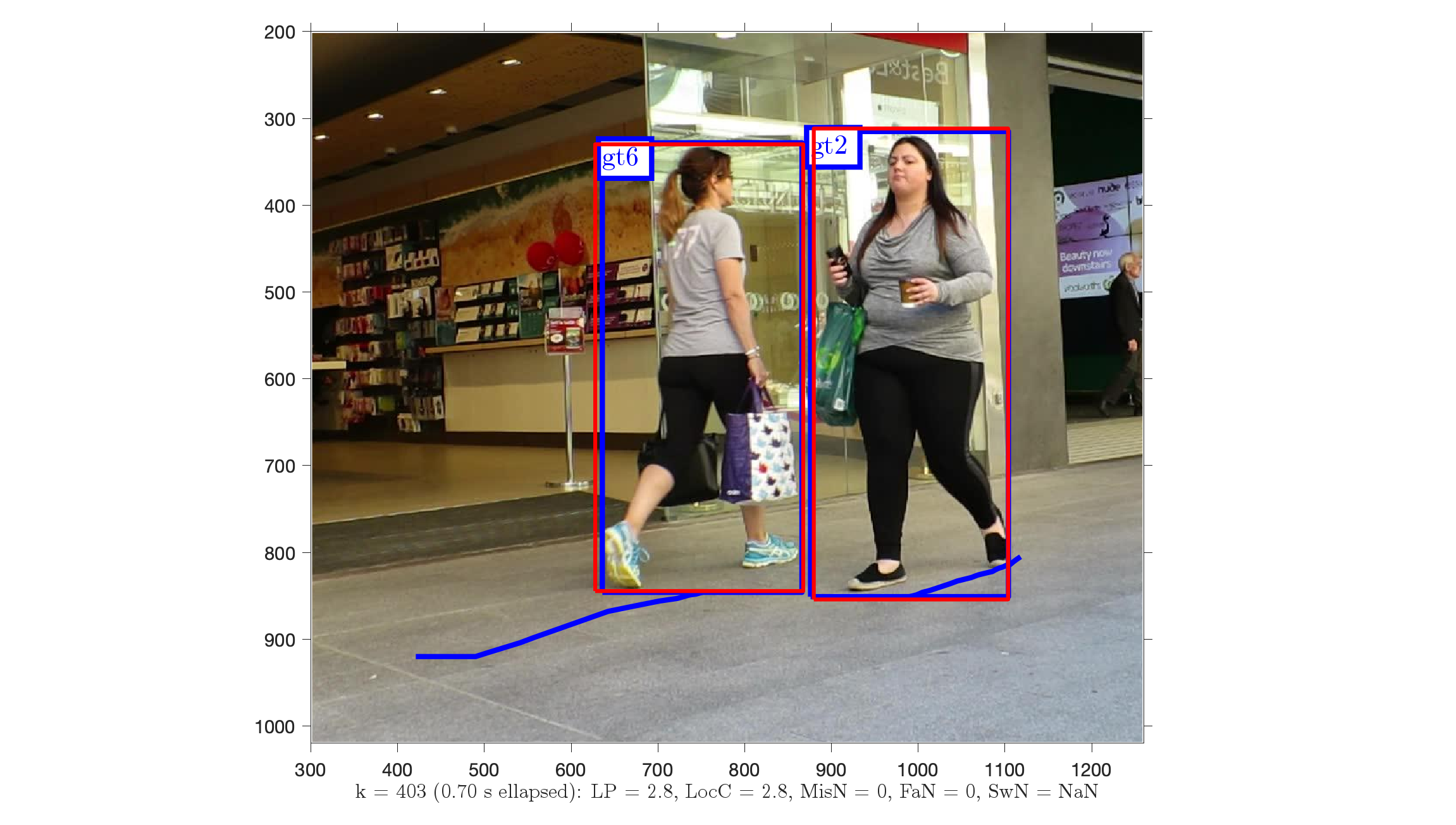}    }
	\subfloat[$k=415$]{ 
		\includegraphics[width=\snapWidth,trim={\cropLeft, \cropBottom, \cropRight, \cropTop},clip]{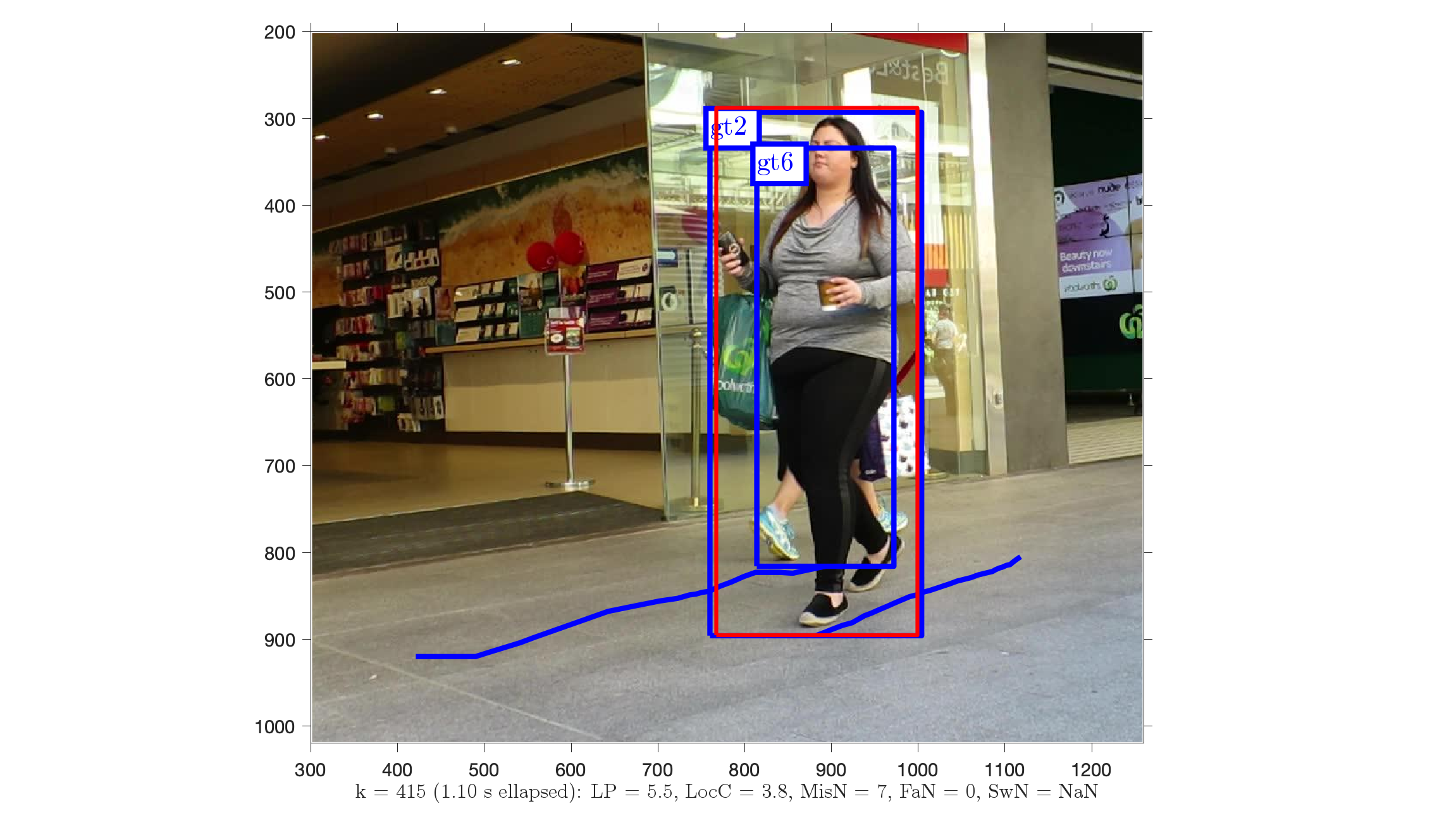}    }
	\caption{FRCNN detector results depicted in red. The detector processes each frame individually, and the estimates are thus not connected over time.}
	\label{fig:FRCNN-detector-bare}
\end{figure}

\textbf{FRCNN detector:}
The Faster R-CNN (FRCNN) detector from~\cite{FasterRCNN:2015}, whose outputs are included in the MOT17 dataset, processes each frame individually.
Consequently, the results from the FRCNN detector are not temporally connected in time and do not form trajectories.
As illustrated in Fig.~\ref{fig:FRCNN-detector-bare}, the detections match the ground truth bounding boxes seemingly well.
However, detections are missing for the occluded pedestrian between time steps $409$ and $421$, i.e., for $13$ time steps.

\begin{figure}
	\vspace{-0.4cm}
	\centering
	\subfloat[$k=403$]{
		\includegraphics[width=\snapWidth,trim={\cropLeft, \cropBottom, \cropRight, \cropTop},clip]{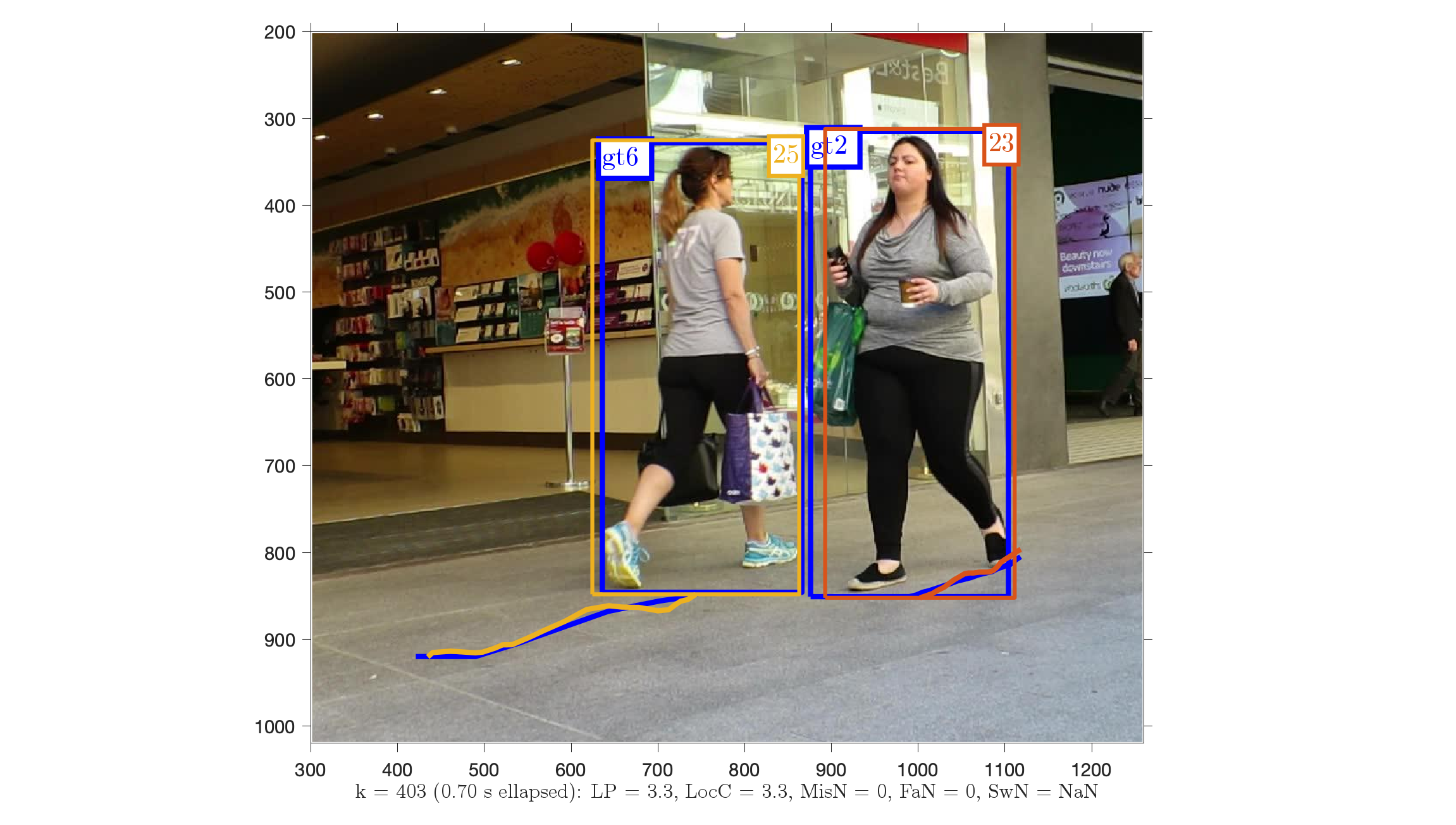}  \label{fig:Tracktor++v2-bare-a}  }
	\subfloat[$k=442$]{ 
		\includegraphics[width=\snapWidth,trim={\cropLeft, \cropBottom, \cropRight, \cropTop},clip]{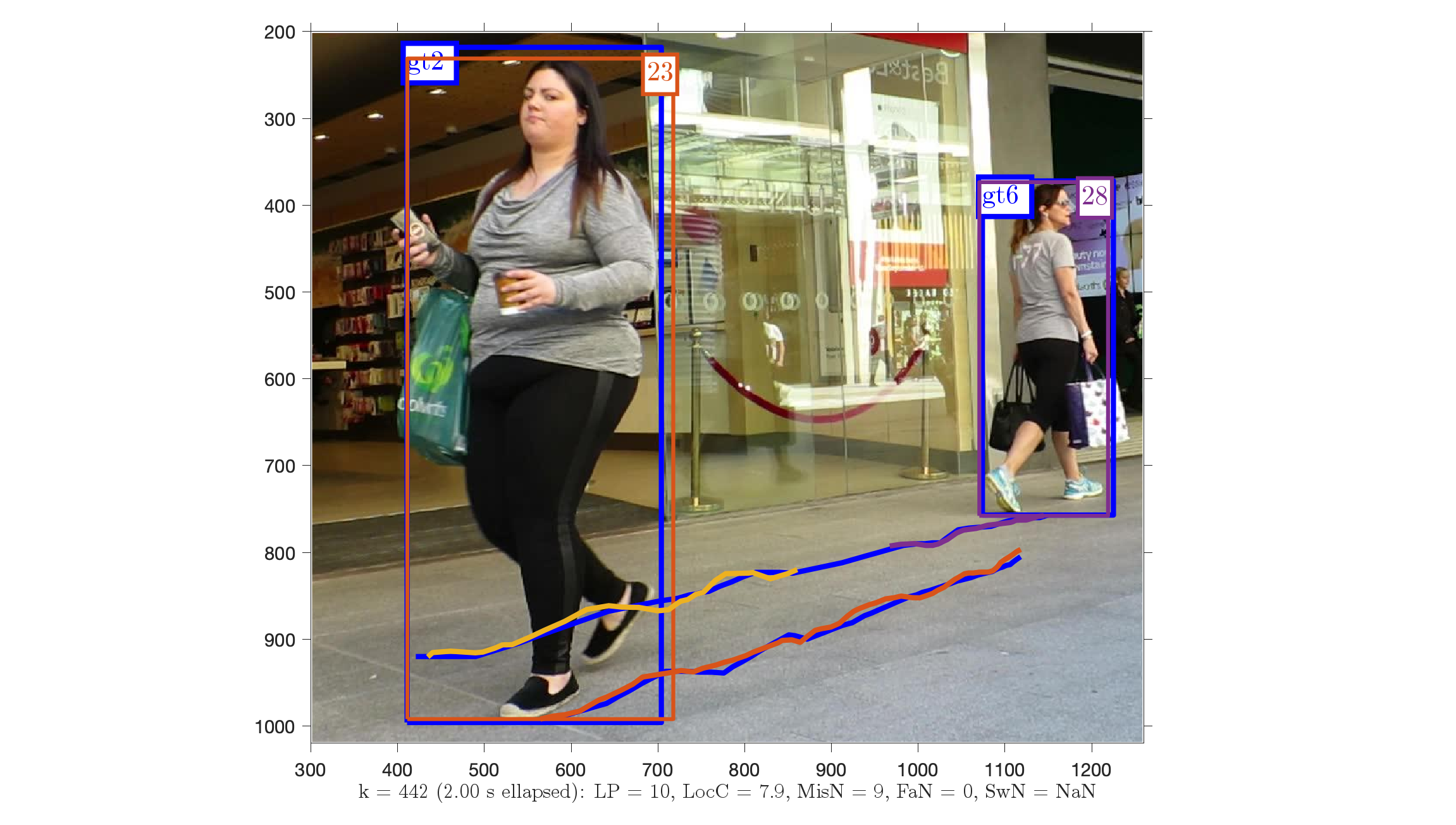}  \label{fig:Tracktor++v2-bare-b}  }
	\caption{Tracktor++v2 tracker results. The pedestrian \emph{gt6} is not tracked when it is occluded, and a new track is initiated afterward.}
	\label{fig:Tracktor++v2-bare}
	\vspace{-0.5cm}
\end{figure}

\textbf{Tracktor++v2 tracker:}
The Tracktor++v2 introduced in~\cite{Tracktor++v2:2019} is evaluated using the MOT17 benchmark, where it processes the FRCNN detections and produces trajectories.
As illustrated in Fig.~\ref{fig:Tracktor++v2-bare}, the occluded pedestrian is not tracked between time steps $412$ and $420$, i.e., for $9$ time steps.
In particular, two different trajectories are produced for the occluded pedestrian \emph{gt6}; the first, marked with ID $25$, is present before the occlusion, while the second, marked with ID $28$, appears after the occlusion.
This situation is called \emph{track fragmentation}.
In general, such behavior is referred to as \emph{long-term \seeminglySwitch{}} in this paper. 
In performance evaluation, a score should be capable of classifying such events as a \emph{switch}.

\begin{figure*}
	\vspace{-0.4cm}
	\centering
	\subfloat[$k=414$]{
		\includegraphics[width=\snapWidth,trim={\cropLeft, \cropBottom, \cropRight, \cropTop},clip]{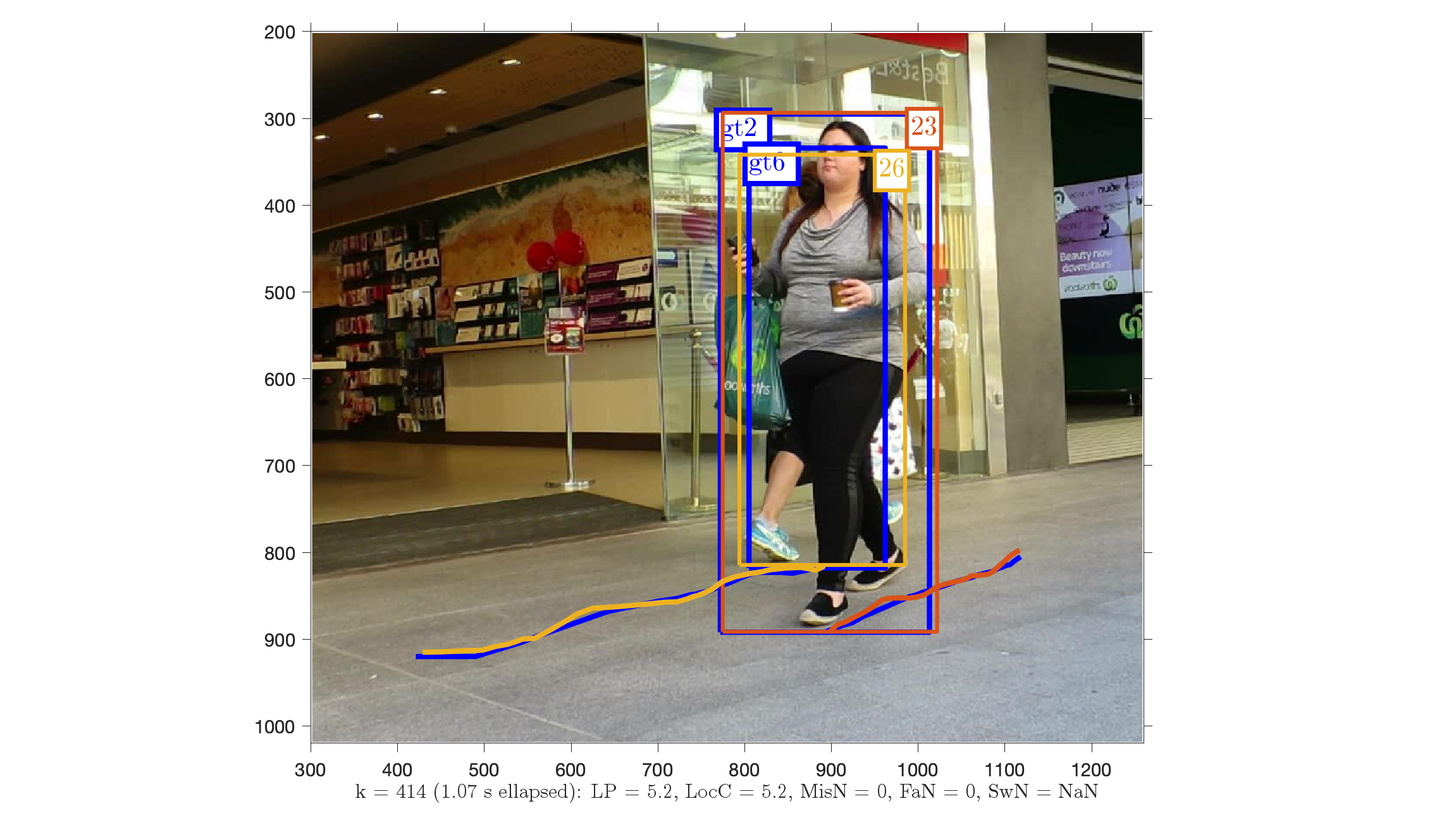} \label{fig:Bot-SORT-bare-a}   }
	\subfloat[$k=415$]{
		\includegraphics[width=\snapWidth,trim={\cropLeft, \cropBottom, \cropRight, \cropTop},clip]{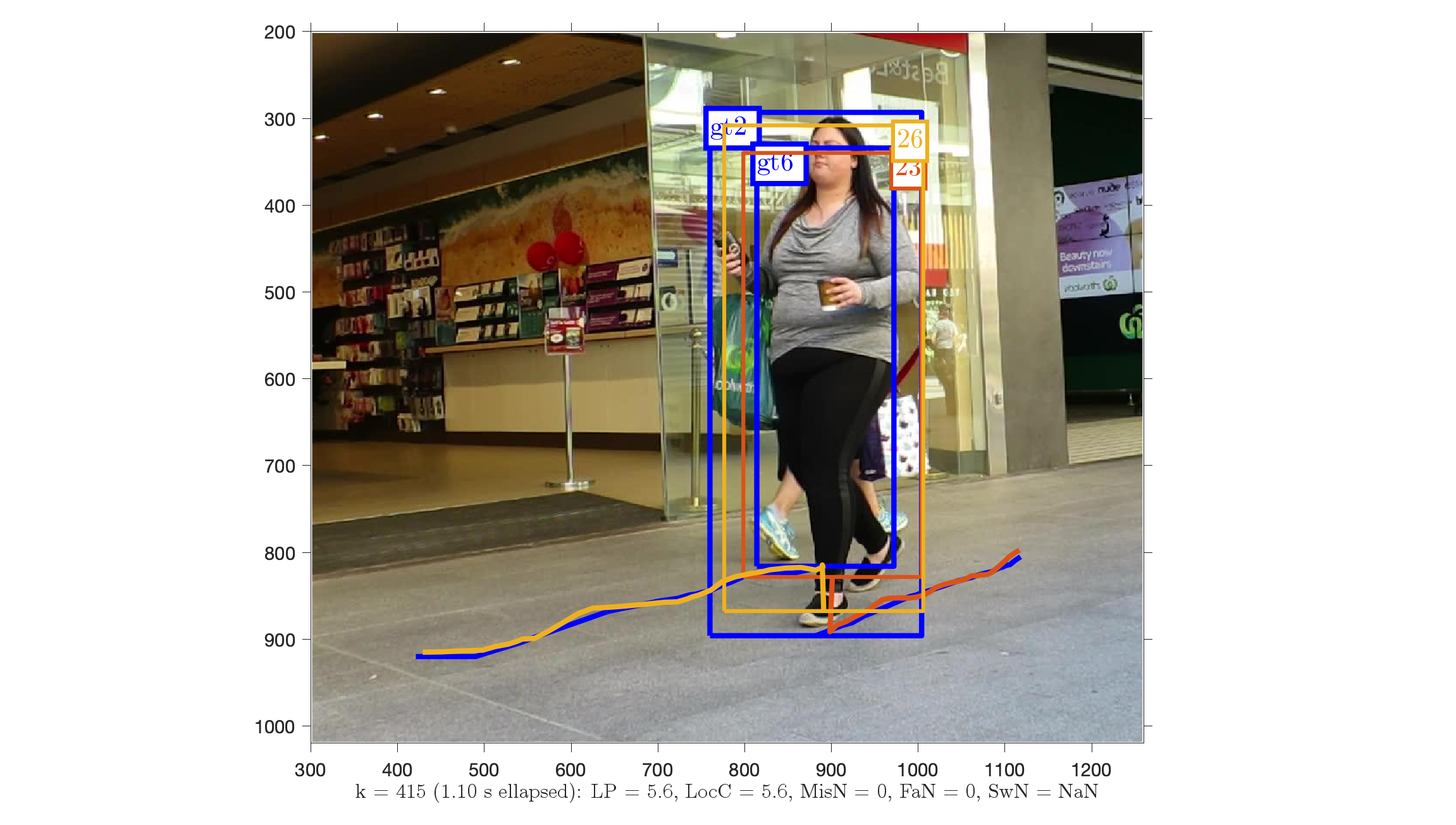} \label{fig:Bot-SORT-bare-b}   }
	\subfloat[$k=416$]{ 
		\includegraphics[width=\snapWidth,trim={\cropLeft, \cropBottom, \cropRight, \cropTop},clip]{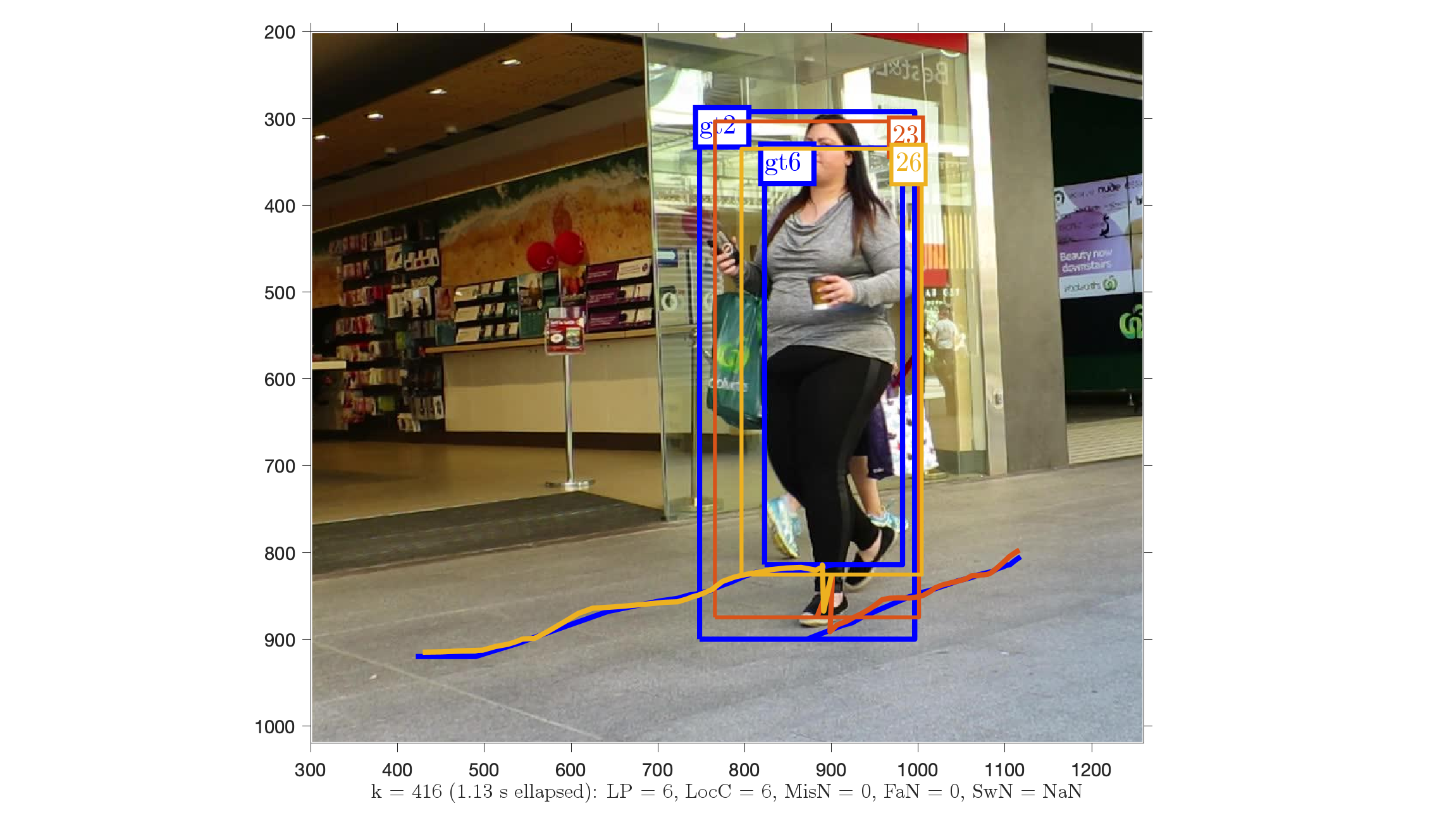} \label{fig:Bot-SORT-bare-c}   }
	\subfloat[$k=442$]{ 
		\includegraphics[width=\snapWidth,trim={\cropLeft, \cropBottom, \cropRight, \cropTop},clip]{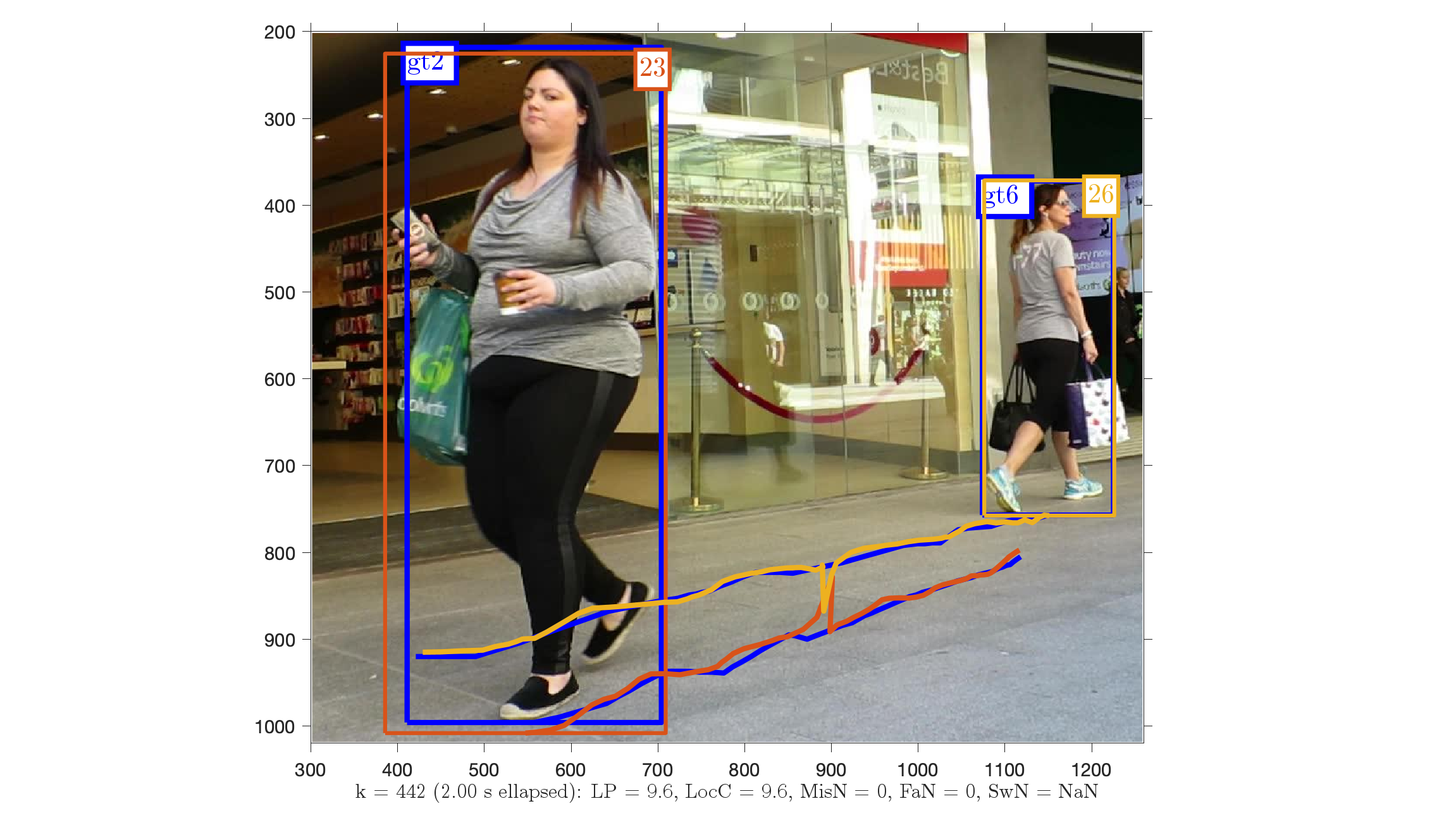} \label{fig:Bot-SORT-bare-d}   }
	\caption{BoT\_SORT tracker results. The estimates switch positions at time step $k{=}415$, see Figure~\ref{fig:Bot-SORT-bare-b}.}
	\label{fig:Bot-SORT-bare}
	\vspace{-0.5cm}
\end{figure*}

\textbf{BoT\_SORT tracker:}
Bag-of-tricks for simple online and real-time tracking (Bot\_SORT) method from~\cite{BoT-SORT:2022} processes custom detections based on a pre-trained YOLOX detector by~\cite{YOLOX:2021}.
The used version of BoT\_SORT employs linear interpolation and is effectively an offline method.
As depicted in Fig.~\ref{fig:Bot-SORT-bare}, both objects are tracked during the entire scenario, except for a single peculiarity appearing during the occlusion at time step $415$.
At that single time step, the two estimated tracks seemingly switch positions as if they swapped the ground truth object they were tracking before and after that time step.
That is, switching seemingly occurs over short time period and it might be caused by an error of the re-ID module combined with linear interpolation employed in the tracker.
Such behavior is referred to as a \emph{short-term interim \seeminglySwitch{}} in this paper.
From Fig.~\ref{fig:Bot-SORT-bare-c}, notice that a considerable misalignment of the estimated bounding boxes w.r.t.~the ground truth bounding boxes appear at time step $k{=}416$.

\begin{figure}
	\vspace{-0.4cm}
	\centering
	\subfloat[$k=416$]{
		\includegraphics[width=\snapWidth,trim={\cropLeft, \cropBottom, \cropRight, \cropTop},clip]{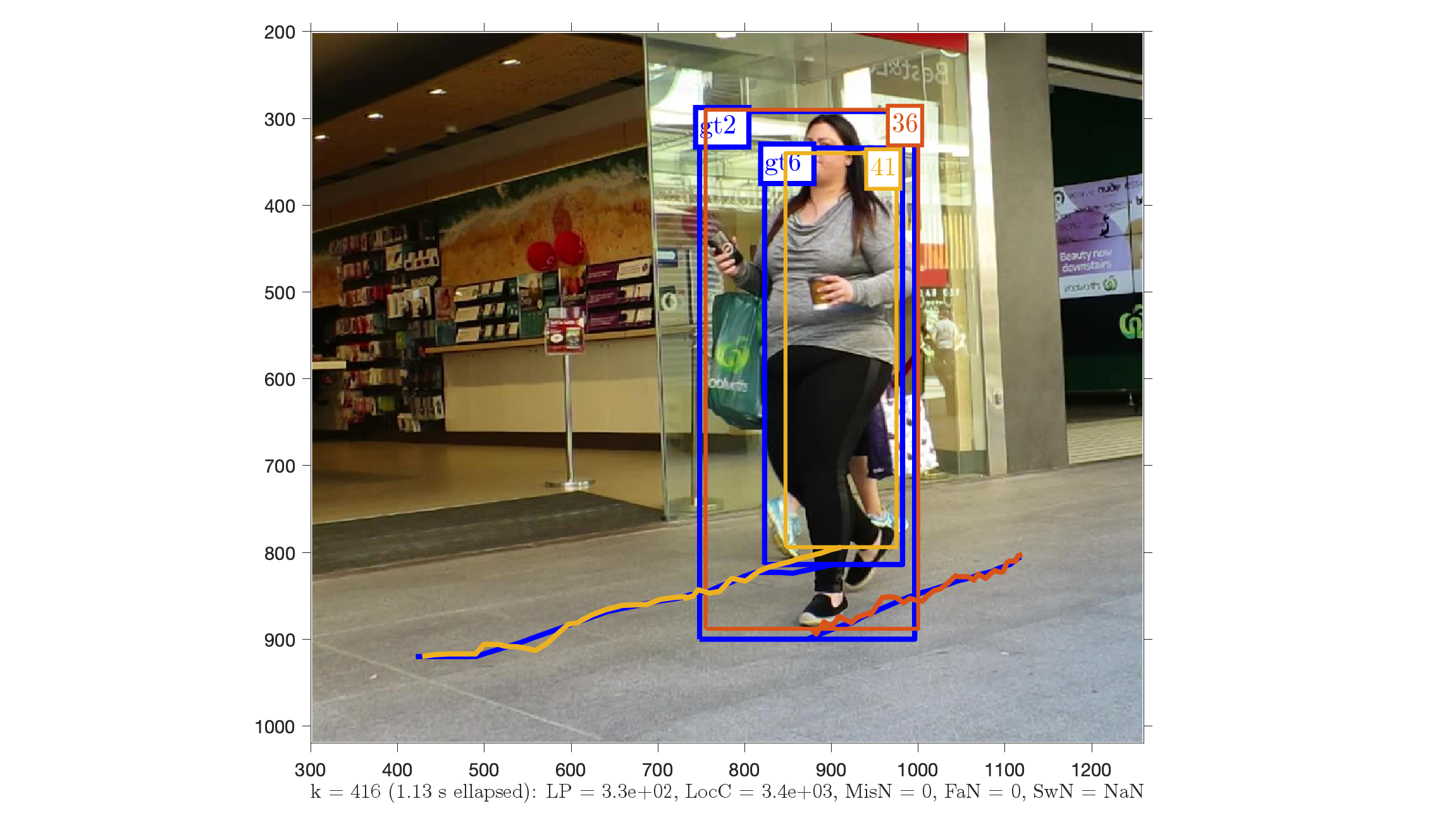} 
		\label{fig:GMPHDOGM17-bare-a}
	}
	\subfloat[$k=442$]{ 
		\includegraphics[width=\snapWidth,trim={\cropLeft, \cropBottom, \cropRight, \cropTop},clip]{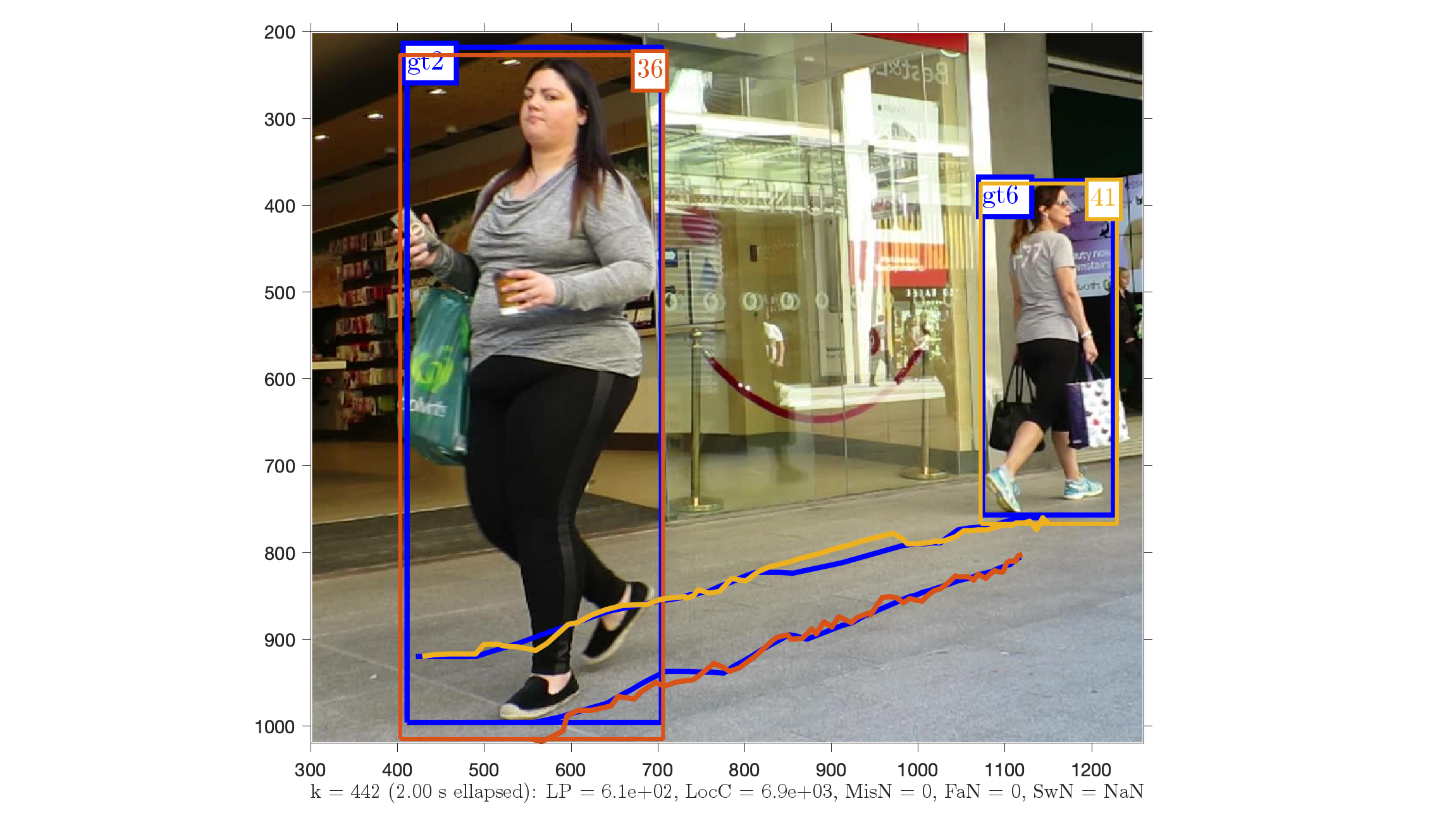} 
		\label{fig:GMPHDOGM17-bare-b}
	}        
	\caption{GMPHDOGM17 tracker results. Estimates are of lower-quality for the pedestrian \emph{gt6} when it is occluded.}
	\label{fig:GMPHDOGM17-bare}
\end{figure}

\textbf{GMPHDOGM17 tracker:}
The online tracker introduced in~\cite{GMPHDOGM17:2019} is based on the Gaussian mixture probability hypothesis density (GM-PHD) filter and employs the occlusion group energy minimization (OGEM).
The tracker processes FRCNN detections and is an online method.
As illustrated in Fig.~\ref{fig:GMPHDOGM17-bare}, all pedestrians are tracked.
During the occlusion of the pedestrian \emph{gt6}, however, the tracker outputs only predictive estimates (with ID $41$).
The predictive boxes may exceed a certain level of error further called as \emph{maximum admissible error} defined by the user for certain applications, e.g., starting at the time step $k{=}403$ (see Fig.~\ref{fig:GMPHDOGM17-bare-a}) until \emph{better} estimates are produced again at time step $442$.

Note that results generated from the above algorithms, except\footnote{
	Results from the BoT\_SORT algorithm were generated by using the publicly available code at \url{https://github.com/NirAharon/BoT-SORT/}{github.com/NirAharon/BoT-SORT/}.
} for BoT\_SORT, were downloaded directly from the MOT17 website~\cite{MOT17-webpage:2023}.
To analyze the particular MOT17-09 occlusion scenario (Fig.~\ref{fig:MO17-09-scenario}), both the ground truth and estimation results were processed by hand to include only the data corresponding to the ground truth IDs $2$ and $6$ between $k_0{=}281$ and $k_F{=}442$.

\subsection{CV Scores Analysis}\label{sec:scores-analysis}
This section motivates the necessity for a performance evaluation score with superior properties compared to those currently used in the CV community.
The need is demonstrated through the above tracking scenario indicating that the CV scores do not meet various requirements.

\begin{table}[]
	\centering
	\caption{Algorithms results evaluated using commonly used scores.
		The higher the value, the better.}
	\label{tab:CV_metrics-results}
	\begin{tabular}{rccc}
		\toprule
		& MOTA & IDF1 & HOTA \\
		\midrule
		FRCNN detector & 0.016 & 0.017 & 0.119  \\
		Tracktor++v2 tracker & 0.918 & 0.774 & 0.789 \\
		BoT\_SORT tracker & 1 & 1 & 0.921 \\
		GMPHDOGM17 tracker & 1 & 1 & 0.942 \\
		\bottomrule
	\end{tabular}
\end{table}

Using the data corresponding to the studied scenario, the MOTA, HOTA, and IDF1 scores for the considered algorithms are given in Table~\ref{tab:CV_metrics-results}.
It can be seen that MOTA and IDF1 scores fail to show any difference between the BoT\_SORT and GMPHDOGM17 algorithms. This outcome is undesirable because the scores should reflect the different results of the algorithms differently.
Nevertheless, the HOTA score could sort the algorithms based on their performance.
While HOTA works well in this scenario, its weaknesses follow.

HOTA can be understood as combining two separate scores for detection and association. 
For clarity, its definition is included in Appendix~\ref{sec:hota}. 
In \cite{HOTA:2021}, HOTA compares favorably with scores such as MOTA and IDF1, addressing their various drawbacks.
However, as we will demonstrate, HOTA still exhibits several undesirable behaviors.
First, HOTA is not, as claimed in \cite{HOTA:2021}, a mathematically well-defined metric.
A mathematically sound metric should satisfy four properties: 1) the distance from a point to itself is zero, 2) positivity, 3) symmetry, and 4) triangle inequality.
It is clear that HOTA does not satisfy property 1) as it is a score such that the higher the value, the better, and the HOTA between a point to itself is one.
The triangular inequality (which is quite essential for practice) is not met either, see~\cite{Nguyen:Visual-metrics-trustworthy:2023} 
Second, in the HOTA calculation, the ground truth-to-estimate assignment problems are individually solved at each time step (frame). While this lowers the computational complexity, it is a heuristic solution as the 2D assignment problems are sequentially connected due to temporal correlation. Principled solutions should be obtained by (approximately) solving a multi-dimensional assignment problem, see~\cite{TrajectoryGOSPA:2020}.
Third, HOTA does not capture the localization accuracy explicitly, which needs to be represented using another score LocA, see~\cite{HOTA:2021}.
Last, HOTA is calculated by averaging multiple scores over multiple localization thresholds for solving the different 2D assignment problems.
The averaging process was introduced to account for the localization accuracy in~\cite{HOTA:2021}, which is not an elegant solution.

While HOTA measures similarity, it might be tempting to define a function of two sets of trajectories $\mathbf{X}$ and $\mathbf{Y}$ as
\begin{align}
    d_{\text{HOTA}} (\mathbf{X}, \mathbf{Y}) = 1 - \text{HOTA}(\mathbf{X}, \mathbf{Y}) \, ,
    \label{eq:HOTA-induced-maybe-metric}
\end{align}
to measure dissimilarity.
Nevertheless, the funciton $d_{\text{HOTA}} (\mathbf{X}, \mathbf{Y})$~\eqref{eq:HOTA-induced-maybe-metric} 
fails to satisfy the identity and triangle inequality axioms~\cite{Nguyen:Visual-metrics-trustworthy:2023}.
The latter is further illustrated in Appendix~\ref{sec:1-HOTA-counterexample} for both HOTA and $d_{\text{HOTA}} (\mathbf{X}, \mathbf{Y})$~\eqref{eq:HOTA-induced-maybe-metric}, leading to the conclusion that neither of those are (mathematical) metrics.

This section demonstrated that CV scores may have problems distinguishing the performance of several algorithms and do not possess the desirable properties of a metric.
The following section presents a mathematically sound performance evaluation ``score'' that is mathematically a metric and shows how it can be used efficiently to evaluate visual tracking algorithms. 

\section{The TGOSPA Metric}\label{sec:TGOSPA_metric}
The \emph{trajectory generalized optimal sub-pattern assignment} (TGOSPA) is a metric on the space of sets of discrete-time trajectories, originally introduced in~\cite{TrajectoryGOSPA:2020}.
First, the notation is introduced, followed by the definition of the TGOSPA metric.
The TGOSPA metric has several parameters that need to be selected prior to its use, which are discussed next.
After revealing the general meaning of the parameters, their detailed choice for the case of CV applications follows.

\subsection{Notation and TGOSPA metric Definition}	
Let $(\X,\baseMetric)$ be a metric space\footnote{
	Note that this paper uses a slightly more general formulation than that in~\cite{TrajectoryGOSPA:2020,TimeWeightedTGOSPA:2021} where it is assumed that $\X=\mathbb{R}^{n}$.
	It can be seen that TGOSPA has the same properties as derived in~\cite{TrajectoryGOSPA:2020,TimeWeightedTGOSPA:2021} with $(\X,\baseMetric)$ being any metric space.
}. 
Note that $\baseMetric$ is a metric, i.e., a function that assigns the \emph{distance} $\baseMetric(x,y)$ to a pair of elements $x,y\in\X$, such as the Euclidean distance.
The elements of $\X$ are referred to as \elementOfTrajectory{s}, and they represent bounding boxes in the CV setting of this paper.
In particular, bounding boxes can be represented both as geometric entities (axis-aligned rectangle $x\subset\mathbb{R}^{4}$) or as vectors (e.g., using the center point $[\centerPoint{x}{1}\ \centerPoint{x}{2}]\T$, width $\widthCoord{x}$ and height $\heightCoord{x}$ as $x{=}[\centerPoint{x}{1}\ \centerPoint{x}{2}\ \widthCoord{x}\ \heightCoord{x}]\T$).
\begin{minipage}{\linewidth}
	\vspace{1mm}
	\begin{wrapfigure}[5]{l}{0.33\linewidth}
		\centering
		\vspace{-3mm}
		\includegraphics[scale=0.6]{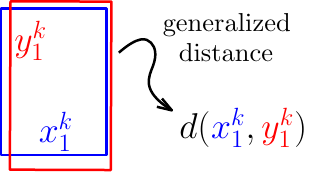}
	\end{wrapfigure}
	A possible choice of the metric $\baseMetric$ in this setting is $\baseMetric(x, y) {=} 1{-}\mathrm{IoU}(x,y)$, where $\mathrm{IoU}(x,y)$ is the intersection over union (IoU) of the two bounding boxes $x$ and $y$, see~\cite{KrKoSt:2023_FUSION}.
	\vspace{1mm}
\end{minipage}

Let $k{=}0$ be the initial (i.e., first) time step and $K{>}0$ be the final time step.
A trajectory $X{\in}\mathcal{T}(\X)$ corresponding to some possibly moving object is a sequence of elements of $\X$ together with time steps that indicate when the elements are present.
For instance, a trajectory that is comprised of a single \emph{segment} can have the form $X {=} \big(k_s, [x^{k_s}\ x^{k_s\!+\!1}\ \dots\ x^{k_s\!+\!\nu\!-\!1}]\big)$, where $k_s$ is the start-time with $0{\leq} k_s {\leq} K$, $\nu$ is the duration (length) and $[x^{k_s}\ x^{k_s\!+\!1}\ \dots\ x^{k_s\!+\!\nu\!-\!1}]$ is the sequence of consecutive elements of $\X$ (i.e., the \elementOfTrajectory{s}) that are indexed by the time step $k{\in}\{k_s,k_s\!+\!1,\dots,k_s\!+\!\nu\!-\!1\}$.
In general, trajectories can have \emph{gaps}, i.e., the \elementOfTrajectory{s} need not appear consecutively in time.
This can be addressed straightforwardly by appending several segments together, see~\cite[Sec.~II.A]{TrajectoryGOSPA:2020} for details.
To access individual elements of a trajectory composed of a single segment, let $\tau^k$ be the set-valued function that returns the set with the element at time step $k$ if it exists, or the empty set as 
\begin{align}
	\mathbf{x}^k = \tau^k(X) {=} \begin{cases}
		\{ x^k \} & \text{if}\ k_s \leq k \leq k_s{+}\nu{-}1, \\
		\emptyset & \text{otherwise}.
	\end{cases}
	\label{eq:tau^k:function}
\end{align}

Multiple trajectories are modeled as a set of trajectories $\mathbf{X} {=}\{ X_1, \dots, X_{|\mathbf{X}|}\} {\in} \FT$, where $\mathcal{F}(\cdot)$ denotes the collection of all finite subsets of the input set, with $|\cdot|$ denoting the cardinality.
To access the set of object instances within $\mathbf{X}$ that are present at time step $k$, $\tau^k$ is generalized to sets of trajectories as
\begin{align}
	\tau^k(\mathbf{X}) = \bigcup_{X\in \mathbf{X}} \tau^k( X ).
\end{align}

Indeed, the TGOSPA metric is a metric on $\FT$~\cite[Appendix~B.A]{TrajectoryGOSPA:2020}, i.e., it formalizes the distance between two sets of trajectories $\mathbf{X} {=} \{X_1, \dots, X_{|\mathbf{X}|}\}$ and $\mathbf{Y} {=} \{Y_1, \dots, Y_{|\mathbf{Y}|}\}$.
For performance evaluation, one of the sets (e.g., $\mathbf{X}$) contains ground truth data, while the other (e.g., $\mathbf{Y}$) contains estimated trajectories.

\begin{figure}
	\centering
	\includegraphics[width=0.47\textwidth]{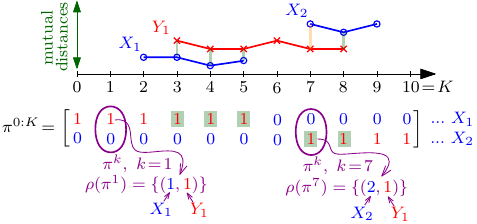}
	\caption{Example assignment matrix $\pi^{0:K}$ for two sets of trajectories $\mathbf{X}=\{X_1, X_2\}$ depicted in blue and $\mathbf{Y}=\{ Y_1\}$ depicted in red.
		For instance, the trajectories $X_2$ and $Y_1$ are assigned at $k=7$, which is indicated by $\rho(\pi^7)=\{(2,1)\}$.
		Similarly, the trajectories $X_1$ and $Y_1$ are assigned at $k=1$, which is indicated by $\rho(\pi^1)=\{(1,1)\}$, although neither of the objects are present at $k=1$.
	}
	\label{fig:assignments:example}
	\vspace{-0.5cm}
\end{figure}

In the computation of TGOSPA, trajectories from $\mathbf{X}$ and $\mathbf{Y}$ are assigned to each other at each time step, for which auxiliary notation is needed.
Let $\Pi_{\mathbf{X}, \!\mathbf{Y}}$ be the set of all assignment vectors between the index sets $\{1,\dots,|\mathbf{X}|\}$ and $\{0,\dots,|\mathbf{Y}|\}$ that maps trajectories to each other at each time step as follows.
At any time step $k$, an assignment vector $\pi^k {=} [\pi_1^k, \dots, \pi_{|\mathbf{X}|}^k]\T$ describes the assignment of each trajectory in $\mathbf{X}$ to a trajectory in $\mathbf{Y}$ at time step $k$, with the index $\pi_i^k\in\{0,\dots,|\mathbf{Y}|\}$.
The value $\pi_i^k {=} 0$ means that the trajectory $i$ is unassigned at time step $k$ and $\pi_i^k {=} j {>} 0$ means that the trajectory $i$ is assigned to trajectory $\mathbf{Y}_j$ at time step $k$.
At each time step, each trajectory in $\mathbf{X}$ can be assigned to at most one trajectory $\mathbf{Y}$, which is expressed by the implication $(\pi_i^k{=}\pi_{j}^k{>}0) \Rightarrow (i{=}j)$.
Let $\pi^{0:K}{=}[\pi^0,\dots,\pi^K]\in\{0,\dots,|\mathbf{Y}|\}^{|\mathbf{X}| \times K}$ be the matrix containing the assignments vectors across all time steps.
To directly access the indices of the trajectories that are paired at time step $k$, let $\rho\big(\pi^k\big)$ denote the set of pairs $(i,j)\in\rho\big(\pi^k\big)$, such that that trajectory $i$ is assigned to trajectory $j$ at time step $k$.
Note that two trajectories can be assigned to each other at any time step, i.e., even at time steps when one (or both) of the trajectories have no object instance present (e.g., did not start yet, or has already ended).
The assignments in $\pi^{0:K}$ being encoded with $\rho$ are \emph{trajectory-level} assignments, and their temporal changes are key to assessing \seeminglySwitch{s} present in the data.
Example assignments are given in Fig.~\ref{fig:assignments:example}.

If two trajectories $X_i$ and $Y_j$ are assigned at a particular time step $k$, their mutual distance at that time step is computed as follows.
First, the \elementOfTrajectory{s} at the time step $k$ are extracted from the trajectories with
\begin{subequations}\label{eq:sers:xy}
	\begin{align}
		\mathbf{x}_i^k &= \tau^k(X_i), & |\mathbf{x}_i^k|&\leq 1 \label{eq:set:x_i}\\
		\mathbf{y}_j^k &= \tau^k(Y_j), & |\mathbf{y}_j^k|&\leq 1. \label{eq:set:y_j}
	\end{align}
\end{subequations}
Then, for $1{\leq} p {<} {+}\infty$ and cut-off parameter $c{>}0$, the distance between the sets $\mathbf{x}_i^k$ and $\mathbf{y}_j^k$ is computed by
\begin{align}
	&\GOSPAatMostOne \!\big( \mathbf{x}_i^k, \mathbf{y}_j^k \big)  \notag \\
	&=\begin{cases}
		\min\!\big(c, \baseMetric(x_i^k ,y_j^k) \big) & \mathbf{x}_i^k {=} \{x_i^k\}, \mathbf{y}_j^k {=} \{y_j^k\}, \\
		0 & \mathbf{x}_i^k {=} \mathbf{y}_j^k {=} \emptyset, \\
		\frac{c}{ \sqrt[p]{2} } & \text{otherwise}.
	\end{cases}
	\notag\\[-1.0cm]
	\label{eq:GOSPAatMostOne}
	\\[-0.1cm]\notag
\end{align}
In fact, $\GOSPAatMostOne$ in \eqref{eq:GOSPAatMostOne} is a special case of the GOSPA metric from~\cite{GOSPA:2017} between the sets $\mathbf{x}_i^k$ and $\mathbf{y}_j^k$ that both have at most one element~\cite[Sec.~II.B]{TrajectoryGOSPA:2020}. 
Note that the first case of the definition of $\GOSPAatMostOne$ in \eqref{eq:GOSPAatMostOne}, i.e., $\GOSPAatMostOne(\{ x_i^k \},\{ y_j^k \}) {=} \min\!\big(c, \baseMetric(x_i^k, y_j^k)\big)$ is a \emph{cut-off} metric.

The following definition of the TGOSPA metric emphasizes that any two trajectories can be assigned at any time step, regardless either of them exists or not at that time step.
Such a definition is beneficial for understanding how the metric assesses \seeminglySwitch{s}.
The use of $\rho$ in the following definition leads to a slightly different notation compared to the original TGOSPA metric definition in~\cite{TrajectoryGOSPA:2020}.

\begin{definition}[TGOSPA metric]
	Let $1 {\leq}p{<}{+}\infty$, cut-off parameter $c{>}0$ and switching penalty $\gamma {>} 0$ be given real numbers (the TGOSPA parameters).
	The TGOSPA metric between two sets of trajectories $\mathbf{X},\mathbf{Y}$ is defined by
	\begin{align}
		\!\!\!\!\! d_p^{(c,\gamma)}(\mathbf{X},\!\mathbf{Y}) \! = \!\!\!
		\min_{ \pi^{0:K} \in \Pi_{\mathbf{X},\!\mathbf{Y}}^{K\ksp+\!1} } \!
		\Bigg( \overbrace{ A\big(\mathbf{X},\!\mathbf{Y},\pi^{0:K}\big) }^{ \mathbf{X},\mathbf{Y}\text{--\emph{assigned} term} } + \notag\\[-1mm]
\!\!\!		  \underbrace{ \gamma^p S\big(\pi^{0:K}\big) }_{ \text{\emph{switch} term} } \!+\!\!\!\!\!
		\hspace{1mm}\underbrace{ \frac{c^p}{2} U\big(\mathbf{X},\pi^{0:K}\big) \!}_{ \mathbf{X} \text{--\emph{unassigned} term} }
		 \!\!+\!\!  \underbrace{ \frac{c^p}{2} U\big(\mathbf{Y},\pi^{0:K}\big) }_{ \mathbf{Y} \text{--\emph{unassigned} term} } \!\!
		\Bigg)^{\!\!\!\sfrac{1}{p}}\!\!\!\!\!, \!\!\!\!
		\label{eq:TGOSPA:definition}
	\end{align}
	where, respectively,
	\begin{subequations}
		\begin{align}
			\! A\big(\mathbf{X},\!\mathbf{Y},\pi^{0:K}\big) \! &= \notag\\
			&\hspace{-0.4cm} \textstyle \sum_{k=0}^K \ \sum_{(i,j) \in \rho(\pi^k)} \GOSPAatMostOne(\mathbf{x}_i^k, \mathbf{y}_j^k)^p, \!\! \label{eq:TGOSPA:AssignedTerm}\\
			\! S\big(\pi^{0:K}\big) \! &= \! \textstyle \sum_{k=0}^{K\!-\!1} \sum_{i=1}^{|\mathbf{X}|} s\big(\pi_i^k, \pi_i^{k+1}\big), \label{eq:TGOSPA:SwitchTerm}\\
			\! U\big(\mathbf{X},\pi^{0:K}\big) &= \textstyle \sum_{k=0}^K \Big( \big|\tau^k(\mathbf{X})\big|  \!-\! \big|\rho(\pi^k)\big| \Big)\!, \label{eq:TGOSPA:MF}
		\end{align}
	\end{subequations}
	are the $\mathbf{X},\mathbf{Y}$--\emph{assigned} term, number of switches and the number of \elementOfTrajectory{s}\footnote{
		The function \eqref{eq:TGOSPA:MF} counts \elementOfTrajectory{s} within trajectories that are left unassigned over time, not entire trajectories.
	} from $\mathbf{X}$ that are left unassigned (analogously to $\mathbf{Y}$).
	For simplicity, the dependency of $A(\mathbf{X}, \!\mathbf{Y}, \pi^{0:K})$~\eqref{eq:TGOSPA:AssignedTerm} on $p$ and $c$ is omitted.
	For a trajectory in $\mathbf{X}$ with the index $i$, switches are counted based on temporal changes in the associations as
	\begin{align}
		&s\big(\pi_i^k, \pi_i^{k+1}\big)  \notag\\
		&=
		\begin{cases}
			0 & \pi_i^k \!=\! \pi_i^{k+1}, \\
			1 & \pi_i^k \!\neq\! \pi_i^{k+1},\hspace{0.1cm} \pi_i^k \!\neq\! 0,\hspace{0.1cm} \pi_i^{k+1} \!\neq\! 0, \\
			\tfrac{1}{2} & \text{otherwise}.
		\end{cases}
		\label{eq:TGOSPA:s:pi}
	\end{align}
	The second and the third case of the $s(\pi_i^k, \pi_i^{k+1})$~\eqref{eq:TGOSPA:s:pi} definition will be referred to as \emph{full-} and \emph{half-}switches, respectively.
	The symbol $\pi^{0:K}_{\star}$ denotes the argument of minimum of~\eqref{eq:TGOSPA:definition}.
	\hfill$\square$
\end{definition}

The TGOSPA metric computation is an NP-hard problem.
Therefore, approximations are required for large-scale problems involving many trajectories, see~\cite{TrajectoryGOSPA:2020,TGOSPA:entropic}.
In practical examples and the Python and Matlab implementations available at this link\footnote{
	\url{github.com/Agarciafernandez/T-GOSPA-metric-python}
}${}^,$\footnote{
	\url{github.com/Agarciafernandez/MTT}
}, an approximation based on the linear programming (LP) relaxation formulation according to~\cite[Sec.~IV.B]{TrajectoryGOSPA:2020} is used.
The resulting approximation is also a metric, referred to as the \emph{LP metric}, and serves as an accurate lower bound for the TGOSPA metric.
Although the LP metric is not generally guaranteed to yield identical results as the TGOSPA metric~\cite[pp.19-20]{TGOSPA:entropic}, it often does in practice and it did in this paper\footnote{
    The LP metric relaxes the so-called \emph{hard} assignments present in the TGOSPA metric definition to \emph{soft} assignments~\cite{TrajectoryGOSPA:2020}.
    However, all optimal assignments resulting from the LP metric computations performed in this paper were hard, in which case the two metrics are identical.
}.
Therefore, the discussion focuses on the TGOSPA metric instead of the LP metric in the following.

It can be seen that the metric classifies the data $\mathbf{X}$ and $\mathbf{Y}$ into four terms depending on the parameters $c$, $p$, and $\gamma$.
In the following, the classification terms are treated first.
The parameters are explained subsequently.

\subsection{TGOSPA Metric Decomposition}\label{sec:TGOSPA:decomposition}
The four terms the data get classified into by TGOSPA correspond to indices where $\pi^{0:K}$ is non-zero ($\mathbf{X},\mathbf{Y}$--assigned term), zero ($\mathbf{X}$--unassigned and $\mathbf{Y}$--unassigned terms\footnote{
	The $\mathbf{X}$--unassigned term is eventually the number of indices where $\pi^{0:K}$ is zero, multiplied by $\tfrac{c^p}{2}$.
	For each time step, the number where $\pi^{0:K}$ it is nonzero (i.e., $|\rho(\pi^k)|$) is subtracted from the maximum possible number (i.e., $|\tau^k(\mathbf{X})|$).
	The $\mathbf{Y}$--unassigned term is also the number of indices where $\pi^{0:K}$ is zero (multiplied by $\tfrac{c^p}{2}$), but for the case when $\mathbf{X}$ and $\mathbf{Y}$ are interchanged, i.e., for $\pi^{0:K}\in \Pi_{\mathbf{Y},\mathbf{X}}$.
}) and to how $\pi^{0:K}$ changes in time for each row (Switch term). 
To make a good sense of the terminology regarding "missed" and "false" objects in the following, let $\mathbf{X}$ represent the set of ground truth trajectories and $\mathbf{Y}$ the set of estimated trajectories. Note, however, that their roles can be interchanged since TGOSPA is a metric.

\subsubsection{Illustrative Example}
Consider the example given in Fig.~\ref{fig:assignments:example} and assume that the depicted assignment matrix $\pi^{0:K}$ is the argument of the TGOSPA minimum $\pi^{0:K}_{\star}$ for some parameters $p$, $c$ and $\gamma$.
In this case, $A \big( \mathbf{X},\!\mathbf{Y},\pi^{0:K} \big)$~\eqref{eq:TGOSPA:AssignedTerm} is the sum of the distances (to the $p$-th power) highlighted in green (assuming each of them is smaller than $c$) and of $\tfrac{c^p}{2} {\times} 2 {=} c^p $ due to \emph{1)} the trajectory $X_1$ assigned to $Y_1$ at $k{=}2$, where $Y_1$ has no \elementOfTrajectory{}, and \emph{2)} the trajectory  $X_2$ assigned to $Y_1$ at $k{=}9$, where $Y_1$ has no \elementOfTrajectory{} neither.
If moreover \emph{3)} the value of $c$ were such that $ \min \!\big(c, \baseMetric(x_2^7, y_1^7) \big) {=} c$ at $k{=}7$, then the corresponding summand in $A \big( \mathbf{X},\!\mathbf{Y},\pi^{0:K} \big)$~\eqref{eq:TGOSPA:AssignedTerm} would be $c^p$.
In addition, the $\mathbf{X}$--unassigned term is zero and the $\mathbf{Y}$--unassigned term counts $\tfrac{c^p}{2}$ once because the trajectory $Y_1$ is unassigned at $k{=}6$.
Furthermore, there would be two half-switches weighted by $\gamma^p$ in the switch term, i.e., one switch in total.

In the context of performance evaluation, it is more convenient to view the three TGOSPA terms ($\mathbf{X}, \mathbf{Y}$--assigned, $\mathbf{X}$-- and $\mathbf{Y}$--unassigned terms) from the perspective of \emph{properly estimated}, \emph{missed} and \emph{false} \elementOfTrajectory{s} regardless of their assignments.
For the example described above, the properly detected \elementOfTrajectory{s} are those giving rise to the summands in $A \big( \mathbf{X},\!\mathbf{Y},\pi^{0:K} \big)$~\eqref{eq:TGOSPA:AssignedTerm} that are \emph{not} due to \emph{1)}, \emph{2)} and neither \emph{3)}.
The summands in $A \big( \mathbf{X},\!\mathbf{Y},\pi^{0:K} \big)$~\eqref{eq:TGOSPA:AssignedTerm} that are due to \emph{1)}, \emph{2)} and "one half" of \emph{3)}, i.e., $\tfrac{c^p}{2}$ would constitute the missed objects term.
The "second half" of \emph{3)}, i.e., $\tfrac{c^p}{2}$ together with $\mathbf{Y}$--unassigned term would constitute the false alarms terms.
The switch cost would stay the same.

\subsubsection{Decomposition Suitable for Performance Evaluation}\label{subsubsec:TGOSPA:decomposition}
The localization term corresponds to properly estimated objects by counting the actual distances.
Properly estimated \elementOfTrajectory{s} constitute the pairs $x_i^k,y_j^k$ for which the corresponding summands in $A \big( \mathbf{X},\!\mathbf{Y},\pi^{0:K} \big)$~\eqref{eq:TGOSPA:AssignedTerm} are the distances to the $p$-th power $\baseMetric(x_i^k,y_j^k)^p$ which are lower than $c^p$.
To directly access the indices of properly estimated trajectories at time step $k$, let $\theta_k^{(c)}\big( \mathbf{X}, \!\mathbf{Y}, \pi^{k} \big) {\subset} \rho\big(\pi^k\big)$ denote the set of pairs
\begin{align}
	&\!\theta_k^{(c)}\big( \mathbf{X}, \!\mathbf{Y}, \pi^{k} \big) = \\
	&\!\big\{ (i,j) {\in} \rho\big( \pi^{k} \big)\!: \hspace{0.0cm} \mathbf{x}_i^k{=}\{x_i^k\}, \hspace{0.1cm}  \mathbf{y}_j^k{=}\{y_j^k\} \hspace{0.15cm} \text{and} \hspace{0.15cm} \baseMetric(x_i^k, y_j^k) {<} c \big\}.
	\notag
\end{align}
The assignments in $\pi^{0:K}$ that are extracted via $\theta$ are \emph{\elementOfTrajectory{}-level} assignments and are key for the TGOSPA metric decomposition.
The distances, i.e., the localization term, is then 
\begin{align}
	L_p^{(c)} \!\big( \mathbf{X},\!\mathbf{Y},\pi^{0:K} \big) \!=\! \sum_{k=0}^K \ \sum_{(i,j) \in \theta_k^{(c)}\!( \mathbf{X}, \!\mathbf{Y}, \pi^k)} \!\! \GOSPAatMostOne\!(\mathbf{x}_i^k, \mathbf{y}_j^k)^p \!\! . \label{eq:TGOSPA:LocTerm}
\end{align}
The number of properly estimated objects is the number of summands in~\eqref{eq:TGOSPA:LocTerm}, and is denoted as
\begin{align}
	N^{(c)}\big( \mathbf{X}, \!\mathbf{Y}, \pi^{0:K} \big) &= \textstyle \sum_{k=0}^K \big|\theta_k^{(c)}\big( \mathbf{X}, \!\mathbf{Y}, \pi^k\big)\big|  . \label{eq:number-of-properly-estimated-objects}
\end{align}
The remaining summands in $A \big( \mathbf{X},\!\mathbf{Y},\pi^{0:K} \big)$~\eqref{eq:TGOSPA:AssignedTerm} correspond to missed and false \elementOfTrajectory{s}.
The number of such remaining objects are all weighted by $\tfrac{c^p}{2}$ and are to be \emph{split} and added into $U\big(\mathbf{X},\pi^{0:K}\big)$~\eqref{eq:TGOSPA:MF} and $U\big(\mathbf{Y},\pi^{0:K}\big)$~\eqref{eq:TGOSPA:MF}.

Consider the sets of missed and false \elementOfTrajectory{s} stored as tuples, containing the time step and the trajectory index $i$ or $j$ that is either missed or false,
\begin{align}
	\hspace{-0.0cm}\mathcal{M}\big( \mathbf{X}, \!\mathbf{Y}, \pi^{0:K}\big) &= \notag\\
	&\hspace{-2.5cm}\big\{ (k,\!i)\!: \nexists j{:}(i,\!j) {\in} \theta_k^{(c)}\big( \mathbf{X}, \!\mathbf{Y}, \pi^{k} \big), \hspace{0.00cm} \mathbf{x}_i^k{=}\{x_i^k\} \big\},
	\label{eq:TGOSPA:MissObjTerm}\\
	\hspace{-0.0cm}\mathcal{F}\big( \mathbf{X} ,\!\mathbf{Y}, \pi^{0:K}\big) &= \notag\\
	&\hspace{-2.5cm}\big\{ (k,\!j)\!: \nexists i{:}(i,\!j) {\in} \theta_k^{(c)}\big( \mathbf{X}, \!\mathbf{Y}, \pi^{k} \big), \hspace{0.00cm}  \mathbf{y}_j^k{=}\{y_j^i\} \big\},
	\label{eq:TGOSPA:FalseAlarmTerm}
\end{align}
respectively, with $k$ ranging over $\{0,1,\dots,K\}$, $i$ ranging over $\{1,\dots,|\mathbf{X}|\}$ and $j$ ranging over $\{1,\dots,|\mathbf{Y}|\}$.
For simplicity, the dependency of $\mathcal{M}\big( \mathbf{X}, \!\mathbf{Y}, \pi^{0:K}\big) $~\eqref{eq:TGOSPA:MissObjTerm} and $\mathcal{F}\big( \mathbf{X}, \!\mathbf{Y}, \pi^{0:K}\big) $~\eqref{eq:TGOSPA:FalseAlarmTerm} on $p$ and $c$ is omitted.
Indeed, the numbers of properly detected, missed, and false \elementOfTrajectory{s} are the cardinalities of these sets.
With this, the TGOSPA metric can be written according to~\cite{TrajectoryGOSPA:2020} as
\begin{multline}
	\!\!\!\!\! d_p^{(c,\gamma)}(\mathbf{X},\!\mathbf{Y}) \! = 
	\!\!\! \min_{ \pi^{0:K} \in \Pi_{\mathbf{X},\!\mathbf{Y}}^{K\ksp+\!1} } \!\!
	\Bigg( \overbrace{ L_p^{(c)}\big(\mathbf{X},\!\mathbf{Y},\pi^{0:K}\big) }^{ \text{\emph{localization} term} } +  \overbrace{ \gamma^p S\big(\pi^{0:K}\big) }^{ \text{\emph{switch} term} } 
	\\[-1mm] +
	\hspace{1mm}\underbrace{ \frac{c^p}{2} \big|\mathcal{M}\big(\mathbf{X},\!\mathbf{Y},\pi^{0:K}\big)\big| \!}_{ \text{\emph{missed objects} term} }
	\ + \ \underbrace{ \frac{c^p}{2} \big|\mathcal{F}\big( \mathbf{X}, \!\mathbf{Y},\pi^{0:K}\big)\big| }_{ \text{\emph{false alarms} term} }
	\Bigg)^{\!\!\!\sfrac{1}{p}}\!\!. \!\!\!\!
	\label{eq:TGOSPA:withDecomposition}
\end{multline}

To see how TGOSPA classifies into $L_p^{(c)}\big( \mathbf{X},\!\mathbf{Y},\pi^{0:K} \big)$~\eqref{eq:TGOSPA:LocTerm}, $S(\pi^{0:K})$~\eqref{eq:TGOSPA:SwitchTerm}, $\mathcal{M}\big(\mathbf{X}, \!\mathbf{Y},\pi^{0:K}\big)$~\eqref{eq:TGOSPA:MissObjTerm}, $\mathcal{F}\big(\mathbf{X},\!\mathbf{Y}, \pi^{0:K}\big)$~\eqref{eq:TGOSPA:FalseAlarmTerm}, the parameters $p$, $c$ and $\gamma$ need to be explained.

\subsection{General Meaning of TGOSPA Parameters}\label{subsec:TGOSPA-parameters-meaning}	
It can be seen that the distances counted in the localization term $L_p^{(c)}(\mathbf{X},\mathbf{Y},\pi^{1:K})$~\eqref{eq:TGOSPA:LocTerm} are \emph{cut-off} distances, i.e., each term in~\eqref{eq:TGOSPA:LocTerm} is always smaller than $c^p$.
\begin{minipage}{\linewidth}
	\vspace{1mm}
	\begin{wrapfigure}[8]{l}{0.4\linewidth}
		\centering
		\vspace{-3.8mm}
		\includegraphics[scale=1.19]{cutoff.pdf}
	\end{wrapfigure}
	That is, $c$ is the maximum localization error between a ground truth object and its estimate such that the ground truth \emph{can} be counted as properly estimated. 
	If an estimate is at a distance greater than $c$, the ground truth object and the estimate constitute a pair of missed/false \elementOfTrajectory{s}, and both are weighted with $\tfrac{c^p}{2}$.
	If two trajectories are assigned to each other but there is no \elementOfTrajectory{} at a particular time step in one of the trajectories, the corresponding cost $\tfrac{c^p}{2}$ is counted within either the missed objects or false alarms term, depending on which one is missing.
	\vspace{0.1cm}
\end{minipage}

Notice that the weight of both a false alarm and a missing estimate is the same\footnote{
	False and missing estimates are usually weighted the same in the CV scores as well, see~\cite{HOTA:2021}.
} and equal to $\tfrac{c^p}{2}$.
As a result, enlarging the cut-off parameter $c$ enlarges the weight of each false and missed target compared to the distance of any properly estimated object, which might be undesirable. 
\begin{minipage}{\linewidth}
	\vspace{1mm}
	\begin{wrapfigure}[11]{l}{0.55\linewidth}
		\centering
		\vspace{-3.2mm}
		\includegraphics[scale=1.22]{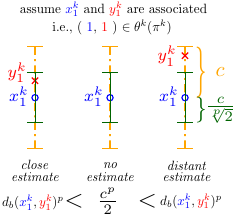}
	\end{wrapfigure}
	On the other hand, a $\it{distant\ estimate}$ (an estimate farther than $\tfrac{c}{ \sqrt[p]{2} }$ that is associated with the ground truth) leads to higher TGOSPA metric value than $\it{no\ estimate}$.
	As a result, setting $c$ too small may lead to preferring algorithms that are prone to missing objects over algorithms returning estimates (although more distant than $\tfrac{c}{ \sqrt[p]{2} }$).
	The borderline value $\tfrac{c}{ \sqrt[p]{2} }$ can be set closer to $c$ by enlarging $p$.
\end{minipage}

The parameter $p {\geq} 1$, in general, penalizes outliers.
That is, $p$ characterizes the discrepancy between a \emph{close} and \emph{distant} estimate in the sense of the metric $\baseMetric$.
If $p{=}1$, localization errors are considered in a \emph{uniform} manner. 
With an increasing value of $p$, estimates that are \emph{close} to the ground truth become more and more indistinguishable relative to estimates that are more \emph{distant} (but still closer than the value of $c$).
With increasing $p$, the missed/false objects term earns a greater impact on the final value of the TGOSPA metric since the term of a pair of a missed and a false target is always larger than any of the terms in the localization term.
If the value of the switching penalty $\gamma$ is larger than $c$, the switch term earns a greater impact on the final value of the TGOSPA metric compared to all other terms with an increasing value of $p$.

The half-switches in~\eqref{eq:TGOSPA:s:pi} ensure symmetry of the metric, and they penalize assignment-to-unassignment temporal changes in rows of $\pi^{0:K}$. 
This implies that the number of switches $S\big(\pi^{1:K}_{\star}\big)$~\eqref{eq:TGOSPA:SwitchTerm} need not be an integer, see~\cite[Appendix~A\textit{3)}]{TrajectoryGOSPA:2020}. 
The switching penalty $\gamma$ sets the weight of a single switch to be $\gamma^p$.
For $\gamma{=}0$, switches are not counted, and TGOSPA can be computed efficiently using the GOSPA metric (with parameter $\alpha{=}2$)~\cite[Sec.~IV.C]{TimeWeightedTGOSPA:2021} at every single time step.
TGOSPA with $\gamma {=} 0$, however, is not a metric on the space of finite sets of trajectories\footnote{
	To see this, consider arbitrary estimation results and \emph{connect} the estimates in time in two different ways to yield two different sets of trajectories.
	As the \emph{connections} are not considered by TGOSPA with $\gamma {=} 0$, the distance between the two sets is zero, although the trajectories are clearly different, violating the identity property of a metric.
	On the other hand, such a choice can be understood as computing the GOSPA metric for any individual time step, which is a metric on the space of finite sets of \elementOfTrajectory{s} introduced in~\cite{GOSPA:2017}.
}.
With an increasing number of $\gamma$, switches that \emph{seemingly} exist in the data may or may not be counted.
For extremely large values of $\gamma {\rightarrow} {+}\infty$, the switches become too costly to be present in $\pi^{0:K}_{\star}$ (i.e., counted as switches in the final TGOSPA value), and the estimates that are responsible for the \seeminglySwitch{s} become counted as false alarms, which may appear as counter-intuitive behavior or $\gamma$.
TGOSPA metric with extremely large $\gamma$ can be computed in a simplified manner in this case as well~\cite[Sec.~IV.C]{TimeWeightedTGOSPA:2021}.
The following Section shows how the (nonzero and finite) value of $\gamma$ can be interpreted geometrically alongside the other parameters.

\subsection{Setup of the Switching Penalty}\label{sec:setup-of-gamma}
In this section, simple rules are derived such that short and long-term \seeminglySwitch{s} are properly found and assessed within the TGOSPA metric as switches.
The rules give rise to two general methods for conveniently selecting $\gamma$.
To ease the notation, let $\mathbf{1}_{n\times m}$ be the matrix of ones and $\mathbf{0}_{n\times m}$ be the matrix  of zeros, both of size $n\!\times\!m$.

\subsubsection{Accounting for Short-term Interim \seeminglySwitchCAPITAL{s}}\label{sec:short-term-switches}
\begin{figure}
	\centering
	\includegraphics[width=0.47\textwidth]{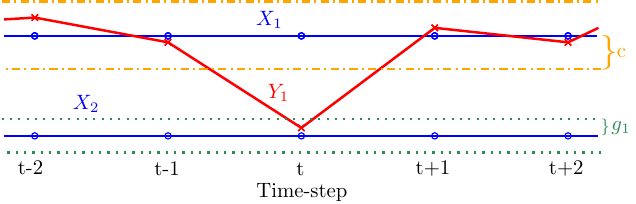}
	\caption{Short-term \seeminglySwitch{} scenario illustration.
    }
	\label{fig:aux_g1}
\end{figure}

Consider a scenario with two ground truth trajectories $X_1$ and $X_2$ and a single estimated trajectory $Y_1$ as shown in Fig.~\ref{fig:aux_g1}.
Seemingly, there are two switches in the scenario.
The first switch is because the trajectory $Y_1$ tracks $X_1$ before the time step $k{=}\kk$ and $X_2$ after that.
The second switch appears because $Y_1$ subsequently switches back to track $X_1$.
Examples of two possible assignments that may optimize the TGOSPA criterion are as follows.

\begin{itemize}
	\item \emph{No switch}: The trajectory $Y_1$ is assigned to trajectory $X_1$ for all time instants, i.e., no switch occurs. The corresponding assignment matrix is $\pi^{0:K}_{\text{no switch}} {=} \left[\begin{smallmatrix} \mathbf{1}_{1\times (K\!+\!1)} \\ \mathbf{0}_{1\times (K\!+\!1)} \end{smallmatrix}\right]$.
	\item \emph{Two switches}: The trajectory $Y_1$ is assigned to trajectory $X_1$ for all time instants \emph{except} the time step $\kk$, at which it is associated with trajectory $X_2$ i.e., two switches occur as explained above. The corresponding assignment matrix is $\pi^{0:K}_{\text{two switches}} {=} \left[\begin{smallmatrix} \mathbf{1}_{1\times \kk\ksp} ,& 0 ,&\mathbf{1}_{1\times (K\!-\ksp \kk)} \\ \mathbf{0}_{1\times \kk\ksp} ,&1 ,&\mathbf{0}_{1\times (K\!-\ksp \kk)}\end{smallmatrix}\right]$ that gives rise to four half-switches and thus $S(\pi^{0:K}_{\text{two switches}}){=}2$.
\end{itemize}
Depending on the value of $\gamma$, assume that one of the above assignments minimizes the TGOSPA criterion, i.e., is equal to $\pi^{0:K}_{\star}$.
The goal is to find the threshold value of $\gamma$ and corresponding geometrical conditions, for which the assignments are equally evaluated by the TGOSPA terms.

As depicted in Fig.~\ref{fig:aux_g1}, assume that the value of $c$ is 
such that\footnote{
	The value of $1{\leq}p{<}{+}\infty$ can be chosen arbitrarily. This holds in the next Section for long-term \seeminglySwitch{s} as well.
} $\baseMetric(x_1^\kk, y_1^\kk) {>} c$, i.e., assigning $Y_1$ to $X_1$ (which is the case for $\pi^{0:K}_{\text{no switch}}$) yields a pair of missed/false objects at the time step $\kk$.
If the assignment matrix $\pi^{0:K}_{\text{no switch}}$ minimizes $\pi^{0:K}_{\star}$ for some $\gamma$, the $p$-th power of TGOSPA is
\begin{align}
	& d_p^{(c,\gamma)}(\mathbf{X},\!\mathbf{Y})^p \!\!=\!\! \textstyle \sum_{k=0}^{\kk\ksp-\!1} \underbrace{ \left( \baseMetric( x_1^{k}, y_1^{k})^p + \tfrac{c^p}{2}\right) }_{ \hspace{-1cm}\substack{ x_1^k \text{ properly estimated with } y_1^k, \\ x_2^k \text{ is missed} }\hspace{-1cm} }
	\ \ + \ \ \overbrace{ \tfrac{c^p}{2} \underbrace{ \cdot \big( 2 + 1 \big)}_{\hspace{-1cm} \substack{ x_1^\kk, x_2^\kk\text{ are missed}, \\ y_1^\kk\text{ is false} } \hspace{-1cm}} }^{ \text{time step } \kk }
	\notag\\[-0.1cm]
	& \hspace{1.85cm}\textstyle +\sum_{k=\kk\ksp+\!1}^{K} \hspace{-0.1cm} \overbrace{ \left( \baseMetric( x_1^{k}, y_1^{k})^p + \tfrac{c^p}{2}\right) }.
	\label{eq:short-term-switch:no-switch}
\end{align}
If, on the contrary, it is the assignment matrix $\pi^{0:K}_{\text{two switches}}$ that minimizes  $\pi^{0:K}_{\star}$ for some other $\gamma$, 
\begin{align}
	& d_p^{(c,\gamma)}(\mathbf{X},\!\mathbf{Y})^p = \textstyle \sum_{k=0}^{\kk\ksp-\!2} \left( \baseMetric( x_1^{k}, y_1^{k} )^p + \frac{c^p}{2}\right)+ 
	\notag\\
	& \hspace{0.5cm} + \overbrace{ \baseMetric( x_1^{\kk\ksp-\!1}, y_1^{\kk\ksp-\!1})^p + \underbrace{ \tfrac{c^p}{2} + \gamma^p }_{ \hspace{-1cm} \substack{ x_2^k \text{ is missed}, \\ \text{two half-switches} } \hspace{-1cm} } }^{\text{time step } \kk\ksp-\!1} + \overbrace{ \baseMetric( x_2^{\kk}, y_1^{\kk})^p + \underbrace{ \tfrac{c^p}{2} + \gamma^p }_{ \hspace{-1cm} \substack{ x_1^\kk \text{ is missed}, \\ \text{two half-switches} } \hspace{-1cm} } }^{\text{time step } \kk}  
	\notag\\
	& \hspace{1.87cm}\textstyle + \sum_{k=\kk\ksp+\!1}^{K} \hspace{-0.1cm} \left( \baseMetric( x_1^{k}, y_1^{k})^p + \frac{c^p}{2}\right), 
	\label{eq:short-term-switch:one-switch}
\end{align}

The threshold for $\gamma$ for which the assignments yield the same TGOSPA metric value is the one for which~\eqref{eq:short-term-switch:no-switch} and~\eqref{eq:short-term-switch:one-switch} are equal, i.e.,
\begin{align}
	\overbrace{\cancel{\text{same terms}} \!+\! \tfrac{3 c^p}{2} }^\text{\emph{No switch} case~\eqref{eq:short-term-switch:no-switch} } =
	\overbrace{\cancel{\text{same terms}} \!+\!  d( x_2^{\kk}, y_1^{\kk})^p \!+\! \tfrac{c^p}{2} \!+\! 2\gamma^p}^\text{\emph{Two switches} case~\eqref{eq:short-term-switch:one-switch} }
	\!,\!\!
\end{align}
where the summation $ \sum_{ \substack{k=0 \\ k\neq\kk} }^{K} \left( \baseMetric( x_1^{k}, y_1^{k})^p + \tfrac{c^p}{2}\right) $ is referred to as ``same terms''.
That is, TGOSPA considers the scenario as a switch if and only if (iff)
\begin{align}
	\gamma < \left(\tfrac{c^p - d( x_2^{\kk}, y_1^{\kk})^p}{2}\right)^{\!\!\sfrac{1}{p}}.
	\label{eq:gamma<grom_dist}
\end{align}
In practice, one can select a threshold distance $g_1{<}c$ for $d( x_2^{\kk}, y_1^{\kk})$ that defines the boundary between the \emph{no switch} and \emph{two switches} assignments.
For a given $g_1$, switching penalty $\gamma$ can be computed as
\begin{align}
	\gamma = \left(\tfrac{c^p - g_1^p}{2}\right)^{\!\!\sfrac{1}{p}}.
	\label{eq:gamma-from-g_1}
\end{align}
From~\eqref{eq:gamma<grom_dist} it follows that whenever $d( x_2^{\kk}, y_1^{\kk}) {<} g_1$, the scenario is considered as a switch in TGOSPA.
Vice-versa, if $\gamma$ is selected such that $\gamma {<} \tfrac{c}{\sqrt[p]{2}}$, there exists $g_1$ such that
\begin{align}
	g_1 = \left( c^p - 2\gamma^p \right)^{\!\!\sfrac{1}{p}} .
	\label{eq:g_1-from-gamma}
\end{align}
An example of $g_1$ is depicted in Fig.~\ref{fig:aux_g1}, for which the scenario is considered a switch in TGOSPA.
Notice that the value of $\gamma$~\eqref{eq:gamma-from-g_1} is rather small when selected via $g_1 {<} c$, i.e., $\gamma\in(0,\tfrac{c}{\sqrt[p]{2}})$.
Accounting for short-term interim \seeminglySwitch{s} by selecting $\gamma$ is thus indicated in this paper with the term \GammaSmall{}.

When $\gamma$ is kept fixed, enlarging the parameter $p$ enlarges the threshold distance $g_1$~\eqref{eq:g_1-from-gamma} up to the (fixed) value of $c$.
When $p$ and $c$ are kept fixed, enlarging $g_1$ lowers the value of $\gamma$~\eqref{eq:gamma-from-g_1} and thus the penalty of a switch.

Setting $g_1$ and computing $\gamma$ using~\eqref{eq:gamma-from-g_1} is a convenient method for selecting $\gamma$ due to the simple graphical interpretation of $g_1$.
However, if there are more estimates/ground truth trajectories and/or the value of $c$ is considerably larger than the one depicted in Fig.~\ref{fig:aux_g1} (or alternatively the ground truth objects are considerably closer to each other), the interpretation of  $\gamma$ using $g_1$ described above is no longer valid.

Note that the subscript $1$ in $g_1$ indicates the concern about assigning the estimated trajectory $Y_1$ to $X_2$ for \emph{one} time step.
The following section considers multiple time steps for a slightly altered scenario where the \seeminglySwitch{} is permanent instead of interim.

Since $\gamma$~\eqref{eq:gamma-from-g_1} is small if set up via $g_1 {<} c$, note that long-term switches discussed next are considered as switches in this case as well.
In many applications, however, it is desirable to penalize switches in the data with a larger weight.
It follows that when setting $\gamma$ so large that $g_1$~\eqref{eq:g_1-from-gamma} no longer exists, no short-term switches will be found in the data.
On the other hand, certain \seeminglySwitch{s} (e.g., due to occlusion) will still be found in the data and will be penalized with the (large) value of $\gamma$.
The next section establishes the details regarding such larger values of $\gamma$.

\subsubsection{Accounting for Long-term \seeminglySwitchCAPITAL{s}}\label{sec:long-term-switches}
\begin{figure}
	\centering
	\includegraphics[width=.47\textwidth]{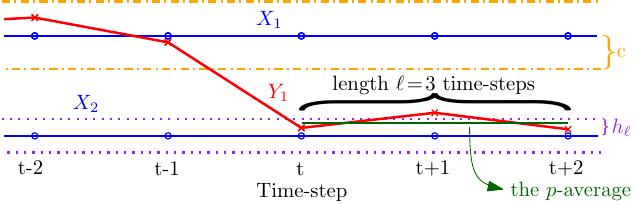}
	\caption{Long-term \seeminglySwitch{} scenario illustration.
	}
	\label{fig:aux_hn}
	\vspace{-0.5cm}
\end{figure}

Consider a scenario with two ground truth trajectories $X_1$ and $X_2$ and a single estimated trajectory $Y_1$ as shown in Fig.~\ref{fig:aux_hn}.
Seemingly, there is a single switch in the scenario because the trajectory $Y_1$ tracks $X_1$ before the time step $k{=}\kk$, but then it switches to track $X_2$.
The trajectory $Y_1$ then tracks $X_1$ for $\ell{=}3$ time steps, after which the trajectory $Y_1$ terminates.
Assuming that $Y_1$ gets assigned to $X_1$ for $k {=} 0,\dots,\kk\ksp-\!1$, it suffices to consider the following two assignments.

\begin{itemize}
	\item \emph{No switch}: The trajectory $Y_1$ is assigned to trajectory $X_1$ for all time instants i.e., no switch occurs. The corresponding assignment matrix is $\pi^{0:K}_{\text{no switch}} {=} \left[\begin{smallmatrix} \mathbf{1}_{1\times (K\!+\!1)} \\ \mathbf{0}_{1\times (K\!+\!1)} \end{smallmatrix}\right]$.
	\item \emph{One switch}: The trajectory $Y_1$ is assigned to trajectory $X_1$ only for $k {=} 0,\dots,\kk{-}1$, after which it gets assigned to the trajectory $X_2$.
	The corresponding assignment matrix is $\pi^{0:K}_{\text{one switch}}{=}\left[\begin{smallmatrix} \mathbf{1}_{1\times \kk\ksp} ,&\mathbf{0}_{1\times (K\!-\ksp \kk\!+\!1)} \\ \mathbf{0}_{1\times \kk\ksp} ,&\mathbf{1}_{1\times (K\!-\ksp \kk\ksp +\!1)}\end{smallmatrix}\right]$.
\end{itemize}
Depending on $\gamma$, it is assumed that either one of the above assignments minimizes TGOSPA.
As depicted in Fig.~\ref{fig:aux_hn}, assume that the value of $c$ is such that $\baseMetric(x_1^\kk, y_1^\kk) {>} c$, i.e., assigning $Y_1$ to $X_1$ (which is the case for $\pi^{0:K}_{\text{no switch}}$) yields a pair of missed/false objects at the time step $\kk$; and assume the same for the forthcoming time steps $\kk{+}1$, $\kk{+}2$ (and so on) as well until $Y_1$ terminates.

Again, the threshold value of $\gamma$, for which the assignments yield the same TGOSPA metric value, is the one for which $\pi^{0:K}_{\text{no switch}}$ and $\pi^{0:K}_{\text{one switch}}$ yield equal (minimum) value of TGOSPA.
Since the derivation follows the same steps as for the interim \seeminglySwitch{s}, only the result is given.
It follows that TGOSPA considers the scenario as a switch iff
\begin{align}
	\gamma < \textstyle \left( \ell\! \cdot\! c^p - \sum_{k=\kk}^{\kk\ksp+\!\ell-\!1} d( x_2^{\kk}, y_1^{\kk})^p \right)^{\!\!\sfrac{1}{p}},
	\label{eq:p-average-vs-h_n}
\end{align}
where $\ell$ is the number of time steps between $\kk$ and the end-time of $Y_1$ (assuming $Y_1$ tracks $X_2$ until it ends), further referred to as the \emph{length of the \seeminglySwitch{}}.
In the example depicted in Fig.~\ref{fig:aux_hn}, the value of $\ell$ is $3$.
It can be seen that one can select a threshold distance $h_\ell{<}c$ that defines the boundary between the \emph{no switch} and \emph{one switch} assignments.
For a given $h_\ell$, the switching penalty $\gamma$ can be computed as
\begin{align}
	\gamma = \left( \ell\!\cdot\!c^p - \ell\!\cdot\!h_\ell^p\right)^{\!\!\sfrac{1}{p}}.
	\label{eq:gamma-from-h_n}
\end{align}
From~\eqref{eq:p-average-vs-h_n} it follows that whenever 
\begin{align}
	\hspace{1cm} \underbrace{ \textstyle \left( \tfrac{1}{\ell}\sum_{k=\kk}^{\kk\ksp+\!\ell-\!1} d( x_2^{\kk}, y_1^{\kk})^p \, \right)^{\!\!\sfrac{1}{p}} }_{ \hspace{-1.5cm} p\text{-\emph{average} loc. error of } Y_1 \text{ w.r.t. } X_2 \text{ (after the \seeminglySwitch{})} \hspace{-1.5cm} } < h_\ell,
	\label{eq:h_n-rule}
\end{align}
the scenario with the corresponding value of $\ell$ will be considered as a switch in TGOSPA.
That is if the p-average localization error of $Y_1$ w.r.t.~$X_2$ (after the switch) is lower than a predefined threshold distance $h_\ell$.
An example $h_\ell{=}h_3$ is depicted in Fig.~\ref{fig:aux_hn}, for which the scenario is considered as a switch in TGOSPA if, moreover, the corresponding $p$-average is such that~\eqref{eq:p-average-vs-h_n} holds.

To account for \seeminglySwitch{s} that are long \emph{enough} only, one can select $n{>}0$ and compute
\begin{align}
	\gamma = \sqrt[p]{n} \!\cdot\! c, \label{eq:gamma-from-n}
\end{align}
so that $h_\ell {=} 0$, for all $\ell {=} 1,2,\dots,n$. 
In other words, any long-term \seeminglySwitch{} that lasts for exactly $\ell{=}n$ time steps or less than $n$ time steps will \emph{not} be considered as a switch in TGOSPA.
On the other hand, \seeminglySwitch{s} that last longer still can be considered as switches in TGOSPA.
In particular, combining~\eqref{eq:gamma-from-n} with~\eqref{eq:gamma-from-h_n} and~\eqref{eq:h_n-rule}, a \seeminglySwitch{} lasting for $\ell{=}n{+}m$ time steps, $m{>}0$, will be considered as a switch in TGOSPA iff
\begin{align}
	\textstyle \left( \! \tfrac{1}{n+m} \! \sum_{k=\kk}^{\kk\ksp+\!n+m-\!1} \!d( x_2^{\kk}, y_1^{\kk})^p  \right)^{\!\!\sfrac{1}{p}} \!\!<\! h_{n+m} \!=\!
	\sqrt[p]{ \tfrac{m}{n+m} } \!\cdot\! c,
	\label{eq:h_n+m_rule}
\end{align}
($n$ is user-defined and $\ell{=}n{+}m$ is length of a real \seeminglySwitch{}).
For a fixed $n$, $c$ and $p$, enlarging $m$ enlarges the threshold distance $h_{n+m}$~\eqref{eq:h_n+m_rule} up to $c$, i.e., \seeminglySwitch{s} that last longer may have larger ($p$-average) localization error to be considered as switches.
When $n$ and $c$ are fixed, enlarging $p$ enlarges $h_{n+m}$ for any $n,m>0$.

Considering that $n$ may be chosen arbitrarily large, the value of $\gamma$ chosen using~\eqref{eq:gamma-from-n} can also be arbitrarily large.
Accounting for long-term \seeminglySwitch{s} by selecting $\gamma$ is thus indicated in this paper with the term \GammaLarge{}.
However, it should be emphasized that the assumption that $Y_1$ gets assigned to $X_1$ for $k{=}0,\dots,\kk{-}1$ is crucial for the validity of the interpretation of~\eqref{eq:gamma-from-n}.
This assumption means that the estimated trajectory $Y_1$ first tracks $X_1$ for \emph{sufficiently} large number of time steps with a \emph{sufficient} accuracy.
Setting $n {>} \tfrac{K+1}{2}$ (note that $K{+}1$ is the total number time steps) and computing $\gamma$~\eqref{eq:gamma-from-n} can be expected to make TGOSPA behave as if $\gamma {\rightarrow} {+}\infty$ since no \seeminglySwitch{} could last for more than $\tfrac{K+1}{2}$ time steps.

Setting $h_n$ and computing $\gamma$ using~\eqref{eq:gamma-from-n} is a convenient method for selecting $\gamma$ due to the simple graphical interpretation of $h_n$.
Regarding short-term interim \seeminglySwitch{s}, note that if there are more estimates/ground truth trajectories and/or the value of $c$ is considerably larger (or the ground truth objects are considerably closer to each other), the interpretation of $\gamma$ using $h_n$ described above is no longer valid.

\begin{figure}[h]
	\centering
	\includegraphics[width=.47\textwidth]{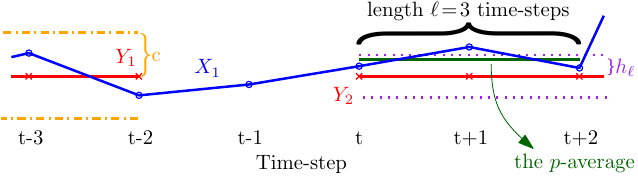}
	\caption{Track fragmentation scenario illustration.
	}
	\label{fig:aux_track-fragmentation}
	\vspace{-0.5cm}
\end{figure}

From the symmetry of the metric, note that the same rules apply if ground truth and estimates switch roles, i.e., for $\mathbf{X} {=}\{Y_1\}$ and $\mathbf{Y} {=}\{X_1,X_2\}$.
Also note that one can \emph{draw} the values of $c$, $g_1$, $h_\ell$, etc.,~relative to the estimates instead of the ground truth.
Hence, it should be emphasized that the same choice of $\gamma$~\eqref{eq:gamma-from-n} applies for track fragmentation and thus for assessing occlusions. 
This is illustrated in Fig.~\ref{fig:aux_track-fragmentation} for one ground truth trajectory and two estimates that lead to one switch in TGOSPA.

To sum up, the switching penalty $\gamma {\in} \Big(0, \sqrt[p]{\tfrac{K+1}{2}} {\cdot} c \Big)$ as explained in this subsection can be selected according to either one of the following two methods.
\begin{itemize}
	\item \GammaSmall{}: \textbf{to find and penalize short-term interim \seeminglySwitch{s}}.
	First, set the threshold distance $g_1$ such that $0{<}g_1{<}c$.
	Then compute $\gamma \!\leq\! \frac{c}{ \sqrt[p]{2} }$ according to~\eqref{eq:gamma-from-g_1}.
	Scenarios similar to those illustrated in Fig.~\ref{fig:aux_g1} will be considered as two switches in TGOSPA whenever the real estimate will be such that  $d( x_2^{\kk}, y_1^{\kk}) {<} g_1$.
	\item \GammaLarge{}: \textbf{to find and penalize long-term \seeminglySwitch{s} lasting for at least $n{+}1$ time steps}.
	Compute $\gamma {\geq} c$ according to~\eqref{eq:gamma-from-n}. 
	Scenarios similar to those illustrated in Fig.~\ref{fig:aux_hn} or~\ref{fig:aux_track-fragmentation} for the case $\ell {=} 3$ will be considered as one switch in TGOSPA whenever the real estimates will be such that~\eqref{eq:h_n+m_rule} holds, i.e., if the $p$-average for the particular length of the switch $\ell{=}n{+}m$ is below $h_{n+m}$~\eqref{eq:h_n+m_rule} for the chosen $n$.
\end{itemize}
It can be seen that given $c$ and $p$, the penalty $\gamma$ allows reflecting user preferences for the assessment of \seeminglySwitch{s} \emph{independently} of the application field (e.g., camera or radar tracking).
The value of $\gamma$ can be set indirectly through setting the parameter $g_1$ in the case of short-term interim \seeminglySwitch{s} and $n$ in the case of long-term \seeminglySwitch{s} using~\eqref{eq:gamma-from-g_1} and~\eqref{eq:gamma-from-n}, respectively.
Moreover, setting $n {>} \tfrac{K}{2}$ results in TGOSPA behaving as if  $\gamma{\rightarrow} {+}\infty$, and thus, no switches will be found in the data.

The following Section proposes a method for conveniently selecting the parameters $p$ and $c$, including the metric~$\baseMetric$, which are specific for a given application.
The CV setting of Section~\ref{sec:motivation} will be used.

\section{Application-dependent Parameters Selection}\label{sec:TGOSPA:CV}
The parameters should be chosen properly to rank different algorithms depending on the application field and the user preferences.
The first step is the selection of the metric $\baseMetric$, which is discussed in the CV setting in this paper.
Note that for TGOSPA to be a valid metric on the space of sets of trajectories, the function $d$ must be a metric on $\X$.
Concise selection of the cut-off $c$ together with the exponent parameter $p$ follows next.

\subsection{Selection of Metric for Bounding Boxes}\label{sec:bounding box-metrics}
As mentioned before, bounding boxes $x,y$ may be represented as vectors (elements of $\mathbb{R}^4$) or as geometrical entities (subsets of $\mathbb{R}^2$).
In the former case, common metrics such as the Euclidean or the maximum metrics can be readily used as in~\cite{KrKoSt:2023_FUSION}.
While the vector representation can be computed efficiently, the metric value depends on the particular chosen description of the bounding box\footnote{
	An element $x{\in}\mathbb{R}^4$ can be comprised of, e.g., the center, top-left or bottom-center point. 
	To capture the extent of the box, the width and height can be used as well as their radius $\tfrac{\text{width}}{2}$ and $\tfrac{\text{height}}{2}$.
	A chosen metric $\baseMetric$ on $\mathbb{R}^4$ leads to different values for different representations of the same boxes.
}.
That is, the user has to choose parameters representing a bounding box for which the estimation error is computed.
The latter case of representation using geometric entities is free of a particular bounding box description\footnote{
	The width and height of a box are both naturally nonnegative, which is immanent to the geometric representation as a set.
	To respect this within the vector representation, however, one should restrict $\mathbb{R}^4$ to some subset.
} and can have favorable geometric interpretations. 

The CV community makes extensive use of the IoU\footnote{
	Generalizations of the IoU exist in the literature, see, e.g., \cite{GIoU:Rezatofighi:2018,Cheng:BIoU:2021,Zheng:GeomFactorsCIoU:2022}.
} that is a \emph{similarity score} defined as
\begin{align}
	\mathrm{IoU}( x,y ) \!=\! \frac{ \text{Area}( x \cap y ) }{ \text{Area}( x \cup y ) },
	\hspace{0.2cm}
	\raisebox{-0.6cm}{
		\includegraphics[width=2.4cm]{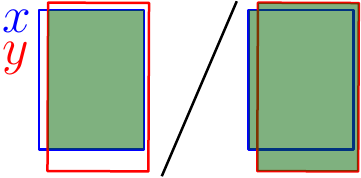}
	}
	\label{eq:IoU-definition}
\end{align}
which is equal to one if the bounding boxes (rectangles containing their interiors) $x$ and $y$ coincide and zero if they have no overlap at all.
Otherwise, the IoU~\eqref{eq:IoU-definition} measures the relative overlap of the sets.
For convenience, the $\mathrm{Area}$ in~\eqref{eq:IoU-definition} is taken as the Lebesgue measure, and the sets $x,y {\subset} \mathbb{R}^2$ are assumed to be non-empty and Lebesgue-measurable.
With this, it is easy to show that the IoU~\eqref{eq:IoU-definition} is scale-invariant\footnote{
	The proof is a simple consequence of the scaling property of the Lebesgue measure~\cite[2.20~Theorem~(e)]{Rudin:Real-and-Complex-Analysis:1987}: take the linear transformation in the theorem to be any nonzero scale.
	The division in~\eqref{eq:IoU-definition} then makes the constant granted by the theorem cancel out.
}.
The function defined in~\cite{KrKoSt:2023_FUSION} as
\begin{align}
	d_{\mathrm{IoU}} (x, y) = 1 - \mathrm{IoU}(x,y)
	\label{eq:IoU-induced-metric-d}
\end{align}
is thus also scale-invariant, it is moreover a metric\footnote{
	Assuming the sets $x,y$ are non-empty and finite, and taking $\mathrm{Area}$ to be the cardinality, the function $d_{\mathrm{IoU}}(x,y)$ was shown to be a metric in~\cite{LEVANDOWSKY:1971:1-IOU}.
	Namely, the triangle inequality was shown to hold for such $d_{\mathrm{IoU}}(x,y)$.
	As the steps in~\cite{LEVANDOWSKY:1971:1-IOU} are valid for taking $\mathrm{Area}$ to be any sigma-finite measure (as far as the sets $x,y$ are measurable and both have \emph{finite} measure), the proof is valid for the Lebesgue measure and bounding boxes.
} and it will be called the IoU-induced metric in this paper.
For its favorable properties, the IoU-induced metric is chosen as the metric $\baseMetric$ in the following considerations.
Two alternative metrics are discussed in Appendix~\ref{appendix:alternative-metrics}, which could be readily used as well.

\subsection{Selection of Cut-off and Exponent Parameters}
Suitable values of $c$ and $p$ naturally depend on the application at hand.
In some cases, the selection of $c$ can be done directly depending on the maximum allowable localization distance such that a ground truth and an estimate can be considered assigned, e.g., for the evaluation in 3D space.
In general, however, the selection may be challenging, e.g., for the evaluation in 2D space where the data are under the effect of perspective projection. 
Although the IoU-induced metric $d_{\mathrm{IoU}}$~\eqref{eq:IoU-induced-metric-d} mitigates the perspective projection effects, the additional selection of $p$ is arguably rather unintuitive.
Although the choices $p{=}1$ or $p{=}2$ seem \emph{natural}, such choices may barely reflect application-dependent user preferences. 

A~method to select $c$ based on data (henceforth referred to as the $c$-selection method) was presented\footnote{
	\cite{KrKoSt:2023_FUSION} dealt with assigning detections to ground truth bounding boxes to estimate the measurement noise covariance matrix.
} in~\cite{KrKoSt:2023_FUSION}.
In the following, the $c$-selection method is extended to fit into the performance evaluation setting of this paper, and a method for joint selection of $c$ and $p$ is proposed.

The main idea of the proposed method is to choose, analyze, and visualize sample data, forming the following three steps:
\begin{itemize}
	\item[\textit{1.}] choose application-relevant sample data,
	\item[\textit{2.}] based on the $c$-selection method, process the data to form example distances in the context of the application,
	\item[\textit{3.}] jointly select $c$ and $p$ based on histogram count and, if possible, visual specimen of the distances.
\end{itemize}

\subsubsection{Application-relevant Sample Data}
First, data based on which the selection of $c$ and $p$ is to be analyzed must be chosen. 
\emph{(i)} The data should include ground truth and estimated trajectories (perhaps from several algorithms and videos), 
\emph{(ii)} The estimated trajectories are \emph{diverse} enough to include both \emph{good} and \emph{bad} estimates (according to the user).
\emph{(iii)} The data should form a relevant sample for the application, for which the evaluation will be done with the selected $c$ and $p$.

The scenario studied in this paper involves 2D bounding boxes resulting from pedestrians walking near a static camera, which is aligned approximately parallel with the ground.
The particular scenario lasts only for 61 frames, involves only two pedestrians, and is taken from the MOT17-09 video for which the same description applies.
It can be assumed that the data from the \emph{entire} video \mbox{MOT17-09} (ground truth and trajectory estimates from all the applied algorithms) fulfills all the above requirements.

\newcommand{\assignmentsOnC}{ \pi^{0:K}_{\star}|_{2}^{(c,0)} }
\subsubsection{Extension of the $c$-selection Method}
Consider ground truth trajectories $\mathbf{X}$ and estimated trajectories $\mathbf{Y}$ produced by an algorithm for a certain video of the chosen data.
Four so-called \emph{guideline} functions were introduced in the $c$-selection method, whose argument is the cut-off $c$.
For simplicity, only two of the guideline functions are considered here, namely \emph{(i)} \emph{total number of assignments} and \emph{(ii)} \emph{sum of the squared distances}.
The guideline functions are constructed upon assignments resulting from computing the GOSPA metric with $p{=}2$ at each single time step for different values of $c$, i.e., the assignments $\assignmentsOnC$ resulting from the computation of $d_{2}^{(c,0)}$~\eqref{eq:TGOSPA:definition}.
In the terminology of this paper, the guideline function \emph{(i)} is the number of properly estimated objects \emph{for given} $c$, which is further shortened as
\begin{align}
	N(c) &= \textstyle N^{(c)}\Big( \mathbf{X}, \!\mathbf{Y}, \assignmentsOnC \Big),
	\label{eq:guideline:N}
\end{align} 
and the guideline function \emph{(ii)} is the localization term \emph{for given } $c$, which is further shortened as
\begin{align}
	L(c) &= \textstyle L_2^{(c)}\Big( \mathbf{X}, \!\mathbf{Y}, \assignmentsOnC \Big).
	\label{eq:guideline:D}
\end{align}

\begin{figure}[h!]
	\centering
	\includegraphics[width=0.45\textwidth]{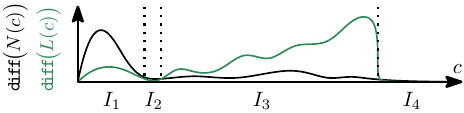}
	\caption{Graphical sketch of expected properties of the increments of $N(c)$~\eqref{eq:guideline:N} and $L(c)$~\eqref{eq:guideline:D} taken from~\cite{KrKoSt:2023_FUSION}.}
	\label{fig:guideline-functions:sketch}
\end{figure}

\cite{KrKoSt:2023_FUSION} argued that four subsequent intervals $I_1$, $I_2$, $I_3$ and $I_4$ of $c\geq0$ can be determined based on the guideline functions, that can be summarized as follows.
\begin{itemize}
	\item[$I_1$] The number of assignments increases rapidly.
	Close estimates get assigned, most of which are seemingly \emph{correct} and minimum are false alarms.
	The function $N(c)$~\eqref{eq:guideline:N} can be expected to increase rapidly in this interval up to a certain level, indicating that most of the correct estimates have been assigned while minimum false alarms have been included, which is the right endpoint of $I_1$.
	In $L(c)$~\eqref{eq:guideline:D}, a large number of small increments is expected for $c {\in} I_1$, and thus its values can be arbitrarily large, offering little information about $I_1$.
	\item[$I_2$] Only correct detections with the largest error get associated, while only a small number of false alarms are used.
	This interval includes convenient distance values that can be used in the evaluation as the cut-off $c$.
	The function $N(c)$~\eqref{eq:guideline:N} should not change much in this interval (and also in the following intervals).
	Similarly, the value of $L(c)$~\eqref{eq:guideline:D} can be expected nearly constant for $c {\in} I_2$.
	\item[$I_3$] A Slow increase in the number of assignments is caused primarily by assigning distant estimates that are seemingly false alarms.
	As a result, $N(c)$~\eqref{eq:guideline:N} should increase only occasionally, and whenever this happens, the distance is expected to be large.
	Therefore, $L(c)$~\eqref{eq:guideline:D} can be expected to have large occasional increments indicating that distant false alarms are being associated.
	\item[$I_4$] There are no more assignments possible in the data.
\end{itemize}
The expected behavior of the guideline functions is illustrated in Fig.~\ref{fig:guideline-functions:sketch}, where $\texttt{d\hspace{-0.5mm}i\hspace{-0.5mm}f\hspace{-0.5mm}f} \big( \, f(c) \big) = f(c+\Delta c)-f(c)$, computes the increments of the function $f$, with $\Delta c>0$ being a user-defined parameter (bin width).

\newcommand{\summandsForHistogram}{ \big\{ \baseMetric(x_i^k, y_j^k) \big\}_{k,i,j} }
\newcommand{\binCenter}{ \texttt{b\hspace{-0.4mm}i\hspace{-0.4mm}n\hspace{-0.4mm}\_\hspace{-0.4mm}c\hspace{-0.4mm}e\hspace{-0.4mm}n\hspace{-0.4mm}t\hspace{-0.4mm}e\hspace{-0.4mm}r} }
It should be pointed out that both functions $\texttt{d\hspace{-0.5mm}i\hspace{-0.5mm}f\hspace{-0.5mm}f} \big( N(c) \big)$~\eqref{eq:guideline:N} and $\texttt{d\hspace{-0.5mm}i\hspace{-0.5mm}f\hspace{-0.5mm}f} \big( L(c) \big)$~\eqref{eq:guideline:D} can be efficiently approximated via computing the histogram of $L(c_{\max})$~\eqref{eq:guideline:D} summands $\summandsForHistogram$ for a single large value $c_{\max}$, as
\begin{align}
	\texttt{d\hspace{-0.5mm}i\hspace{-0.5mm}f\hspace{-0.5mm}f} \big( N(c) \big) &\approx \texttt{h\hspace{-0.5mm}i\hspace{-0.5mm}s\hspace{-0.5mm}t\hspace{-0.5mm}o\hspace{-0.5mm}g\hspace{-0.5mm}r\hspace{-0.5mm}a\hspace{-0.5mm}m} \Big[ \summandsForHistogram \Big](c) , \label{eq:diff-N-hist}\\
	\texttt{d\hspace{-0.5mm}i\hspace{-0.5mm}f\hspace{-0.5mm}f} \big( L(c) \big) &\approx \texttt{d\hspace{-0.5mm}i\hspace{-0.5mm}f\hspace{-0.5mm}f} \big( N(c) \big) \cdot \binCenter{}(c)^2 , \label{eq:diff-L-hist}
\end{align}
where $\binCenter{}(c)$ is the center of the bin of the computed histogram~\eqref{eq:diff-N-hist} closest to the value of $c$.
Moreover, one can use the set of summands $\summandsForHistogram$ collected from multiple algorithms and/or videos chosen for the selection of $c$ and $p$ to compute $\texttt{d\hspace{-0.5mm}i\hspace{-0.5mm}f\hspace{-0.5mm}f} \big( N(c) \big)$~\eqref{eq:diff-N-hist} and $\texttt{d\hspace{-0.5mm}i\hspace{-0.5mm}f\hspace{-0.5mm}f} \big( L(c) \big)$~\eqref{eq:diff-L-hist}.

\begin{figure}[h!]
	\centering
	\includegraphics[width=0.425\textwidth,trim={0.6cm, 0.1cm, 0.6cm, 0.6cm},clip]{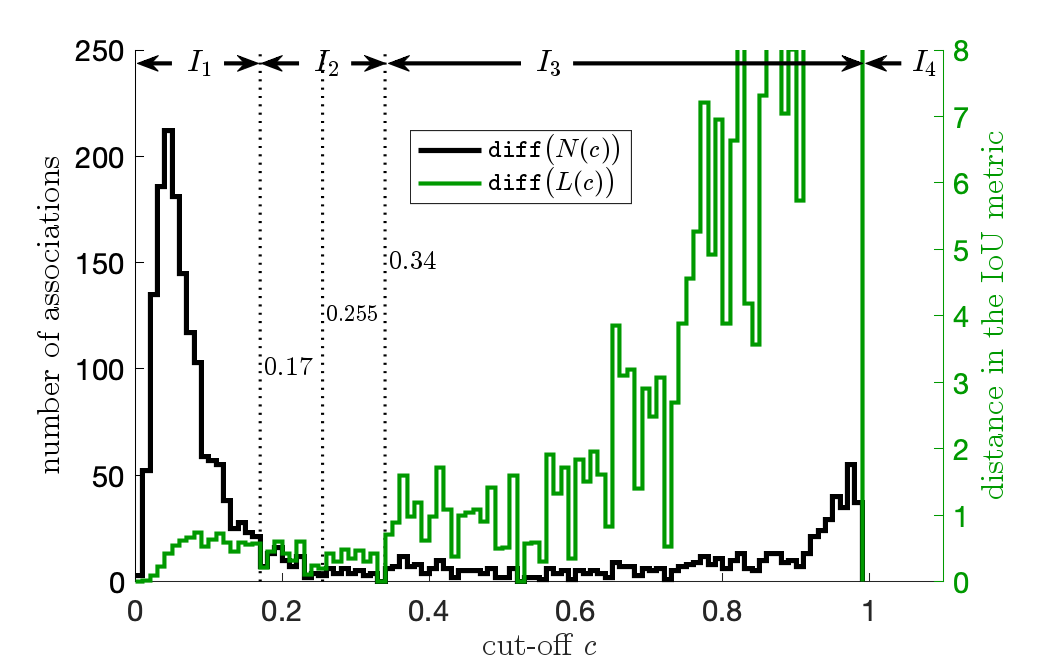}
	\caption{Increments of the guideline functions $\texttt{d\hspace{-0.5mm}i\hspace{-0.5mm}f\hspace{-0.5mm}f} \big( N(c) \big)$~\eqref{eq:diff-N-hist} and $\texttt{d\hspace{-0.5mm}i\hspace{-0.5mm}f\hspace{-0.5mm}f} \big( L(c) \big)$~\eqref{eq:diff-L-hist} for the entire MOT17-09 video and sample distances collected from all the algorithms described in Section~\ref{sec:algorithms}.}
	\label{fig:IoU-histogram-all}
\end{figure}

Using the IoU-induced metric $\baseMetric {=} d_{\mathrm{IoU}}$~\eqref{eq:IoU-definition}, the value $c_{\max} {=} 1$ can readily be used.
The guideline functions computed from all the algorithms applied to the MOT17-09 video are shown in Fig.~\ref{fig:IoU-histogram-all}, where the exemplified intervals $I_1$, $I_2$, $I_3$ and $I_4$ were determined by hand based on the expected behavior (Fig.~\ref{fig:guideline-functions:sketch}).

\begin{figure*}[t]
	\centering
	\begin{tikzpicture}
		\node[anchor=north west] (upper) at (0,0) {\includegraphics[scale=0.42,trim={6cm, 5.9cm, 4cm, 0.4cm},clip]{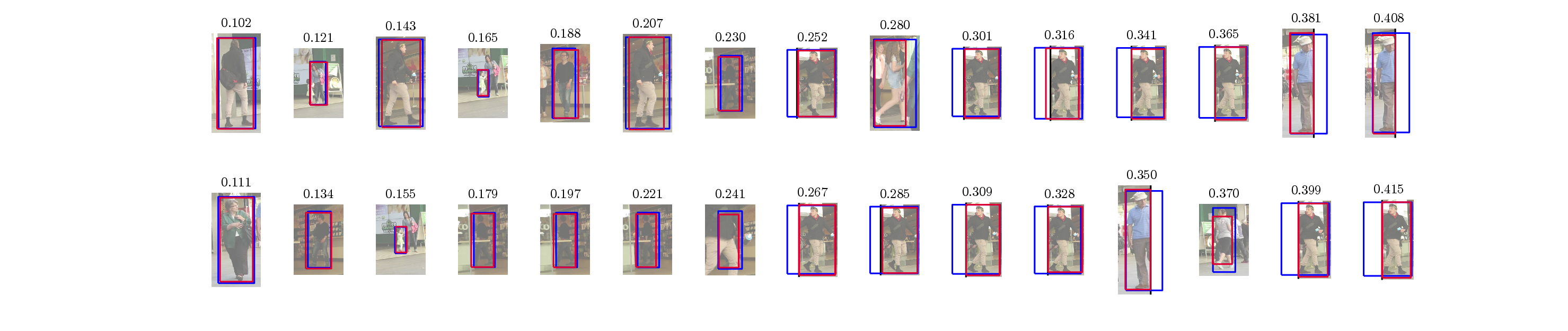}};
		\node[anchor=north,shift={(0,3mm)}] (lower) at (upper.south) {\includegraphics[scale=0.42,trim={6cm, 1.1cm, 6.6cm, 5.2cm},clip]{IoU-bboxes.png}};
		\coordinate (I2Left) at (4.35cm,0);
		\coordinate (I2Middle) at (8.75cm,0);
		\coordinate (I2Right) at (12.65cm,0);
		\draw[dotted, line width=0.2mm] (I2Left) -- (I2Left|-lower.south);
		\draw[dotted, line width=0.2mm] (I2Middle) -- (I2Middle|-lower.south);
		\draw[dotted, line width=0.2mm] (I2Right) -- (I2Right|-lower.south);
		\node[anchor=south west,shift={(0,-0.2cm)},scale=0.6] at (I2Left) {$0.17$};
		\node[anchor=south west,shift={(0,-0.2cm)},scale=0.6] at (I2Middle) {$0.255$};
		\node[anchor=south west,shift={(0,-0.2cm)},scale=0.6] at (I2Right) {$0.34$};
	\end{tikzpicture}
	\caption{Examples of bounding box pairs ordered using the IoU-induced metric. The value of the metric is given above each pair of boxes such that it grows from left to right. Each blue box is a ground truth box, while each red one is an estimate.  The examples are drawn based on Fig.~\ref{fig:IoU-histogram-all}.}
	\label{fig:IoU-bboxes}
\end{figure*}  

At this point, the sample distances $\summandsForHistogram$ from the data (multiple algorithms/videos) have been ordered from the smallest to the largest.
If possible, the distances should be visualized in the context of the application including the bounding box pairs $x_i^k,y_j^k$ giving rise to the distance $\baseMetric(x_i^k,y_j^k)$.
The visualization should respect the ordering and disregard the information about the algorithm that produced the particular estimate.
For the exemplary case, the visualization is given in Fig.~\ref{fig:IoU-bboxes}.

It should be emphasized that the data used for drawing both Fig.~\ref{fig:IoU-histogram-all} and Fig.~\ref{fig:IoU-bboxes} contain boxes from a visual detector and several tracking algorithms.
Visual detectors on their own, however, are likely to yield different histograms and thus lead to different $c$ and $p$ suitable for detector evaluation, and vice versa.
Furthermore, different users may determine different interval edges for the same data depending on the application at hand, especially the right-hand edge of $I_2$ can arguably be selected much larger for Fig.~\ref{fig:IoU-histogram-all}.

To sum up, the extended method analyses distance distributions of estimates and it was exemplified for several conceptually different algorithms from the CV domain.
The presented results (Fig.~\ref{fig:IoU-histogram-all} and~\ref{fig:IoU-bboxes}) thus likely offer enough insight for many applications. 
The results thus can be re-used in practice, especially since the method is rather complicated.

\subsubsection{Joint Selection of $c$ and $p$}
As discussed above, the cut-off $c$ can be viewed as the maximum \emph{possible} error for an estimate to be considered proper.
The convenient value of $c$ for the chosen data should lie in the interval $I_2$, and its particular selection is made by the user (by hand) ideally with the help of the data visualization such as in Fig.~\ref{fig:IoU-bboxes}.

At the same time, the value $a{=}\tfrac{c}{ \sqrt[p]{2} }$ can be understood as a maximum \emph{admissible} error such that the further estimates are penalized in TGOSPA more compared to the case of missing estimate (see Section~\ref{subsec:TGOSPA-parameters-meaning}).
As $a$ is a distance, the user can easily select $a {\in} ( \tfrac{c}{2}, c)$ similarly to selecting $c$ from the data.
The exponent parameter $p {\geq} 1$ can then be computed as
\begin{align}
	p = \tfrac{ \log(2) }{ \log(c) - \log(a) } , \label{eq:select-a-c-and-compute-p}
\end{align}
which concludes the proposed method.

It is important to note that step \textit{3.} may be sufficient on its own for some applications, e.g., for the evaluation in 3D with the Euclidean distance where $c$ and $a$ can be chosen without relying on a particular dataset.

Three possible selections of $c$ and $p$ are discussed in the following section, together with the performance evaluation of algorithms in the discussed CV scenario.
The purpose of the section is to provide intuitive insight into TGOSPA.

\section{Numerical Examples}\label{sec:numerical:examples}
In this Section, TGOSPA is evaluated using the LP metric implementation from~\cite{TrajectoryGOSPA:2020}. 
The set of ground truth trajectories $\mathbf{X}$ includes only \emph{gt2} and \emph{gt6} for the $61$ frames considered (the final time step is $K{=}60$).
The TGOSPA parameters are chosen to elucidate, especially the effect of the switching penalty $\gamma$, and show how it can be used for different purposes.
In particular, the following four key configurations are used and presented in each of the following tables.
\begin{itemize}
	\item \GammaZero: $\gamma {=} 0$ and no switches are assessed.
	This case can be implemented using the simpler GOSPA metric at each time step and can be used for applications where information concerning trajectories is not present or needed, such as for training visual object detectors.
	\item \GammaSmall: $\gamma \in (0, \frac{c}{\sqrt[p]{2}})$ is selected according to the method proposed in Section~\ref{sec:short-term-switches} to detect and penalize short-term \seeminglySwitch{s}.
	The particular value of $\gamma$~\eqref{eq:gamma-from-g_1} has been selected such that the threshold distance $g_1 {=} \tfrac{3}{4}c$ regardless of $p$.
	Such parametrization could be used for applications where any \seeminglySwitch{} matters, such as for assessing and training re-ID modules.
	\item \GammaLarge: $\gamma {=} n^{1/p} c$ with $n {=} 10$ is selected according to the method proposed in Section~\ref{sec:long-term-switches} to detect and penalize long-term \seeminglySwitch{s} lasting for at least $11$~time steps.
	Such parametrization could be used for most practical tracking applications to conveniently assess \seeminglySwitch{s} such as illustrated in Fig.~\ref{fig:aux_hn}, or to assess how tracking algorithms cope with occlusions such as illustrated in Fig.~\ref{fig:aux_track-fragmentation}.
	\item \GammaExtreme: $\gamma {=} n^{1/p} c$  with $n{=}31{>}\tfrac{K+1}{2} {=}30.5$ is selected according to the findings of Section~\ref{sec:long-term-switches} to neglect \emph{any} switches that may arise in the studied scenario consisting of 61 frames. 
	As trajectories are assigned one-to-one, \seeminglySwitch{s} that may be recognized as switches by humans are treated as pairs of missed and false targets within the evaluation.
	This case can be implemented in a simplified manner~\cite[Sec.~IV.C]{TimeWeightedTGOSPA:2021}. 
\end{itemize}

The remaining TGOSPA metric parameters were chosen based on Figures~\ref{fig:IoU-histogram-all} and~\ref{fig:IoU-bboxes}.
The evaluation results are given in Tables~\ref{tab:TGOSPA:IoU:experimentOne},~\ref{tab:TGOSPA:IoU:experimentTwo}, and~\ref{tab:TGOSPA:IoU:experimentThree} where the following three combinations of the parameters $p$ and $c$ values are used:
\newcommand{\ExperimentOne}{\textit{\textbf{Combination A}}}
\newcommand{\ExperimentTwo}{\textit{\textbf{Combination B}}}
\newcommand{\ExperimentThree}{\textit{\textbf{Combination C}}}
\begin{itemize}
	\item \ExperimentOne{}:
	the cut-off $c {=} 0.34$ is chosen as the right endpoint of Interval $I_2$ and $p{=}1$ so that $a{=}\tfrac{c}{ \sqrt[p]{2} } {=} 0.17$ is the left endpoint of $I_2$.
	\item \ExperimentTwo{}:
	the cut-off $c {=} 0.255$ is chosen as the middle point of Interval $I_2$ and $p {=} 1.71$ so that $a {=} 0.17$ is the left endpoint of $I_2$ as before.
	\item \ExperimentThree{}:
	the cut-off $c {=} 0.34$ is chosen as the right endpoint of Interval $I_2$ and $p{=}2.409$ so that $a {=} 0.255$ is the middle endpoint of $I_2$.
\end{itemize}
The IoU-induced metric $\baseMetric_{\mathrm{IoU}}$ for bounding boxes was used in all the examples.
Every cell of each table~\ref{tab:TGOSPA:IoU:experimentOne},~\ref{tab:TGOSPA:IoU:experimentTwo} and~\ref{tab:TGOSPA:IoU:experimentThree} shows the value of the metric and its decomposition as shown in Fig.~\ref{fig:cell-description}.
For a given algorithm results contained in $\mathbf{Y}$ and $\pi^{0:K}_{\star}$ resulting from solving~\eqref{eq:TGOSPA:definition} (or~\eqref{eq:TGOSPA:withDecomposition}), the value 
\begin{align}
	\avgLocError\big(\mathbf{X},\!\mathbf{Y},\pi^{0:K}_{\star} \big) = \Big( \tfrac{ L_p^{(c)}(\mathbf{X},\!\mathbf{Y},\pi^{0:K}_{\star} ) }{ N^{(c)}( \mathbf{X}, \!\mathbf{Y}, \pi^{0:K}_{\star}) } \Big)^{\!\!\sfrac{1}{p}} \!\!, \label{eq:p-average-loc-error} 
\end{align}
is the $p$-average localization error.

\begin{figure*}[h]
	\centering
	\includegraphics[width=0.87\linewidth,trim={9.8cm, 2.5cm, 10.8cm, 3.5cm},clip]{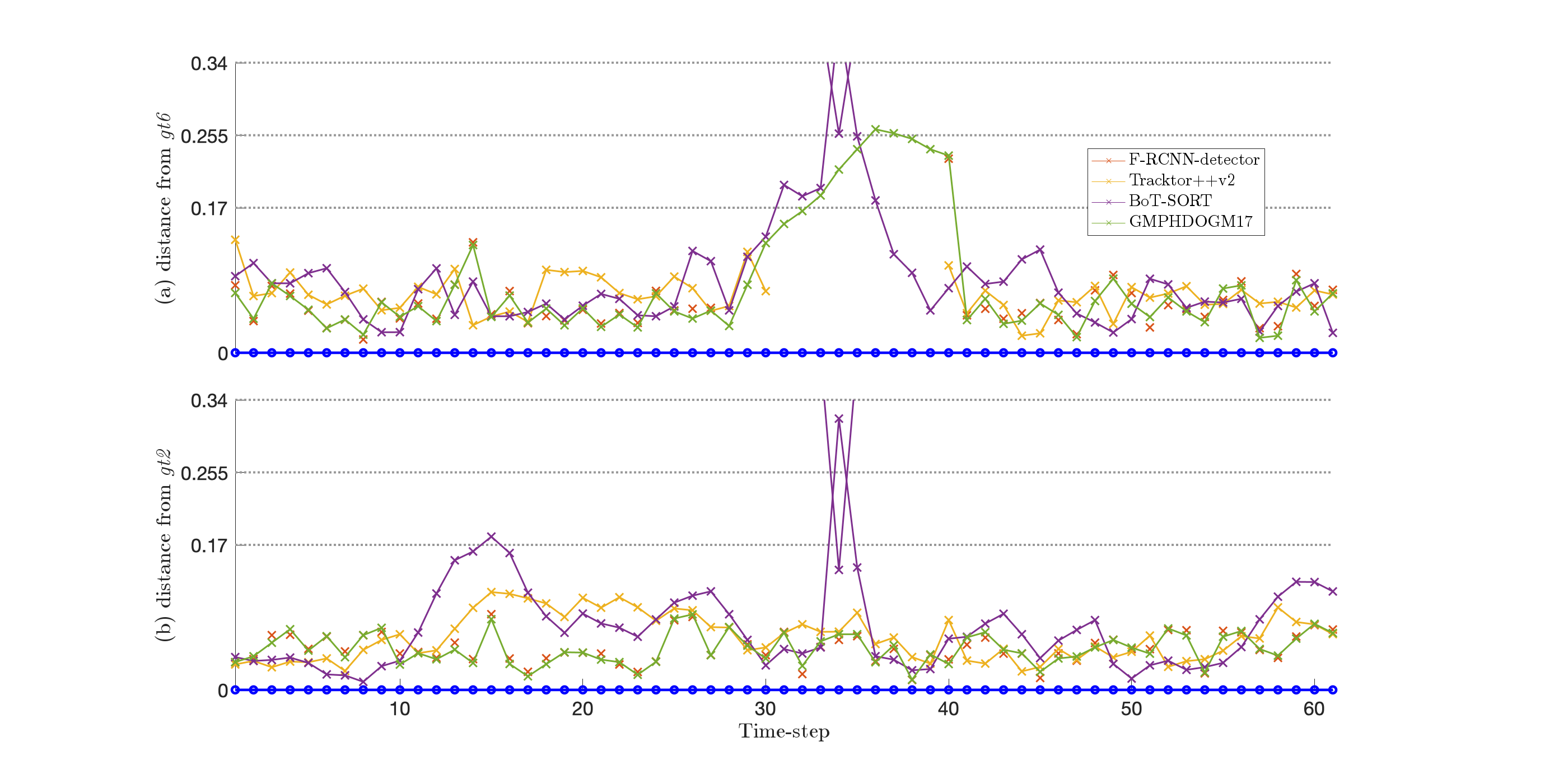}
	\caption{The IoU-induced distance of individual estimates from each ground truth bounding box.
		Note that the F-RCNN detections depicted in red are not connected in time, and each detection is treated as a single trajectory containing a single \elementOfTrajectory{}.
		There are two different trajectories for Tracktor++v2 in Subfigure (a), depicted in yellow.}
	\label{fig:gtdistance_plot}
	\vspace{-0.5cm}
\end{figure*}

To visualize the data, distances between the estimates and ground truth bounding boxes are shown in Fig.~\ref{fig:gtdistance_plot}.
It should be noted that the studied data do not contain any \emph{distant} false alarms or estimates that might be associated when increasing the value of $c$ beyond $0.34$.
It can be seen that most estimates have errors lower than $0.17$ in the IoU-induced metric (except for BoT\_SORT at the $16$-th time step), and there are significant errors merely during the occlusion of \emph{gt6} either due to missing estimates (FRCNN, Tracktor++v2) or slightly more distant estimates (BoT\_SORT, GMPHDOGM17).
For the track fragmentation appearing in Tracktor++v2, the length of the corresponding \seeminglySwitch{} is $\ell{=}22$ as defined in Section~\ref{sec:long-term-switches} (i.e., assuming its first estimated trajectory is assigned to \emph{gt6}).

Next, several observations are pointed out based on the results from Fig.~\ref{fig:gtdistance_plot}, and Tables~\ref{tab:TGOSPA:IoU:experimentOne},~\ref{tab:TGOSPA:IoU:experimentTwo}, and~\ref{tab:TGOSPA:IoU:experimentThree}.

\begin{figure*}[h]
	\centering
	\scalebox{1.6}{
		\createcellannotated{{1.305}{1.128}{0.435}{\color{lightgray}0}{0.013}{113}{9}{{\color{lightgray}0}}{\hspace{0.1cm}1}{0.068}{2}}
	}
	\caption{Description of a single cell of Tables~\ref{tab:TGOSPA:IoU:experimentOne},~\ref{tab:TGOSPA:IoU:experimentTwo} and~\ref{tab:TGOSPA:IoU:experimentThree}. The example is taken from Table~\ref{tab:TGOSPA:IoU:experimentTwo}: Tracktor++v2.}
	\label{fig:cell-description}
	\vspace{-0.5cm}
\end{figure*}

\newcommand{\tabSmall}[1]{ \hspace{17mm} \scalebox{0.82}{\it #1}}
\newcommand{\nameInTabFRCNN}{FRCNN \tabSmall{temporarily disconnected} \tabSmall{estimates}}
\newcommand{\nameInTabTracktor}{Tracktor++v2 \tabSmall{$1{\times}$long-term \seeminglySwitch{}}}
\newcommand{\nameInTabBoTSORT}{BoT\_SORT \tabSmall{$2{\times}$short-term interim} \tabSmall{\seeminglySwitch{}}}
\newcommand{\nameInTabGMPHDOGM}{GMPHDOGM17 \tabSmall{no \seeminglySwitch{}}}

\begin{table*}[]
	\centering
	\caption{\ExperimentOne{}: evaluation using the IoU metric with TGOSPA parameters $p{=}1$ and $c{=}0.34$ ($a{=}0.17$).}
	\label{tab:TGOSPA:IoU:experimentOne}
	\begin{tabular}{p{32mm}cccc}
		\toprule
		& \footnotesize \GammaZero & \footnotesize \GammaSmall & \footnotesize \GammaLarge & \footnotesize \GammaExtreme \\
		& \footnotesize No switch matter &\footnotesize Any switch matter &\footnotesize Only switches lasting for &\footnotesize One-to-one trajectory \\
		& &\footnotesize (the more, the worse) &\footnotesize $\scriptstyle \ell>10$ time steps matter &\footnotesize matching \\
		& $\scriptstyle \gamma=0$ & $\scriptstyle \gamma=0.043$ &$\scriptstyle \gamma=3.4$ , ($\scriptstyle n=10$)& ``$\scriptstyle \gamma \rightarrow \infty$'' \\
		&\footnotesize (GOSPA) & $\scriptstyle g_1 = 0.255$ & $\scriptstyle h_{10}=0$, $\scriptstyle h_{11}=0.0309$ & $\scriptstyle h_{\ell} = 0$, $\scriptstyle \forall \ell \leq \tfrac{\text{no. frames}}{2}$ \\
		\midrule
		{ \nameInTabFRCNN }&
		\createcell{{7.828}{5.788}{2.04}{\color{lightgray}0}{\color{lightgray}0}{110}{12}{\color{lightgray}0}{\color{lightgray}-}{0.053}{1}}&
		\createcell{{12.418}{5.788}{2.04}{\color{lightgray}0}{4.59}{110}{12}{\color{lightgray}0}{108}{0.053}{4}}&
		\createcell{{38.787}{0.027}{20.4}{18.36}{\color{lightgray}0}{2}{120}{108}{0}{0.014}{4}}&
		\createcell{{38.787}{0.027}{20.4}{18.36}{\color{lightgray}0}{2}{120}{108}{0}{0.014}{4}}
		\\
		{ \nameInTabTracktor } &
		\createcell{{8.791}{7.261}{1.53}{\color{lightgray}0}{\color{lightgray}0}{113}{9}{\color{lightgray}0}{\color{lightgray}-}{0.064}{3}}&
		\createcell{{8.833}{7.261}{1.53}{\color{lightgray}0}{0.043}{113}{9}{\color{lightgray}0}{1}{0.064}{2}}&
		\createcell{{12.191}{7.261}{1.53}{\color{lightgray}0}{3.4}{113}{9}{\color{lightgray}0}{1}{0.064}{3}}&
		\createcell{{14.929}{5.919}{5.27}{3.74}{\color{lightgray}0}{91}{31}{22}{0}{0.065}{3}}
		\\
		{ \nameInTabBoTSORT } &
		\createcell{{9.28}{9.28}{\color{lightgray}0}{\color{lightgray}0}{\color{lightgray}0}{122}{\color{lightgray}0}{\color{lightgray}0}{\color{lightgray}-}{0.076}{4}}&
		\createcell{{9.45}{9.28}{\color{lightgray}0}{\color{lightgray}0}{0.17}{122}{\color{lightgray}0}{\color{lightgray}0}{4}{0.076}{3}}&
		\createcell{{9.541}{9.201}{0.17}{0.17}{\color{lightgray}0}{121}{1}{1}{0}{0.076}{2}}&
		\createcell{{9.541}{9.201}{0.17}{0.17}{\color{lightgray}0}{121}{1}{1}{0}{0.076}{2}}
		\\
		{ \nameInTabGMPHDOGM } &
		\createcell{{7.867}{7.867}{\color{lightgray}0}{\color{lightgray}0}{\color{lightgray}0}{122}{\color{lightgray}0}{\color{lightgray}0}{\color{lightgray}-}{0.064}{2}}&
		\createcell{{7.867}{7.867}{\color{lightgray}0}{\color{lightgray}0}{\color{lightgray}0}{122}{\color{lightgray}0}{\color{lightgray}0}{0}{0.064}{1}}&
		\createcell{{7.867}{7.867}{\color{lightgray}0}{\color{lightgray}0}{\color{lightgray}0}{122}{\color{lightgray}0}{\color{lightgray}0}{0}{0.064}{1}}&
		\createcell{{7.867}{7.867}{\color{lightgray}0}{\color{lightgray}0}{\color{lightgray}0}{122}{\color{lightgray}0}{\color{lightgray}0}{0}{0.064}{1}}
		\\
		\bottomrule
	\end{tabular}
\end{table*}
\begin{table*}[]
	\centering
	\caption{\ExperimentTwo{}: evaluation using the IoU metric with TGOSPA parameters $p{=}1.71$ and $c{=}0.255$ ($a{=}0.17$).}
	\label{tab:TGOSPA:IoU:experimentTwo}
	\begin{tabular}{p{32mm}cccc}
		\toprule
		&\footnotesize \GammaZero &\footnotesize \GammaSmall &\footnotesize \GammaLarge &\footnotesize \GammaExtreme \\
		&\footnotesize No switch matter &\footnotesize Any switch matter &\footnotesize Only switches lasting for &\footnotesize One-to-one trajectory \\
		& &\footnotesize (the more, the worse) &\footnotesize $\scriptstyle\ell>10$ time steps matter &\footnotesize matching \\
		& $\scriptstyle \gamma=0$ & $\scriptstyle \gamma=0.079$ &$\scriptstyle\gamma=0.981$ , ($\scriptstyle n=10$)& ``$\scriptstyle\gamma \rightarrow \infty$'' \\
		&\footnotesize (GOSPA) & $\scriptstyle g_1 = 0.2125$ & $\scriptstyle h_{10}=0$, $\scriptstyle h_{11}=0.0627$ & $\scriptstyle h_{\ell} = 0$, $\scriptstyle\forall \ell \leq \tfrac{\text{no. frames}}{2}$ \\
		\midrule
		{ \nameInTabFRCNN }&
		\createcell{{1.213}{0.811}{0.58}{\color{lightgray}0}{\color{lightgray}0}{110}{12}{\color{lightgray}0}{\color{lightgray}-}{0.057}{1}}&
		\createcell{{1.822}{0.811}{0.58}{\color{lightgray}0}{1.398}{110}{12}{0}{108}{0.057}{4}}&
		\createcell{{4.072}{0.001}{5.803}{5.222}{\color{lightgray}0}{2}{120}{108}{0}{0.014}{4}}&
		\createcell{{4.072}{0.001}{5.803}{5.222}{\color{lightgray}0}{2}{120}{108}{0}{0.014}{4}}
		\\
		{ \nameInTabTracktor } &
		\createcell{{1.299}{1.128}{0.435}{\color{lightgray}0}{\color{lightgray}0}{113}{9}{\color{lightgray}0}{\color{lightgray}-}{0.068}{3}}&
		\createcell{{1.305}{1.128}{0.435}{\color{lightgray}0}{0.013}{113}{9}{\color{lightgray}0}{1}{0.068}{2}}&
		\createcell{{1.721}{1.128}{0.435}{\color{lightgray}0}{0.967}{113}{9}{\color{lightgray}0}{1}{0.068}{3}}&
		\createcell{{2.08}{0.933}{1.499}{1.064}{\color{lightgray}0}{91}{31}{22}{0}{0.069}{3}}
		\\
		{ \nameInTabBoTSORT } &
		\createcell{{1.423}{1.73}{0.048}{0.048}{\color{lightgray}0}{121}{1}{1}{\color{lightgray}-}{0.083}{4}}&
		\createcell{{1.44}{1.73}{0.048}{0.048}{0.039}{121}{1}{1}{3}{0.083}{3}}&
		\createcell{{1.45}{1.695}{0.097}{0.097}{\color{lightgray}0}{120}{2}{2}{0}{0.083}{2}}&
		\createcell{{1.45}{1.695}{0.097}{0.097}{\color{lightgray}0}{120}{2}{2}{0}{0.083}{2}}
		\\
		{ \nameInTabGMPHDOGM } &
		\createcell{{1.271}{1.314}{0.097}{0.097}{\color{lightgray}0}{120}{2}{2}{\color{lightgray}-}{0.071}{2}}&
		\createcell{{1.271}{1.314}{0.097}{0.097}{\color{lightgray}0}{120}{2}{2}{0}{0.071}{1}}&
		\createcell{{1.271}{1.314}{0.097}{0.097}{\color{lightgray}0}{120}{2}{2}{0}{0.071}{1}}&
		\createcell{{1.271}{1.314}{0.097}{0.097}{\color{lightgray}0}{120}{2}{2}{0}{0.071}{1}}
		\\
		\bottomrule
	\end{tabular}
\end{table*}
\begin{table*}[]
	\centering
	\caption{\ExperimentThree{}: evaluation using the IoU metric with TGOSPA parameters $p{=}2.409$ and $c{=}0.34$ ($a{=}0.255$).}
	\label{tab:TGOSPA:IoU:experimentThree}
	\begin{tabular}{p{32mm}cccc}
		\toprule
		&\footnotesize \GammaZero &\footnotesize \GammaSmall &\footnotesize \GammaLarge &\footnotesize \GammaExtreme \\
		&\footnotesize No switch matter &\footnotesize Any switch matter &\footnotesize Only switches lasting for &\footnotesize One-to-one trajectory \\
		& &\footnotesize (the more, the worse) &\footnotesize $\scriptstyle\ell>10$ time steps matter &\footnotesize matching \\
		& $\scriptstyle\gamma=0$ & $\scriptstyle\gamma=0.149$ &$\scriptstyle\gamma=0.884$, ($\scriptstyle n=10$)& ``$\scriptstyle\gamma \rightarrow \infty$'' \\
		&\footnotesize (GOSPA) & $\scriptstyle g_1 = 0.2975$ & $\scriptstyle h_{10}=0$, $\scriptstyle h_{11}=0.1257$ & $\scriptstyle h_{\ell} = 0$, $\scriptstyle\forall \ell \leq \tfrac{\text{no. frames}}{2}$ \\
		\midrule
		{ \nameInTabFRCNN }&
		\createcell{{0.797}{0.133}{0.446}{\color{lightgray}0}{\color{lightgray}0}{110}{12}{\color{lightgray}0}{\color{lightgray}-}{0.062}{4}}&
		\createcell{{1.241}{0.133}{0.446}{\color{lightgray}0}{1.104}{110}{12}{\color{lightgray}0}{108}{0.062}{4}}&
		\createcell{{2.428}{0.000}{4.46}{4.014}{\color{lightgray}0}{2}{120}{108}{0}{0.014}{4}}&
		\createcell{{2.428}{0.000}{4.46}{4.014}{\color{lightgray}0}{2}{120}{108}{0}{0.014}{4}}
		\\
		{ \nameInTabTracktor } &
		\createcell{{0.765}{0.19}{0.334}{\color{lightgray}0}{\color{lightgray}0}{113}{9}{\color{lightgray}0}{\color{lightgray}-}{0.071}{3}}&
		\createcell{{0.771}{0.19}{0.334}{\color{lightgray}0}{0.01}{113}{9}{\color{lightgray}0}{1}{0.071}{3}}&
		\createcell{{1.103}{0.19}{0.334}{\color{lightgray}0}{0.743}{113}{9}{\color{lightgray}0}{1}{0.071}{3}}&
		\createcell{{1.369}{0.16}{1.152}{0.818}{\color{lightgray}0}{91}{31}{22}{0}{0.072}{3}}
		\\
		{ \nameInTabBoTSORT } &
		\createcell{{0.699}{0.422}{\color{lightgray}0}{\color{lightgray}0}{\color{lightgray}0}{122}{\color{lightgray}0}{\color{lightgray}0}{\color{lightgray}-}{0.095}{2}}&
		\createcell{{0.726}{0.422}{\color{lightgray}0}{\color{lightgray}0}{0.041}{122}{\color{lightgray}0}{\color{lightgray}0}{4}{0.095}{2}}&
		\createcell{{0.758}{0.439}{0.037}{0.037}{\color{lightgray}0}{121}{1}{1}{0}{0.097}{2}}&
		\createcell{{0.758}{0.439}{0.037}{0.037}{\color{lightgray}0}{121}{1}{1}{0}{0.097}{2}}
		\\
		{ \nameInTabGMPHDOGM } &
		\createcell{{0.67}{0.381}{\color{lightgray}0}{\color{lightgray}0}{\color{lightgray}0}{122}{\color{lightgray}0}{\color{lightgray}0}{\color{lightgray}-}{0.091}{1}}&
		\createcell{{0.67}{0.381}{\color{lightgray}0}{\color{lightgray}0}{\color{lightgray}0}{122}{\color{lightgray}0}{\color{lightgray}0}{0}{0.091}{1}}&
		\createcell{{0.67}{0.381}{\color{lightgray}0}{\color{lightgray}0}{\color{lightgray}0}{122}{\color{lightgray}0}{\color{lightgray}0}{0}{0.091}{1}}&
		\createcell{{0.67}{0.381}{\color{lightgray}0}{\color{lightgray}0}{\color{lightgray}0}{122}{\color{lightgray}0}{\color{lightgray}0}{0}{0.091}{1}}
		\\
		\bottomrule
	\end{tabular}
\end{table*}

\subsection{Observations and Discussion}

\ObservationX[\seeminglySwitch{s} are found]{
	Whenever using \GammaZero{}, no \seeminglySwitch{s} are assessed.
	Using \GammaSmall{}, both short-term interim and long-term \seeminglySwitch{s} are found and penalized for. 
	That is, short-term interim \seeminglySwitch{s} cannot be found without counting long-term ones at the same time.
	With increasing $\gamma$ further to \GammaLarge{}, only long-term \seeminglySwitch{s} are found and penalized.
	Using \GammaExtreme{}, no switches arise, and all the \seeminglySwitch{s} are assessed as missed and false estimates, which can be seen in the TGOSPA decomposition.
}

\ObservationX[algorithm with no \seeminglySwitch{s}]{
	The GMPHDOGM17 tracker has no \seeminglySwitch{s}, and the value of TGOSPA is thus independent of the choice of $\gamma$ among all Tables~\ref{tab:TGOSPA:IoU:experimentOne},~\ref{tab:TGOSPA:IoU:experimentTwo} and~\ref{tab:TGOSPA:IoU:experimentThree}.
	The TGOSPA values for the other algorithms thus increase with increasing $\gamma$.
}

\ObservationX[detector evaluation is meaningful only with $\gamma {=} 0$]{
	As FRCNN outputs are temporarily disconnected, switches arise when \GammaSmall{} is used.
	Increasing $\gamma$ further makes any switches too costly, and all (but the closest estimate to each ground truth trajectory) are treated as false estimates.
	Thus, for detector training, only \GammaZero{} is recommended.
}

\ObservationX[TGOSPA metric is non-decreasing with increasing value of $\gamma$]{
	Values in each row in Tables~\ref{tab:TGOSPA:IoU:experimentOne},~\ref{tab:TGOSPA:IoU:experimentTwo} and~\ref{tab:TGOSPA:IoU:experimentThree} increase (or stay the same) from left to right.
}

That is, although no switches are counted in the decomposition for \GammaExtreme{} on the one hand, the corresponding TGOSPA metric values are largest among the different $\gamma$ setups.
It can be seen that \seeminglySwitch{s} that are present in the data are penalized using \GammaExtreme{} with the maximum possible yield (the longer the \seeminglySwitch{}, the higher the value), but TGOSPA does \emph{not} show this fact in the decomposition which is undesirable for understanding the results and making further decisions.

Observing that both \GammaLarge{} and \GammaExtreme{} setups lead to the same \emph{ordering} of the algorithms (regardless of the parameters $p$ and $c$), one can use \GammaLarge{} instead of \GammaExtreme{} and observe the decomposition such that the found switches correspond to \seeminglySwitch{} with minimal length of $\ell {=} n{+}1 {=} 11$.
Considering the meaning of the \seeminglySwitch{} \emph{length} and thus the \emph{kind} of \seeminglySwitch{} that are penalized (Section~\ref{sec:long-term-switches}), the use of \GammaExtreme{} in practice is not recommended.

\ObservationX[algorithm ordering based on the type of \seeminglySwitch{}]{
	The lengths of \seeminglySwitch{s} clearly matter, which can be seen for BoT\_SORT and Tracktor++v2 in \ExperimentOne{} and \ExperimentTwo{}: BoT\_SORT is worse than Tracktor++v2 using \GammaSmall{}, but it is better for \GammaLarge{}.
	On the other hand, according to Table~\ref{tab:TGOSPA:IoU:experimentThree} for \ExperimentThree{}, the algorithms are given the same ordering regardless of $\gamma$.
	The ordering thus does not necessarily reflect the number of switches or their length, as the final TGOSPA metric value considers all the different error types jointly based on the chosen parameters. 
	It can be seen from the corresponding decomposition that the switches found in BoT\_SORT have little effect on the final value of the metric for \GammaSmall{} using \ExperimentThree{}, which is mainly due to the large value of $p{=}2.409$ selected.
}

\ObservationX[increasing $p$]{
	From Table~\ref{tab:TGOSPA:IoU:experimentThree}, the impact of localization error is mitigated with the large $p$ in favor of the missed/false cost.
	Furthermore, switches can have even larger impact on the final TGOSPA metric value depending on the relative value of $\gamma$ compared to $c$: for $\gamma {>} \tfrac{c}{\sqrt[2]{p}}$ (e.g., for \GammaLarge{}), even small number of switches has considerably larger impact to the final TGOSPA metric value compared to missed/false estimates and vice versa.
}

\ObservationX[the effect of non-admissible estimates]{
	Consider the FRCNN detector and the GMPHDOGM17 tracker in Tables~\ref{tab:TGOSPA:IoU:experimentOne} and~\ref{tab:TGOSPA:IoU:experimentThree} for \ExperimentOne{} and \ExperimentThree{}, respectively, with \GammaZero{}.
	From Fig.~\ref{fig:gtdistance_plot}, it follows that while the pedestrian \emph{gt6} is visible (not occluded), estimates from both FRCNN and GMPHDOGM17 algorithms have both errors lower than $0.17$ in the IoU-induced metric.
	During the occlusion of \emph{gt6}, estimates are missing for the FRCNN detector, while most estimates of the GMPHDOGM17 tracker have errors larger than $0.17$.
	In Table~\ref{tab:TGOSPA:IoU:experimentOne}, the TGOSPA metric values with \GammaZero{} for the two algorithms are nearly the same: the $12$ missed objects of the FRCNN detector are slightly \emph{better} than the \emph{non-admissible} (larger than $a{=}0.17$) estimates appearing in the GMPHDOGM17 tracker.
	In Table~\ref{tab:TGOSPA:IoU:experimentThree}, however, the fine localization of the FRCNN detector evaluated with \GammaZero{} has a negligible effect ($0.133$) compared to the missed detection cost ($0.446$) corresponding to 12 missed estimates, which is the effect of large $p$.
}

\ObservationX[the effect of smaller $c$]{
	Although $p$ in \ExperimentTwo{} has increased relative to \ExperimentOne{}, it can be seen that the cost of missed/false estimate $\tfrac{c^p}{2} {\approx} 0.048$ has decreased, and thus the localization error plays a dominant role especially for \GammaZero{}.
	Note that $c{=}0.255$ cuts off two estimates in the GMPHDOGM17 and one from BoT\_SORT that are treated as false estimates.
}

\ObservationX[two short-term interim \seeminglySwitch{s} resulting into three switches only]{
	For \ExperimentTwo{} using \GammaSmall{}, BoT\_SORT is found to have only three switches instead of the expected four switches for the \emph{two} short-term interim \seeminglySwitch{s} (see Section~\ref{sec:short-term-switches}).
	This is due to the estimate with error larger than $c{=}0.255$ at the $34$-th time step being cut off:
	denoting with index $1$ the corresponding BoT\_SORT trajectory that tracks \emph{gt2}, the resulting assignment for \emph{gt6} is $[\dots,1, 0,1,\dots ] {=} [\pi_{\star}^{...33,34,35,...}]_{i=\text{"\emph{gt6}"}} $, which leads to two subsequent half-switches.
}

\ObservationX[BoT\_SORT estimate is larger than $g_1{=}0.255$ but results into switches anyway]{
	For \ExperimentOne{} using \GammaSmall{}, the BoT\_SORT estimate discussed above is larger than $g_1$, and a total of four switches were found.
	The reason is that the assumptions introduced in Section~\ref{sec:short-term-switches} do not apply in this particular case since the value of $c$ is larger (the distance from \emph{gt2} to the estimate is smaller than $c$, i.e., $d(x_{\text{"\emph{gt2}"}}^{34}, y_1^{34}) {<} c$).
	It turns out that the threshold (for the distance so that a switch results) for such a case is smaller than $g_1{=}0.255$.
}
\\[-0.2cm]

Note that for algorithms yielding similar TGOSPA metric values, its decomposition can be used to explain the efficiency of the algorithms\footnote{
    The decomposition itself is not recommended to be used for deciding whether some algorithm is better than another.
}.
In particular, the TGOSPA metric decomposition can be easily defined over time
offering further insight into the algorithm's behavior.
This is not the case for the HOTA score that uses averaging over different threshold values.

The TGOSPA metric and HOTA score are compared in Appendix~\ref{sec:HOTA:vs:TGOSPA} using several toy examples to get even better insight on the differences.
The following Section continues exploring numerical examples in more practical settings than before.

\section{Practical Evaluation}\label{sec:practical-evaluation}
The short CV scenario discussed above was used to get an in-depth understanding of the TGOSPA metric.
To observe its practical use, we evaluate several algorithms on the entire MOT17-09 video. 

\subsection{Recommended Parameters}
For the CV domain, we continue to select $d$ as the $\mathrm{IoU}$-induced metric $d_{\mathrm{IoU}}(x,y)$~\eqref{eq:IoU-induced-metric-d} for its favorable properties (cf. Appendix~\ref{appendix:alternative-metrics}).
Given that, it is clear that different values of $c$, $p$, and $\gamma$ are suitable for different applications.
Based on the discussion given so far, three particular combinations are recommended in this paper for: 
\begin{itemize}
    \item (visual) detector training,
    \item online surveillance,
    \item offline scene understanding.
\end{itemize}

\subsubsection{TGOSPA for Detector Training}
\emph{Recommendation for combination of parameter values: $c {=} 0.255$, $p {=} 1.71$, $\gamma {=} 0$.}

This selection corresponds to \ExperimentTwo{} with \GammaZero{} discussed before, and the corresponding maximum admissible distance (in $d_{\mathrm{IoU}}$) is $a {=} 0.17$.
This setup encourages visual detectors to output estimates within $0.83 = 1{-}0.17$ in IoU.
Estimates with errors ranging from $0.17$ to $0.255$ are preferable to be omitted as these are likely "predictive" estimates (e.g., GMPHDOGM17 in Fig~\ref{fig:gtdistance_plot}).
Estimates with errors larger than $0.255$ are treated as false.
If estimates form trajectories over time, they are neglected by using $\gamma {=} 0$ and TGOSPA itself is no longer a metric.

Since $p {>} 1$, note that $a^p {=} \tfrac{c^p}{2} {=} 0.048$ is the cost same for \emph{1)} the utmost estimate with the distance $0.17$, \emph{2)} a missing estimate and \emph{3)} a false estimate.
Furthermore, the TGOSPA metric values are likely to be driven by the number of false/missed estimates for the same reason.

In Table~\ref{tab:detectors-evaluation}, we evaluated the FRCNN, SDP, and DPM visual detectors that are publicly available within the MOT17 dataset.
The metric values are mostly driven by the large number of missed/false estimates, and they indicate superior performance of the FRCNN detector.
While the SDP detector contains more proper estimates than the FRCNN detector, the estimates contain larger localization error.
The DPM detector contains the largest number of estimates among the considered ones, but they are too far to be considered even proper, using this TGOSPA parameterization.
Note that a dummy detector outputting no estimates at all would lead to the TGOSPA metric value equal to $(5325 \cdot \tfrac{c^p}{2})^{\sfrac{1}{p}} = 62.409$, with $5325$ being the number of considered ground truth objects (pedestrians in the MOT17-09 video).

\begin{table}[h]
    \centering
    \caption{\textbf{detector training} MOT17 public detectors evaluation using the IoU metric with TGOSPA parameters $p{=}1.71$, $c{=}0.255$ ($a{=}0.17$) and $\gamma {=} 0$.}
    \label{tab:detectors-evaluation}
    \begin{tabular}{p{32mm}p{32mm}}
    \toprule
        FRCNN & \createcell{{20.077}{38.083}{120.308}{10.251}{\color{lightgray}0}{2837}{2488}{212}{\color{lightgray}-}{0.080}{1}} \\[0.5cm]
        SDP & \createcell{{23.854}{107.69}{100.917}{17.843}{\color{lightgray}0}{3238}{2087}{369}{\color{lightgray}-}{0.137}{2}} \\[0.5cm]
        DPM & \createcell{{37.152}{94.22}{178.624}{210.104}{\color{lightgray}0}{1631}{3694}{4345}{\color{lightgray}-}{0.189}{3}} \\
    \bottomrule
    \end{tabular}
\end{table}

\subsubsection{TGOSPA for Tracking: Online Surveillance}
\emph{Recommendation: $c {=} 0.5$, $p {=} 1.8$, $\gamma {=} 0.31$.}

For this selection, the maximum admissible distance is $a {=} 0.34$ and $\gamma$ was selected as \GammaSmall{} using~\eqref{eq:gamma-from-g_1} with the distance $g_1 {=} 0.17$ (for details refer to Section~\ref{sec:short-term-switches}).
This setup encourages tracking algorithms to output filtering and predictive estimates with IoU less than $0.66 = 1{-}0.34$, i.e., even estimates containing larger errors (such as those of the GMPHDOGM17 in Fig~\ref{fig:gtdistance_plot}) are considered desirable.
Estimates with errors larger than $0.5$ are treated as false.
Switches encapsulate both short-term interim and long-term \seeminglySwitch{s} appearing in the data.

Note that since $p {>} 1$, $a^p {=} \tfrac{c^p}{2} {=} 0.144$ is the same cost of \emph{1)} the utmost estimate with the distance $0.34$, \emph{2)} a missing estimate and \emph{3)} a false estimate.
As in the previous combination, the TGOSPA metric values are likely to be driven by the number of false/missed estimates for the same reason.

In the penultimate column of Table~\ref{tab:online-surveillance}, we evaluated the first five tracking algorithms currently leading the MOT17 Public leader board at the webpage~\cite{MOT17-webpage:2023}, which have a reference paper indicated.
That is, the algorithms claimed they used the public FRCNN detections to track pedestrians in the MOT17-09 video: FLWM~\cite{FLWM:2025}, FeatureSORT~\cite{FeatureSORT:2024}, PermaTrack~\cite{PermaTrack:2021}, MOTer~\cite{MOTer:2022}, and PixelGuide~\cite{PixelGuide:2022}.
The algorithms ordering according to the TGOSPA metric values matches neither MOTA, IDF1, nor HOTA, and the TGOSPA metric decomposition gives a detailed explanation.
It can be seen that switch costs are rather negligible relative to the localization costs and that the metric values are reasonably sensitive to missed/false estimates. 
The numbers of switches, however, provide insight into the total number of \seeminglySwitch{s} in the data.
Note that a dummy tracking algorithm outputting no estimates at all would lead to the TGOSPA metric value equal to $(5325 \cdot \tfrac{c^p}{2})^{\sfrac{1}{p}} = 40.251$.

\begin{table*}[h]
    \centering
    \caption{\textbf{online surveillance} and \textbf{offline scene understanding} tracking algorithms evaluation using the IoU metric, \mbox{MOT17-09} video processing the public FRCNN detections.
    }
    \label{tab:online-surveillance}
    \begin{tabular}{p{35mm}ccccc}
    \toprule
        & MOTA ($\uparrow$) & IDF1 ($\uparrow$) & HOTA ($\uparrow$) & TGOSPA ($\downarrow$) & TGOSPA ($\downarrow$) \\
        & & & & online setup & offline setup \\
        & & & & \small $c{=}0.5$, $p{=}1.8$, $\gamma {=} 0.31$ & \small $c{=}0.5$, $p{=}1$, $\gamma {=} 5$ \\
    \midrule 
        FLWM & \createoutlinecell{{0.917}{1}} & \createoutlinecell{{0.666}{2}} & \createoutlinecell{{0.744}{1}} & \createcell{{19.606}{150.965}{38.698}{15.968}{4.742}{5056}{269}{111}{38.5}{0.142}{1}} & \createcell{{807.14}{596.64}{78.75}{39.25}{92.5}{5010}{315}{157}{18.5}{0.119}{3}} \\[0.5cm]
        FeatureSORT & \createoutlinecell{{0.897}{2}} & \createoutlinecell{{0.592}{4}} & \createoutlinecell{{0.646}{4}} & \createcell{{20.219}{153.928}{46.322}{17.119}{4.865}{5003}{322}{119}{39.5}{0.144}{4}} & \createcell{{890.91}{623.16}{94.25}{43.5}{130}{4948}{377}{174}{26}{0.126}{0}} \\[0.5cm]
        PermaTrack & \createoutlinecell{{0.738}{0}} & \createoutlinecell{{0.499}{0}} & \createoutlinecell{{0.555}{0}} & \createcell{{24.374}{110.008}{189.748}{3.884}{7.328}{4006}{1319}{27}{59.5}{0.135}{0}} & \createcell{{1001.3}{446.763}{376.25}{53.25}{125}{3820}{1505}{213}{25}{0.117}{0}} \\[0.5cm]
        MOTer & \createoutlinecell{{0.712}{0}} & \createoutlinecell{{0.572}{0}} & \createoutlinecell{{0.510}{0}} & \createcell{{25.067}{105.809}{212.477}{2.158}{6.589}{3848}{1477}{15}{53.5}{0.135}{0}} & \createcell{{1018.0}{423.045}{415.25}{49.75}{130}{3664}{1661}{199}{26}{0.115}{0}} \\[0.5cm]
        PixelGuide & \createoutlinecell{{0.830}{4}} & \createoutlinecell{{0.743}{1}} & \createoutlinecell{{0.660}{3}} & \createcell{{20.185}{91.768}{121.272}{5.754}{2.771}{4482}{843}{40}{22.5}{0.115}{3}} & \createcell{{740.74}{429.993}{223.25}{22.5}{65}{4432}{893}{90}{13}{0.097}{1}} \\
        Tracktor++v2 & \createoutlinecell{{0.634}{0}} & \createoutlinecell{{0.493}{0}} & \createoutlinecell{{0.546}{0}} & \createcell{{25.571}{58.073}{273.329}{4.028}{3.510}{3425}{1900}{28}{28.5}{0.103}{0}} & \createcell{{882.04}{287.04}{482.75}{14.75}{97.5}{3394}{1931}{59}{19.5}{0.085}{4}} \\[0.5cm]
        {BoT\_SORT \hspace{1cm} (uses private detector)} & \createoutlinecell{{0.882}{3}} & \createoutlinecell{{0.655}{3}} & \createoutlinecell{{0.737}{2}} & \createcell{{19.850}{131.903}{65.167}{12.803}{5.234}{4872}{453}{89}{42.5}{0.134}{2}} & \createcell{{798.94}{543.94}{131.75}{40.75}{82.5}{4798}{527}{163}{16.5}{0.113}{2}} \\[0.5cm]
        GMPHDOGM17 & \createoutlinecell{{0.622}{0}} & \createoutlinecell{{0.527}{0}} & \createoutlinecell{{0.614}{0}} & \createcell{{27.749}{100.96}{272.753}{15.249}{3.633}{3429}{1896}{106}{29.5}{0.141}{0}} & \createcell{{956.05}{344.55}{500.75}{53.25}{57.5}{3322}{2003}{213}{11.5}{0.104}{0}} \\
    \bottomrule
    \end{tabular}
\end{table*}

\subsubsection{TGOSPA for Tracking: Offline Scene Understanding}
\emph{Recommendation: $c {=} 0.5$, $p {=} 1$, $\gamma {=} 5$.}

For this combination, the maximum admissible distance $a {=} 0.25$ was selected lower compared to the previous combination, especially to encourage tracking algorithms to output \emph{smoothed} (interpolated) estimates within $0.75 = 1 {-} 0.25$ in IoU.
The switching penalty $\gamma$ was selected as \GammaLarge{} using~\eqref{eq:gamma-from-n} with the number $n {=} 10$ to assess track fragmentations/occlusions for which the tracks changes last for at least 10 time steps (for details refer to Section~\ref{sec:long-term-switches}) to encourage algorithms that form trajectories without long-term \seeminglySwitch{s}.

Since $p {=} 1$, this setup places more emphasis on the precision of localization than on the number of missed/false estimates compared to the previous combination suitable for online surveillance.
The TGOSPA metric value becomes a direct sum of the decomposed costs, and $a {=} \tfrac{c}{2} {=} 0.25$ directly becomes the cost same for \emph{1)} the utmost estimate with the distance $0.25$, \emph{2)} a missing estimate and \emph{3)} a false estimate.

The same algorithms as in the previous case were evaluated with this setup, and the results are given in the last column of Table~\ref{tab:online-surveillance}.
The corresponding algorithms' ordering matches neither MOT, IDF1, nor HOTA, and differs from the ordering for the combination used previously.
It can be seen that the number of switches is lower, while the corresponding cost has a considerably larger influence on the final TGOSPA metric values compared to the online surveillance evaluation discussed previously.
Since many \seeminglySwitch{s} are no longer considered as switches in this combination, the corresponding estimates are no longer assigned to a ground truth and thus contribute to the false estimates, while missed \elementOfTrajectory{s} increases accordingly.
Note that a dummy tracking algorithm outputting no estimates at all would lead to the TGOSPA metric value equal to $(5325 \cdot \tfrac{c^p}{2})^{\sfrac{1}{p}} = 1331.25$.

\section{Conclusion}\label{sec:Conclusion}
This paper indicated that having hyper-parameters for performance evaluation is beneficial, and that their proper application-specific selection is crucial and indeed possible.
This paper proposed to use the trajectory generalized optimal sub-pattern assignment (TGOSPA) metric in the context of computer vision (CV) and showed how to select its parameters conveniently using simple example evaluations.
In particular, this paper focused primarily on the effects of the switching penalty, whose direct selection was found to be counterintuitive.
Its indirect selection for particular purposes has been proposed along with the selection of other parameters.

While this paper proposed a method for selecting the TGOSPA metric parameters and suggested some particular values, it is ultimately the user who should decide the values for their particular application. 
It should be emphasized that the derived rules are independent of the CV application and can be readily employed within the signal-processing community.

Nevertheless, it is also possible to relieve the user from having to select the parameters.
Similarly to the HOTA score, one could average the results from using several TGOSPA metric parameterizations to yield a single metric value.
Observing that the TGOSPA metric values (e.g., Tables~\ref{tab:TGOSPA:IoU:experimentOne}-\ref{tab:TGOSPA:IoU:experimentThree}) have completely different scales and decompositions for different parameterizations, however, 
the resulting average value would no longer be useful for the specific application in question, and is thus not recommended.

As the evaluation is based solely on comparing the actual estimates with ground truth data, the evaluation metric can be used to evaluate any (visual) tracking method.
On the other hand, many algorithms provide covariance matrices or entire probability distributions along with the (point-)estimates that are not considered in this paper.
This and other aspects, such as computational demands, numerical stability, etc., could form a topic for future research.


\section*{Acknowledgements}
The authors would like to thank Ji\v{r}\'i Vysko\v{c}il for providing the results from the BoT-SORT algorithm by adopting it from \url{github.com/NirAharon/BoT-SORT/}.


\appendices

\section{HOTA Score Definition}\label{sec:hota}
For convenience, the definition of the HOTA score is given in the notation of this paper, which is introduced in Section~\ref{sec:TGOSPA_metric}.
As mentioned before, HOTA computes the assignments of ground truth with the estimates at each time step (frame) individually.
It follows that the \emph{trajectory-level} and \emph{\elementOfTrajectory{}-level} assignments for the HOTA score coincide and the function $\rho$ suffices for the definition\footnote{
	For the HOTA score, $\theta$ does not depend on $c$, and it coincides with $\rho$.
}.
With this, the set of all properly estimated pairs of \elementOfTrajectory{s} stored as tuples containing the time step and the indices $i$ and $j$ of the assigned trajectories, further called \emph{links}, is
\begin{align}
	\mathcal{N}\big( \pi^{0:K}\big) =
	\big\{ (k,i,j) \! : \hspace{0.1cm} (i,j) {\in} \rho\big( \pi^{k} \big)\big\}.
	\label{eq:mathcalN-auxiliary}
\end{align}
with $k$ ranging over $\{0,1,\dots,K\}$.
Indeed, for the HOTA score, the number of properly estimated objects is the cardinality of the set $\mathcal{N}\big( \pi^{0:K}\big)$~\eqref{eq:mathcalN-auxiliary}.
Similarly, consider the sets of missed and false \elementOfTrajectory{s} stored as links containing the time step and the trajectory indices $i$ or $j$ that are either missed or false,
\begin{align}
	\!\!\mathcal{M}\big( \mathbf{X}, \pi^{0:K}\big) &= \big\{ (k,i)\!:\hspace{0.1cm} \pi^{k}_{i} {=} 0, |\mathbf{x}_i^k|{=}1  \big\} , \\
	\!\!\mathcal{F}\big( \mathbf{Y}, \pi^{0:K}\big) &= \big\{ (k,j)\!:\hspace{0.1cm} \nexists i {:} (i,j) {\in} \rho\big( \pi^{k} \big), |\mathbf{y}_j^k|{=}1  \big\} ,
\end{align}
respectively, where $k$ ranges over $\{0,1,\dots,K\}$, $i$ ranges over $\{1,\dots,|\mathbf{X}|\}$ and $j$ ranges over $\{1,\dots,|\mathbf{Y}|\}$.
Indeed, for the HOTA score, the numbers of properly detected, missed, and false \elementOfTrajectory{s} are the cardinalities of these sets.

\begin{figure}[h]
	\centering
	\includegraphics[width=0.47\textwidth]{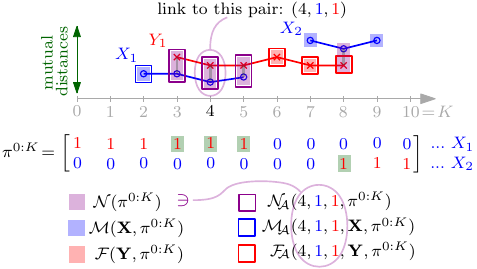}
	\caption{Illustration of sets involved in the HOTA score computation.
		The sets on the left concern all trajectories, but the sets on the right regard only the two trajectories to which the input link $(k,i,j)$ belongs.}
	\label{fig:HOTA-assignments-illustration}
	\vspace{-0.5cm}
\end{figure}

The HOTA score accounts for \seeminglySwitch{s} via sets
\begin{align}
	\!\mathcal{N}_{\!\!\mathcal{A}}&\big(k,i,j, \pi^{0:K} \big) \!=\! \!\big\{ (k'\!\!,i'\!,j'){\in}\mathcal{N}\big(\pi^{0:K}\big) \!\!:\hspace{0.0cm} j {=} j', i{=}i'  \big\} \! , \!\!\! \\
	\!\!\mathcal{M}_{\!\!\mathcal{A}}&\big(k,i,j, \mathbf{X}, \pi^{0:K} \big) \!=\! \!\big\{ (k'\!,i'){\in}\mathcal{M}\big( \mathbf{X}, \pi^{0:K}\big) \!\!:\hspace{0.0cm} i{=}i' \big\} \notag\\
	&\hspace{1.55cm} \cup\ \big\{ (k'\!,i'\!,j'){\in}\mathcal{N}\big(\pi^{0:K}\big)\!\! :\hspace{0.0cm} j {\neq} j', i{=}i' \big\} \! , \!\!\! \\
	\!\mathcal{F}_{\!\!\mathcal{A}}&\big(k,i,j, \!\mathbf{Y}, \pi^{0:K} \big) \!=\! \!\big\{ (k',j'){\in}\mathcal{F}\big( \mathbf{Y}, \pi^{0:K}\big)\!\!:\hspace{0.0cm} j{=}j' \big\} \!\! \notag\\
	&\hspace{1.55cm} \cup\ \big\{ (k'\!,i'\!,j'){\in}\mathcal{N}\big(\pi^{0:K}\big) \!\!:\hspace{0.0cm} j {=} j', i {\neq} i' \big\} \!, \!\!\! 
\end{align}
that are \emph{relative} to the proper estimated pair $(k,i,j){\in}\mathcal{M}\big(\pi^{0:K}\big)$.
Namely, $\mathcal{N}_{\!\!\mathcal{A}}(k,i,j, \pi^{0:K} )$ is the set of links to the estimates out of the trajectory $j$ that properly track the particular ground truth trajectory $i$ among the time steps.
The set $\mathcal{M}_{\!\!\mathcal{A}}(k,i,j, \mathbf{X}, \pi^{0:K} )$ contains links to ground truth objects out of the trajectory $i$ that \emph{are not} properly tracked by the particular estimated trajectory $j$ among the time steps.
Furthermore, the set $\mathcal{F}_{\!\!\mathcal{A}}(k,i,j, \mathbf{Y}, \pi^{0:K} )$ contains links to estimates out of the trajectory $j$ that \emph{do not} properly track the particular ground truth trajectory $i$ among the time steps.
An illustration of the above-defined sets is given in Fig.~\ref{fig:HOTA-assignments-illustration}.

With this notation, according to~\cite{HOTA:2021}, the HOTA score is defined as follows.

\begin{definition}[HOTA score]
    Given two trajectories $\mathbf{X}$ and $\mathbf{Y}$, the HOTA score is defined and approximated as
	\begin{subequations}\label{eq:HOTA-definition}
		\begin{align}
			\text{HOTA}(\mathbf{X}, \mathbf{Y}) &= \textstyle \int_{0}^{1} \text{HOTA}^{\!(\alpha)}(\mathbf{X}, \!\mathbf{Y})\, \mathrm{d}\alpha \\
			&\approx \textstyle \tfrac{1}{19} \sum_{l = 1}^{19} \text{HOTA}^{\!( 0.05 \cdot l )}(\mathbf{X}, \!\mathbf{Y}),
		\end{align}
	\end{subequations}
	where HOTA for the \emph{threshold}\footnote{
		Notice that the threshold parameter $\alpha$ is analogous to the cut-off parameter in TGOSPA with $c{=}1{-}\alpha$.
	} $\alpha{>}0$ is 
	\begin{align}
		&\text{HOTA}^{\!(\alpha)}(\mathbf{X}, \!\mathbf{Y}) = \notag\\
		&\hspace{0.5cm} \sqrt{ \frac{ \sum_{ (k,i,\, j) \in \mathcal{N}(\pistarHOTA) } \mathcal{A}_{\mathbf{X}, \!\mathbf{Y}} \big( k,i, j, \pistarHOTA \big) }{ \big| \mathcal{N}\big( \pistarHOTA \big) \big| {+} \big| \mathcal{M}\big( \mathbf{X}, \pistarHOTA \big) \big| {+} \big| \mathcal{F}\big( \mathbf{Y}, \pistarHOTA \big) \big| } },
	\end{align}
	where $\mathcal{A}_{\mathbf{X}, \!\mathbf{Y}} \big( k,i, j, \pistarHOTA \big)$ is the \emph{assignment score} 
	\begin{align}
		\label{eq:HOTA-assignment-score-definition}
		\!\!\!&\mathcal{A}_{\mathbf{X}, \!\mathbf{Y}} \big( k,i,j, \pistarHOTA \big) = \hspace{0.0cm} \frac{ |\mathcal{N}_{\!\!\mathcal{A}}( k,\!i,\!j,\! \pistarHOTA )| }{\mathcal{U} }, 
	\end{align}
    where $\mathcal{U}= |\mathcal{N}_{\mathcal{A}}( k,i,j,\pistarHOTA )| + |\mathcal{M}_{\mathcal{A}}( k,i,j, \mathbf{X},\! \pistarHOTA )| + |\mathcal{F}_{\mathcal{A}}( k,i,j, \mathbf{Y}, \pistarHOTA )|$.
	The assignments $\pistarHOTA {=} [ \pi^0_{*} {(\alpha)}, \dots, \pi^K_{*} {(\alpha)} ] $ are computed individually for each time step (using the Hungarian algorithm), which can be written as
	\begin{align}
		& \pi^{k}_{*} {(\alpha)} \! = \! \textstyle \underset{ \pi^{k} }{\arg\min}
		\sum_{(i,j) \in \rho(\pi^k) } \text{MS}^{(\alpha)}(\mathbf{x}_i^k, \mathbf{y}_j^k, \pi^k)
	\end{align}
	where $\text{MS}^{(\alpha)}(\mathbf{x}_i^k, \mathbf{y}_j^k, \pi^k)$ is 
	the \emph{scoring function for potential matches} 
	defined as
	\begin{align}
		&\text{MS}^{(\alpha)}(\mathbf{x}_i^k, \mathbf{y}_j^k, \pi^k) \!=\! \tfrac{1}{\epsilon} \!+\! \mathcal{A}_{\max}^{(\alpha)}( i,j ) \!+\! \epsilon {\cdot} \mathcal{S}( x_i^k, y_j^k), 
	\end{align}
	if $\big( \mathbf{x}_i^k {=} \{x_i^k\},$ $\mathbf{y}_j^k {=} \{ y_j^k\}$ and $\mathcal{S}( x_i^k, y_j^k) {>} \alpha \big)$ and zero otherwise; 
	the number $\epsilon$ is a "small number such that the components have different magnitudes," and $\mathcal{S}$ is a chosen \emph{similarity score} for \elementOfTrajectory{s} such as the IoU~\eqref{eq:IoU-definition} for bounding boxes.
	The function $\mathcal{A}_{\max}^{(\alpha)}(i,j)$ is the maximum assignment score possible for the particular pair $(k,i,j)$, which is equal to 
	\begin{align}
		\mathcal{A}_{\max}^{(\alpha)}(i,j) &= \mathcal{A}_{\mathbf{X}, \!\mathbf{Y}}( k,i,j, \mu^{0:K}_{(i,j)} (\alpha) ),
	\end{align}
	where the assignment $\mu^{0:K}_{(i,j)} (\alpha)$ is designed to assign trajectories $i$ and $j$ for all time steps wherever they both exist, and their similarity score is higher than $\alpha$, i.e., it can be defined as a matrix of zeros and $j$'s (in the $i$-th row) as
	\begin{align}
		\! [\mu^{0:K}_{(i,j)} (\alpha) ]_{(k, i')} \!=\!
		\begin{cases}
			j & \text{if} \hspace{0.2cm} i=i',\ \mathbf{x}_i^{k} {=} \{x_i^{k}\},\ \mathbf{y}_j^{k} {=} \{ y_j^{k}\} \\
			& \hspace{0.45cm} \text{and} \hspace{0.2cm} \mathcal{S}( x_i^{k}, y_j^{k}) {>} \alpha, \\
			0 & \text{otherwise}.
		\end{cases} \notag\\[-1cm]
		\\\notag
	\end{align}
	where $[\cdot]_{(k',i')}$ is the $(k',i')$-th element of the input matrix.
    \hfill$\square$
\end{definition}

From Fig.~\ref{fig:HOTA-assignments-illustration}, notice that the so-called assignment score $\mathcal{A}_{\mathbf{X}, \!\mathbf{Y}} \big( k,i,j, \pistarHOTA \big)$~\eqref{eq:HOTA-assignment-score-definition} can be understood as an \emph{intersection over union} between the two trajectories to which the given link $(k,i,j)$ "belongs".

\section{HOTA and Triangle Inequality}\label{sec:1-HOTA-counterexample}

\begin{proposition}[HOTA does not satisfy the triangle inequality]
Consider three sets of trajectories $\mathbf{X} {=} \{X_1\}$, $\mathbf{Y} {=} \{X_1\}$ and $\mathbf{Z} {=} \emptyset$, where $X_1$ is an arbitrary trajectory.
In this case, we obtain $\text{HOTA}(\mathbf{X},\mathbf{Y}) {=} 1$, $\text{HOTA}(\mathbf{X},\mathbf{Z}) {=} 0$, and $\text{HOTA}(\mathbf{Y},\mathbf{Z}) {=} 0$.
As a result, 
\begin{align}
    \underbrace{ \text{HOTA}(\mathbf{X},\mathbf{Y}) }_{1} \cancel{\leq} \underbrace{ \text{HOTA}(\mathbf{X},\mathbf{Z}) }_{0} + \underbrace{ \text{HOTA}(\mathbf{Z},\mathbf{Y}) }_{0},
\end{align}
which means that HOTA does not satisfy the triangle inequality.
\hfill$\square$
\end{proposition}

\begin{proposition}[1-HOTA does not satisfy the triangle inequality]

	Consider three sets of trajectories $\mathbf{X} {=} \{A\}$, $\mathbf{Y} {=} \{A,B\}$ and $\mathbf{Z} {=} \{ B\}$, where $A {=} (0,a)$ and $B {=} (0,b)$ are two trajectories present at time step $k{=}0$ only, with $a$ and $b$ being arbitrary \elementOfTrajectory{s} (bounding boxes) having zero overlap $\mathrm{IoU}(a,b) {=} 0$.
    Since the value of $\alpha$ in HOTA definition (Appendix~\ref{sec:hota}) plays no role in this case, it follows that
    \begin{subequations}
    \begin{align}
        \mathrm{HOTA}(\mathbf{X}, \mathbf{Z}) &= \sqrt{ \tfrac{0}{ 0 + 1 + 1} } &&= 0 \,, \\
        \mathrm{HOTA}(\mathbf{X}, \mathbf{Y}) &= \sqrt{ \frac{ \frac{1}{1 + 0 + 0} }{ 1 + 0 + 1 } } &&\approx 0.707 \,, \\
        \mathrm{HOTA}(\mathbf{Y}, \mathbf{Z}) &= \sqrt{ \frac{ \frac{1}{1 + 0 + 0} }{ 1 + 1 + 0 } } &&\approx 0.707 \,.
    \end{align}
    \end{subequations}
	As a result, 
	\begin{align}
		\!\!
        \underbrace{ d_{\text{HOTA}}(\mathbf{X},\!\mathbf{Z}) }_{=1}
        \cancel{\leq}
        \underbrace{ d_{\text{HOTA}}(\mathbf{X},\mathbf{Y}) }_{ \approx 0.293 }
        {+}
        \underbrace{ d_{\text{HOTA}}(\mathbf{Y},\!\mathbf{Z}) }_{ \approx 0.293}
        , \!\!
	\end{align}
	which means that $d_{\text{HOTA}} (\mathbf{X}, \mathbf{Y})$~\eqref{eq:HOTA-induced-maybe-metric} does not satisfy the triangle inequality. 
	\hfill$\square$
\end{proposition}

\section{HOTA vs. TGOSPA}\label{sec:HOTA:vs:TGOSPA}
Consider five toy examples, Ex1-Ex5, illustrated in Fig.~\ref{fig:mainfig}. 
The corresponding rankings iduced by HOTA and TGOSPA are summarized in Table \ref{tab_tgospa_hota}.
Note that perfect localization is considered in the toy examples, so the threshold parameter $\alpha$ in HOTA plays no role.
Furthermore, $c$ is assumed \emph{small} for the TGOSPA metric so that the outlying estimates Ex2, Ex3, and Ex5 result in false alarms.
In addition, two choices of $\gamma$ for the TGOSPA metric are used so that the \seeminglySwitch{s} in Fig.~\ref{fig:subfig-a} either are considered as switches in the metric (which is desirable for the perfect localization) or not, respectively.

\newcommand{\ExWidthScale}{0.095}
\begin{figure}[h!]
	\centering
	\subfloat[Ex1]{%
		\includegraphics[width=\ExWidthScale\textwidth]{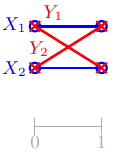} 
		\label{fig:subfig-a}
	}
	\subfloat[Ex2]{%
		\includegraphics[width=\ExWidthScale\textwidth]{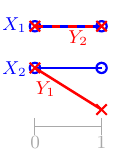} 
		\label{fig:subfig-b}
	}
	\subfloat[Ex3]{%
		\includegraphics[width=\ExWidthScale\textwidth]{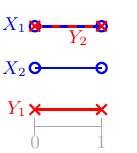} 
		\label{fig:subfig-c}
	}
	\subfloat[Ex4]{%
		\includegraphics[width=\ExWidthScale\textwidth]{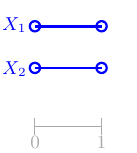} 
		\label{fig:subfig-d}
	}
	\subfloat[Ex5]{%
		\includegraphics[width=\ExWidthScale\textwidth]{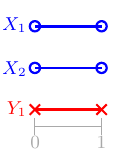} 
		\label{fig:subfig-e}
	}
	\caption{Illustrative toy examples.}
	\label{fig:mainfig}
	\vspace{-0.5cm}
\end{figure}

\textbf{Ex1 (perfect localization and two track switches):}
The TGOSPA metric with $\gamma {<} c$ is $d_p^{(c,\gamma)}(\mathbf{X},\!\mathbf{Y}) {=} 2\gamma$ and two switches are found.
For $\gamma {>} c$, however, it is $d_p^{(c,\gamma)}(\mathbf{X},\!\mathbf{Y}) {=} 4 {\times} \tfrac{c}{2} {=} 2c$ and two pairs of missed and false estimates resut.
The HOTA score for Ex1 is
\begin{align}
	\text{HOTA}(\mathbf{X},\!\mathbf{Y})
	&= \sqrt{\frac{4 \times \frac{1}{1 + 1 + 1}}{2 + 1 + 1}} \approx 0.577.
\end{align}

\textbf{Ex2 (perfect localization, one missed, and one false):}
For this and the following Ex3, Ex4, and Ex5, the TGOSPA metric is independent of the choice of~$\gamma$.
In this case, TGOSPA: $d_p^{(c,\gamma)}(\mathbf{X},\!\mathbf{Y}) {=} 2 {\times} \tfrac{c}{2} {=} c$, whereas
\begin{align}
	\text{HOTA}(\mathbf{X},\!\mathbf{Y}) \!=\! \sqrt{\frac{\frac{1}{1 + 0 + 0} {+} \frac{1}{1 + 0 + 0} {+} \frac{1}{1 + 1 + 1}}{3 + 1 + 1}} \!\approx\! 0.683.
\end{align}

\textbf{Ex3 (perfect localization, two missed, and two false):}
TGOSPA: $d_p^{(c,\gamma)}(\mathbf{X},\!\mathbf{Y}) {=} 4 {\times} \tfrac{c}{2} {=} 2c$, whereas
\begin{align}
	\text{HOTA}(\mathbf{X},\!\mathbf{Y}) = \sqrt{\frac{\frac{1}{1 + 0 + 0} + \frac{1}{1 + 0 + 0}}{2 + 2 + 2}} \approx 0.577.
\end{align}

\textbf{Ex4 (all missed):}
TGOSPA: $d_p^{(c,\gamma)}(\mathbf{X},\!\mathbf{Y}) {=} 4 {\times} \tfrac{c}{2} {=} 2c$, $\text{HOTA} {=} 0$.

\textbf{Ex5 (all missed and two false):}
TGOSPA: $d_p^{(c,\gamma)}(\mathbf{X},\!\mathbf{Y}) {=} 6 {\times} \tfrac{c}{2} {=} 3c$, $\text{HOTA} {=} 0$.

\begin{table}[h]
	\centering
	\caption{Performance ordering of TGOSPA and HOTA for the five toy examples illustrated in Fig.~\ref{fig:mainfig}.}
	\label{tab_tgospa_hota}
	\begin{tabular}{cc}
		\toprule
        & Performance ordering \\ 
		& worse \hspace{0.4cm}$<$\hspace{0.4cm} better \\ \midrule  
		TGOSPA with $\gamma {<} c$ &  Ex5 $<$ Ex4 $=$ Ex3 $<$ Ex2 $<$ Ex1 \\
		TGOSPA with $\gamma {>} c$ &  Ex5 $<$ Ex4 $=$ Ex3 $=$ Ex1 $<$ Ex2 \\
        HOTA    &  Ex5 $=$ Ex4 $<$ Ex3 $=$ Ex1 $<$ Ex2 \\
		\bottomrule
	\end{tabular}
\end{table}

From the HOTA scores in different examples, we can observe that the considered tracker:
\begin{enumerate}
	\item has the same performance in Ex1 and Ex3.
	\item has the best performance in Ex2.
	\item has the worst performance in Ex4 and Ex5.
\end{enumerate}

The ranking of tracking performance based on HOTA in these examples may not always be desirable for different applications. It makes sense that the case with two missed and two false estimates (Ex3) is worse than the case with only one missed and one false estimate (Ex2).
However, \emph{should the case with two \seeminglySwitch{s} (Ex1) be worse than the case with only one missed and one false detection (Ex3)}?
This should be application dependent, and for this to be possible, hyper-parameters should be introduced such that missed detection, false detection, and track switches can be penalized differently.
This is employed in the TGOSPA metric to some level\footnote{
	Remind that missed and false \elementOfTrajectory{s} are both penalized with the same value $\tfrac{c^p}{2}$ in TGOSPA.
}: for small $\gamma{<}c$ Ex1 is better than Ex3 and for large $\gamma{>}c$, Ex1 performs the same as Ex3.
Moreover, we can observe that HOTA yields counterintuitive results in Ex4 and Ex5, as it does not penalize the additional false detections in Ex5.

\section{Alternative Bounding Box Metrics}\label{appendix:alternative-metrics}
To be a valid metric, the function $d$ in TGOSPA must be a metric.

\subsection{Hausdorff Metric}
Consider the sets $x,y {\subset} \mathbb{R}^2$ being non-empty and compact (i.e., closed and bounded).
In general, the collection of non-empty compact subsets of the metric space $(\mathbb{R}^n, d_{\mathbb{R}^n})$ can be made into a metric space\footnote{
	The resulting metric space is also consistent with the \emph{hit-or-miss} topology used in the theory of random closed/finite sets~\cite[p.~138]{Mahler:1997} commonly adopted in the multi-object estimation community.
} by using the Hausdorff metric~\cite[p.~6]{Kendall:1995},\cite[pp.~137-138]{Mahler:1997},
denoted as $d_{\mathrm{H}}( x,y )$.
The Hausdorff metric generalizes the metric $d_{\mathbb{R}^n}$ on $\mathbb{R}^n$ straightforwardly, as for any $\xi,\eta {\in} \mathbb{R}^n$, it is $d_{\mathrm{H}}( \{\xi\}, \{\eta\} ) {=} d_{\mathbb{R}^n} ( \xi, \eta )$.
Using the maximum metric $d_{\mathbb{R}^2}( \xi, \eta ) {=} d_{\infty}( \xi, \eta )$, the Hausdorff metric can be computed easily for bounding boxes (rectangles containing their interiors) as~\cite{Hausdorff:boxes:2004}
\begin{minipage}{\linewidth}
	\begin{align}
		d_{\mathrm{H}}( x,y ) =
		\max\!\Big\{ &\max\{ |\leftEndPoint{x}{1} - \leftEndPoint{y}{1}|, |\rightEndPoint{x}{1} - \rightEndPoint{y}{1}| \}, \notag\\
		&\max\{ |\leftEndPoint{x}{2} - \leftEndPoint{y}{2}|, |\rightEndPoint{x}{2} - \rightEndPoint{y}{2}| \}\, \Big\},
		\label{eq:Hausdorff-metric-BBoxes}
	\end{align}
	\begin{wrapfigure}[7]{l}{0.3\linewidth}
		\centering
		\vspace{-13.0mm}
		\includegraphics[scale=0.83]{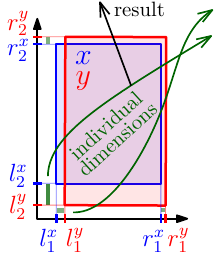}
	\end{wrapfigure}
	where $\leftEndPoint{x}{\dimIndex}$ and $\rightEndPoint{x}{\dimIndex}$ are the left and right end-points of the set $x$ projected to the dimension $\dimIndex$, respectively (analogically for $y$).
	The resulting Hausdorff metric~\eqref{eq:Hausdorff-metric-BBoxes} focuses solely on the mutual discrepancy between the \emph{edges} of the bounding boxes and may thus ill-consider their overall geometric relationship. 
\end{minipage}

\subsection{Wasserstein Metric}
Consider the sets $x,y {\subset} \mathbb{R}^2$ being non-empty and measurable.
Such sets can be understood as supports of probability density functions (PDFs), say, $p_x(\xi)$ and $p_y(\eta)$ with probability measures $P_x$ and $P_y$, respectively.
One can define the Wasserstein metric $d_{\mathrm{W}}(P_x, P_y)$ between the measures, which has a convenient interpretation: \emph{the mean distance needed to transport the mass under one PDF curve into the other}, see~\cite{OptimalTransportBook:2003}.
\begin{minipage}{\linewidth}
	\vspace{1mm}
	\begin{wrapfigure}[8]{l}{0.25\linewidth}
		\centering
		\vspace{-3mm}
		\includegraphics[scale=0.9]{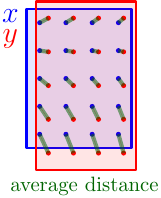}
	\end{wrapfigure}
	Taking uniform distributions and the Euclidean metric $d_{\mathbb{R}^2}(\xi,\eta) {=} d_2(\xi,\eta)$ on $\mathbb{R}^2$, the Wasserstein distance can be computed easily for bounding boxes (rectangles containing their interiors) as derived in~\cite{IrpinoVerde:WassDistBox:2008}.
	The computation is the same as the Euclidean distance for the vector representation $x {=} [\centerPoint{x}{1}\ \centerPoint{x}{2}\ \tfrac{1}{\sqrt{12}}\widthCoord{x}\ \tfrac{1}{\sqrt{12}}\heightCoord{x}]\T$, where $[\centerPoint{x}{1}\ \centerPoint{x}{2}]\T$ is the center-point, $\widthCoord{x}$ is the width and $\heightCoord{x}$ is the height of the bounding box $x$ (analogically for $y$), i.e., as
\end{minipage}
\begin{align}
	\! d_{\mathrm{W}} ( P_x, P_y ) 
	\!=\! \Big[ &\big(\centerPoint{x}{1} - \centerPoint{y}{1} \big)^2 + \tfrac{1}{3} \big( \tfrac{\widthCoord{x}}{2} - \tfrac{\widthCoord{y}}{2} \big)^2 \notag\\
	+ \big(\centerPoint{x}{2} &- \centerPoint{y}{2} \big)^2 + \tfrac{1}{3} \big( \tfrac{\heightCoord{x}}{2} - \tfrac{\heightCoord{y}}{2} \big)^2\ \Big]^{\! \sfrac{1}{2} } \!\!. \!
\end{align}

The above metrics depend on the \emph{shapes}, as well as \emph{sizes} (scales) of the input bounding boxes to the metric.
That is, two boxes that are small (presumably in the background of the image) can be expected to have \emph{smaller} metric value than two boxes that are large (presumably in the foreground) even though both pairs would \emph{look} the same if scaled to the same width and height.
This inconvenience is solved by the IoU-induced metric $d_{\mathrm{IoU}} (x, y)$~\eqref{eq:IoU-induced-metric-d} discussed in Section~\ref{sec:bounding box-metrics}.



\bibliographystyle{ieeetr}
\bibliography{references}

\end{document}